\documentclass[%
 reprint,
superscriptaddress,
 amsmath,amssymb,
 aps,prx
]{revtex4-2}

\usepackage{graphicx}
\usepackage{textalpha}
\usepackage{dcolumn}
\usepackage{bm}
\usepackage[separate-uncertainty = true]{siunitx}
\usepackage{xr} 
\usepackage{titlesec}

\newcommand{\citeSMSoftCoreRepNoBrackets}{Sec.~\ref{secSM:BoltEntropyMaximization}} 
\newcommand{\citeSMMainDerivationsNoBrackets}{Sec.~\ref{secSM:ContinuousInteractions}} 
\newcommand{\citeSMGaussFormNoBrackets}{Sec.~\ref{secSM:particlefluctuations}} 
\newcommand{\citeSMMetropolisHastings}{[Sec.~\ref{secSM:MetropolisSampling}]} 
\newcommand{\citeSMMetropolisHastingsNoBrackets}{Fig.~\ref{figSM:localConditionalProbability} and Sec.~\ref{secSM:MetropolisSampling}} 
\newcommand{\citeSMSelectBioExamplesNoBrackets}{Sec.~\ref{secSM:ParticleLikeStructures}} 
\newcommand{\citeSMMutualInformation}{[Sec.~\ref{secSM:MutualInfoDerivation}]} 
\newcommand{\citeSMMutualInformationNoBrackets}{Sec.~\ref{secSM:MutualInfoDerivation}} 
\newcommand{\citeSMSecAreaFrac}{Sec.~\ref{secSM:influenceOfBindingSiteDensity}} 
\newcommand{\citeSMfBS}{[Sec.~\ref{secSM:influenceOfBindingSiteDensity}]} 
\newcommand{\citeSMFigTwo}{[Fig.~\ref{figSM:fBScomparison}]} 
\newcommand{\citeSMOneD}{\citeSMfBS} 
\newcommand{\citeSMEMestimatesAndTableNoBrackets}{Sec.~\ref{secSM:ImageAnalysisMethods} and Tab.~\ref{tabSM:lestimatesfromEM}} 
\newcommand{\citeSMFigNPCnobrackets}{Fig.~\ref{figSM:FitResultsAllClusters}} 
\newcommand{\citeSMFigNPC}{[Fig.~\ref{figSM:FitResultsAllClusters}]} 
\newcommand{\citeSMTabbioExamplesNoBracket}{Tab.~\ref{tabSM:biologicalExamples}} 
\newcommand{\citeSMTabFitVals}{[Tab.~\ref{tabSM:FittedData}]} 
\newcommand{\citeSMEqMeanFieldPotential}{[Eq.~\eqref{eqSM:FermiDiracE}]} 
\newcommand{\citeSMSecBa}{[Sec.~\ref{secSM:ShortRangeRepInteractions}]} 
\newcommand{\citeSMUxEM}{[Sec.~\ref{secSM:UExM}]} 
\newcommand{\citeSMParamFitting}{[Sec.~\ref{secSM:NPC-MTinteractionEnergy}]} 
\newcommand{\citeSMMTNPCInteractionsandFigThreeNoBrackets}{Fig.~\ref{figSM:FitResultsAllClusters} and Sec.~\ref{secSM:DistNPC}} 

\newcommand{\kB}{k_{\mathrm{B}}}

\newcommand{\epp}{E}
\newcommand{\epe}{\epsilon}
\usepackage{orcidlink}
\usepackage{hyperref}
\hypersetup{
    colorlinks,
    linkcolor={blue},
    citecolor={blue},
    urlcolor={blue}
}
\let\oldparagraph\paragraph
\makeatletter
\renewcommand{\paragraph}{\@startsection{paragraph}{4}{\z@}%
  {-3.25ex \@plus -1ex \@minus -0.2ex}%
  {0ex}%
  {\normalfont\normalsize\itshape}}
\makeatother

\begin{document}
\preprint{APS/123-QED}

\title{Repulsive particle interactions enable selective information processing at cellular interfaces}

\author{J. Elliott\orcidlink{0000-0002-3703-4234}}
\affiliation{Cell Biology and Biophysics Unit, European Molecular Biology Laboratory, Meyerhofstraße 1, 69117 Heidelberg, Germany}
\affiliation{Department of Physics and Astronomy, Heidelberg University, 69120 Heidelberg, Germany}
\author{H. Shah\orcidlink{0000-0002-8143-8244}}
\affiliation{Cell Biology and Biophysics Unit, European Molecular Biology Laboratory, Meyerhofstraße 1, 69117 Heidelberg, Germany}
\author{R. Belousov\orcidlink{0000-0002-8896-8109}}
\affiliation{Cell Biology and Biophysics Unit, European Molecular Biology Laboratory, Meyerhofstraße 1, 69117 Heidelberg, Germany}
\author{G. Dey\orcidlink{0000-0003-1416-6223}}
\affiliation{Cell Biology and Biophysics Unit, European Molecular Biology Laboratory, Meyerhofstraße 1, 69117 Heidelberg, Germany}
\author{A. Erzberger\orcidlink{0000-0002-2200-4665}}%
\email{erzberge@embl.de}
\affiliation{Cell Biology and Biophysics Unit, European Molecular Biology Laboratory, Meyerhofstraße 1, 69117 Heidelberg, Germany}
\affiliation{Department of Physics and Astronomy, Heidelberg University, 69120 Heidelberg, Germany}
\date{\today}

\begin{abstract}

Living systems relay information across membrane interfaces to coordinate compartment functions. We identify a physical mechanism for selective information transmission that arises from the sigmoidal response of surface-bound particle densities to spatial features in adjacent external structures through a non-uniform binding energy. This mechanism implements a form of spatial thresholding, enabling the binary classification of external cues. Expansion microscopy measurements of nuclear pore complex distributions in \emph{S. arctica} show signatures of such physical thresholding.
\end{abstract}

\maketitle

Living systems process information to adapt and respond to their environment~\cite{adamalaPresentFutureSynthetic2024,shannonNeumannsContributionsAutomata1958, farnsworthLivingInformationProcessing2013,Graf2024,Dullweber2023}. 
Theoretical principles underlying biological information processing have been identified in gene regulation~\cite{udaApplicationInformationTheory2020,estradaInformationIntegrationEnergy2016}, biochemical signaling~\cite{collinetProgrammedSelforganizedFlow2021,ganesanHowCellsProcess2012,cheongInformationTransductionCapacity2011,brennanHowInformationTheory2012,kramar2024singlecellsresolvegraded,Nandan2022}, and cell-fate patterning ~\cite{falascoInformationThermodynamicsTuring2018,avanziniNonequilibriumThermodynamicsNonIdeal2024,brucknerInformationContentOptimization2023,wolpertPositionalInformationPattern1994,dubuisPositionalInformationBits2013,tkacikManyBitsPositional2021,mcgoughFindingLastBits2024,flowersRememberingWhereWe2020}. Yet beyond molecular circuits, \emph{physical} interactions, mechanical properties, and geometrical relations also impact how living materials process information~\cite{zieskeReconstitutionSelforganizingProtein2014,wigbersHierarchyProteinPatterns2021,wettmannMinproteinOscillationsEscherichia2018,hauptHowCellsSense2018,Rangamani2013,Evans2024,Mijatovic2025,BrazTeixeira2024,Banerjee2025,Rombouts2023}. Physical degrees of freedom influence how cells navigate complex environments~\cite{Sitarska2023,Gomes2005,Vercruysse2024,dAlessandro2021,Ron2024}, internalize pathogens~\cite{Han2025,Baranov2021,Stow2016}, or coordinate subcellular processes~\cite{Cho2017,wigbersHierarchyProteinPatterns2021,burkartControlProteinbasedPattern2022}. For example, migrating cells respond to gradients in the elastic properties of structures in their environment (durotaxis \cite{Shellard2020,Clark2022}), or the density of binding sites for specific adhesion molecules (haptotaxis, \cite{CARTER1967,Luo2020,Shellard2020,Fortunato2024}) using biophysical interactions between adhesion molecules, cell membrane, cytoskeleton, and external binding sites to read out information about external spatial heterogeneities. However, the principles underlying such physically mediated signal processing are not well understood~\cite{Leiphart2018,Ji2023}. In particular, how physical interactions shape the encoding and transmission of  spatially-resolved mechanical signals is unclear.

\begin{figure}[!t]
\includegraphics[width=\columnwidth]{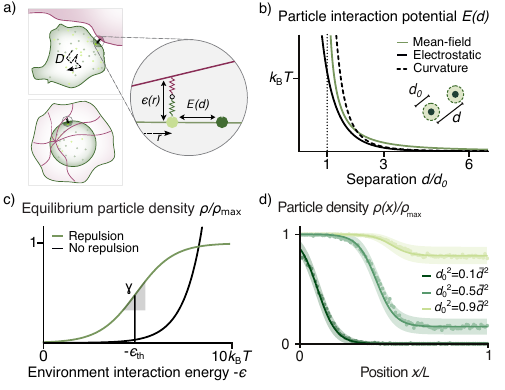}
\caption{\label{fig:ModelIntro} The density of repulsive particles responds nonlinearly to changes in binding energy. a) Particles on the surface of a membrane-enclosed compartment interact via a potential $\epp(d)$ that depends on their average separation distance $d$. Each particle can be either bound (light green circles) to an external structure (purple) or unbound (dark green circles), with binding rates set by a non-uniform interaction energy field $\epe$.
b) The mean-field potential $\epp\bigl(d\bigr)$ captures short-range repulsion due to steric effects between particles of diameter $d_0$, similar to curvature-mediated interactions and shielded electrostatic repulsion~\cite[\citeSMSoftCoreRepNoBrackets{}]{Yolcu2014,Karal2023,Liboff1959,Rowlinson1989,Milo2015Avesizeprotein,Zhdanov2009,Hu2012,Wennerstrm2020}. c) Repulsive particle interactions lead to a sigmoidal mapping between the particle-environment interaction energy $\epe$ and the particle density $\rho$, approaching the maximum density $\rho_{\text{max}}$ in contrast to ideal, non-repulsive particles. Vertical line and shaded triangle show the threshold $\epe_{\rm{th}}$ and gain $\gamma$ of the sigmoid. d) For an interaction energy $\epe(x) = \kB T(20 x/L-10)$ on a domain of length $L$, this mapping produces sigmoidal equilibrium density fields (points from Metropolis-Hastings sampling~\citeSMMetropolisHastings{}, shading denotes two standard deviations).}
\end{figure}

In this letter, we focus on physical interactions of surface-embedded particles at the interfaces of biological, membrane-enclosed systems [Fig.~\ref{fig:ModelIntro}(a)]. We investigate how the spatial distribution of such particles selectively encodes information stored in the features of surrounding external structures, e.g. the proximity of binding sites, which can subsequently change internal cellular states such as polarity ~\cite{Rappel2017}. We identify how nonlinear relations due to particle interactions control the flow of information across the interface, whereby the interface itself functions as an information-processing layer in a hierarchical architecture.

Specifically, we find that particle interactions of the simplest type---short-range repulsion---create sigmoidal mappings between an input potential field and output particle distribution, providing a physical mechanism for \emph{binary thresholding} reminiscent of thresholding filters in computer vision~\cite{Seelaboyina2023}. By deriving explicit expressions for the \textit{gain} and \textit{threshold} parameters of the filter, we identify how the system's physical properties, i.e. the effective size of the surface-embedded particles and their typical concentrations, control the transfer of information. We identify an optimal information-processing regime and find that diverse interface-associated protein complexes and macromolecular structures in cells operate within this regime. In particular, we experimentally measure the predicted sigmoidal distribution of nuclear pore complexes in the envelope of \textit{Sphaeroforma arctica} nuclei~\cite{Dudin2019}, which form in response to interactions with the surrounding cytoskeleton~\cite{Shah2024}. Our results indicate that physical thresholding could facilitate information transmission in the coordination of biological processes across cellular and subcellular interfaces.

\paragraph*{\label{sec:level1}Repulsive particle interactions binarize binding energy profiles.}---We consider diffusive particles on a membrane, which interact with the environment by attaching to and detaching from adjacent binding sites. The densities of particles $\rho_{i}$ in the bound ($i=\mathrm{b}$) and unbound ($i = \mathrm{u}$) states, with state energies $\epe_i$, evolve according to mass-conserving reaction-diffusion equations~\cite{Halatek2018,Brauns2020,Gardiner2010-zc}
\begin{equation}\label{eq:Generaldiffusion}
    \partial_t \rho_i =\nabla\cdot\left[
        D_i\nabla\rho_i
        +\beta D_i\rho_i\nabla(\epp+\epe_i)
    \right]+ \mathcal{R}_i
\end{equation}
driven by a mean-field interaction potential $\epp(\rho)$, which depends on the total density $\rho = \rho_{\rm{b}}+\rho_{\rm{u}}$, with the inverse thermal energy scale $\beta = (\kB T)^{-1}$ and, in general, density-dependent diffusion coefficients $D_i$~\cite{Belousov2022}. The reaction terms describe transitions between the two states according to $\mathcal{R}_{\rm{u}} = -\mathcal{R}_{\rm{b}} =  k_{\rm{off}} \rho_{\rm{b}} -k_{\rm{on}} \rho_{\rm{u}}$. To investigate how spatial heterogeneities in the environment influence the particle distribution, we consider nonuniform attachment rates $k_{\rm{on}} = k_{\rm off}e^{-\beta \epe}$, in which $\epe(\mathbf{r})=\epe_{\rm b}(\mathbf{r})-\epe_{\rm u}$ is a particle-environment binding energy field, with $\epe_{\rm u}$ generally assumed spatially uniform.

From Eq.~\eqref{eq:Generaldiffusion} and assuming no-flux boundary conditions, we obtain the \textit{equilibrium} total density~\cite[\citeSMMainDerivationsNoBrackets{}]{Carter2000}
\begin{equation}\label{eq:EquilibriumBolzmannSolution}
    \rho =\frac{1}{l^2} e^{-\beta\epp(\rho)}\Big(1+e^{-\beta\epe(\mathbf{r})}\Big),
\end{equation}
where the integration constant $l$ is a length-scale set by the conservation of the total particle number $N$ over the whole membrane area $A$ according to
\begin{equation}\label{eq:normalization}
    N = \int_A \mathrm{d}A\, \rho.
\end{equation}
This constraint determines the average separation distance between particles 
across the whole surface ${\bar{d}=\sqrt{A/N}}$. 
Equation~\eqref{eq:EquilibriumBolzmannSolution} describes how the distribution of particles on the membrane surface responds to heterogeneities in mechanical and/or chemical properties of external structures, which give rise to a non-uniform $\epe(\mathbf{r})$. For example, an effective Hookean interaction energy could characterize the interactions of membrane-associated particles with external structures at a variable separation distance such that the non-uniform interaction energy captures spatial variations in binding site proximity. 

Interparticle interactions influence how the density~\eqref{eq:EquilibriumBolzmannSolution} responds to such external heterogeneities. Focusing on steric interactions~\citeSMSecBa{}, we introduce an effective particle size $d_0$, imposing that each particle occupies an area $d_0^2$, and obtain the mean-field potential \citeSMEqMeanFieldPotential{}
\begin{equation}\label{eq:epp}
    \epp(\rho)=- k_B T\ln(1-d_0^2\rho)
\end{equation}
which corresponds to the chemical potential difference between a volume-excluding lattice gas and the ideal gas ~\cite{Landau1996-qq,Hill1987-fu}. By construction, the length-scale $d_0$ determines the maximum possible density of the particles $\rho_{\text{max}} = 1/d_0^2 $. Although Equation~\eqref{eq:epp} assumes a volume-excluding lattice, the resulting potential also approximates well short-range repulsion arising e.g. from curvature-mediated or shielded electrostatic interactions [Fig.~\ref{fig:ModelIntro}(b)]. 

The equilibrium particle distribution resulting from Eqs.~\eqref{eq:EquilibriumBolzmannSolution} and \eqref{eq:epp} 
\begin{equation}\label{eq:EquilibriumFermiDiracSolution}
    \rho(\mathbf{r}) = \frac{1+e^{-\beta\epe(\mathbf{r})}}{l^2+d_0^2(1+e^{-\beta\epe(\mathbf{r})})},
\end{equation} 
corresponds to a sigmoidal mapping between the binding energy field $\epe(\mathbf{r})$ and the density $\rho(\mathbf{r})$ [Fig.~\ref{fig:ModelIntro}(c)].
In fact, we recover a Fermi-Dirac-like distribution, where the particle density at different positions is akin to the occupation number of energy levels, which are resolved along the surface according to $\epe(\mathbf{r})$: particles occupy the most energetically favorable spatial positions, given steric interactions and subject to mass-conservation, leading to regions of high and low density.  Therefore, the resulting particle distribution sigmoidally \emph{filters} external heterogeneities, providing a physical mechanism for binarizing the binding-energy field. A cellular or subcellular interface could thereby read out spatial variations in, for example, the proximity of nearby structures.

The sigmoidal relation Eq.~\eqref{eq:EquilibriumFermiDiracSolution} is characterized by the gain, $\gamma := \text{max}\left( \left\|\partial \rho/\partial \epe \right\|\right) = \beta l^2/[4d_0^2(d_0^2+l^2)]$, and the threshold, $\epe_{\rm th}:= \text{argmax}\left( \left\|\partial \rho/\partial \epe\right\|\right) =k_BT\ln{\left[d_0^2/(d_0^2+l^2)\right]}$, [Fig.~\ref{fig:ModelIntro}(c)],
which depend on the two relevant length scales, the effective particle size  $d_0$ and the average particle separation $\bar{d}$, set by the total number of particles $N$. 
Note that through the integral over the entire membrane surface,  constraint Eq.~\eqref{eq:normalization} incorporates a global dependence of the particle distribution on the interaction energy field into the normalization constant $l$, whose explicit expression follows $l \propto \sqrt{\bar{d}^2-d_0^2}$ for uniform $\epe$.

\paragraph*{Particle size and total number tune information transmission.}---Could cells use Eq.~\eqref{eq:EquilibriumFermiDiracSolution} to sense the proximity of nearby structures? 
To investigate how the distance between a membrane patch and an adjacent set of binding sites can be conveyed by the particle density, we analyze the transmission of information in the presence of fluctuations. 

In general, a noisy sigmoidal response can amplify an input signal close to the threshold, while rendering small energy variations unresolvable far from the threshold [Fig.~\ref{fig:ModelIntro}(d)]. Repulsive particle interactions may thereby enable the \emph{selective} transmission of information about interaction energies close to the threshold. When suitably optimized, such an information bottleneck can pick up specific task-relevant features of the input~\cite{Tishby2000,Bauer2023,Bauer2022,Kleinman2023}. By controlling the gain and threshold, the biophysical parameters $\bar{d}$ and $d_0$ determine the amplified signal range, similar to the tunable resistors that shape the nonlinear mapping between input and output voltages in electronic audio processing~\cite{Mancini2003}.

\begin{figure}[!t]
\includegraphics[width=0.5\textwidth]{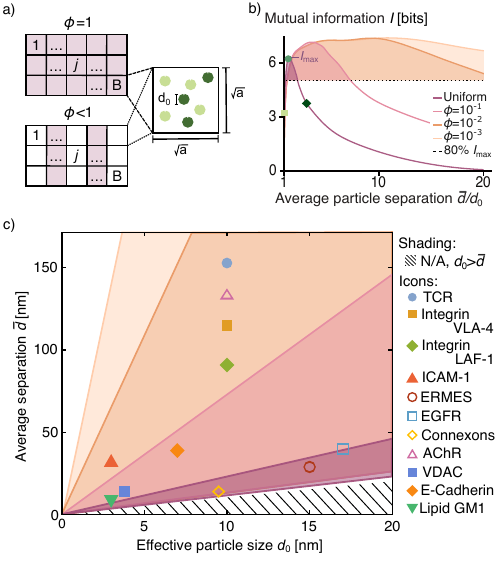}
\caption{\label{fig:SimAndContinuousResults} Repulsive particle interactions threshold binding energy profiles. a) We discretize the surface into $B$ boxes of area $a$ either with (purple shading) or without (no shading) binding sites, with $\phi$ the area fraction of the binding region. 
b) The mutual information between the external field and the particle density profile shows a maximum as a function of $\bar{d}/d_0$, where the gain is high and the threshold energy is kept within the input range (computed for a system discretized into $B=5$ boxes with $a=10d_0^2$ as described in \citeSMMutualInformation). The optimal information transmission regime---in which $I$ is within 80\% of its maximum---becomes larger as $\phi$ is decreased. Icons label parameter values used in Fig.~\ref{fig:ModelIntro}(d). c) 
The optimal information transmission regime is much larger for an actin-like area fraction $\phi= 10^{-2}$~\cite[\citeSMSelectBioExamplesNoBrackets{}]{Lembo2023} (dark orange) than for uniform 2D sheets (dark purple). 
Macromolecular complexes in cells (colored icons, values from~\cite[\citeSMTabbioExamplesNoBracket{}]{Robert2021,Ekerdt2013,TruongQuang2013,Janeway2001,Jiang2020,Swamy2008,Ma2022,Bui2020,KurzIsler1992,Goodenough1970,Kuntze2020,Mulhall2023,Rizzo2003,Thomsen2002,Mannella2021,Hiller2010,Wozny2023,Lyu2024,Mojumdar2019,Evans1987,McMahon1994,Lo1982,Geng2009,Kerntke2020,Wacleche2018,Saji1999,Zhang2015,Abulrob2010,Horzum2014,TruongVo2017,Horzum2014,Erickson2009}) indeed often associate with filamentous structures, such as the actin cytoskeleton (e.g. AChR, E-Cadherin). 
}
\end{figure}

To investigate how these parameters influence the compression of the original input, we first analyze the channel noise arising due to the microscopic particle dynamics. Discretizing space into boxes of area $a$ indexed by $j=1,2,...,B$ allows representing the particle and energy fields as random vectors which take realizations $\{\rho_j\}$ and $\{\epe_j\}$ from a finite set of density values and energy states [Fig.~\ref{fig:SimAndContinuousResults}(a)]. Considering the limit of large particle numbers in a coarse-graining area element $a$, we approximate the conditional probability $P(\rho_j|\{\epe_k\})$ of a realization $\rho_j$ given a set of binding energies by a Gaussian distribution with mean $\bar{\rho}_j\bigl(\{\epe_k\}\bigr)=\rho\bigl(\epe_j,l(\{\epe_k\})\bigr)$ given by Eq.~\eqref{eq:EquilibriumFermiDiracSolution}, and standard deviation $\sigma_\rho(\bar{\rho}_j(\{\epe_k\})) = \sqrt{(1-d_0^2\bar{\rho}_j(\{\epe_k\}))\bar{\rho}_j(\{\epe_k\})/a}$, such that the probability is conditioned on a realization of the entire binding energy vector through the parameter $l$~\cite[\citeSMGaussFormNoBrackets{}]{Touchette2009}. Metropolis-Hastings sampling of particle distributions numerically confirm these results~\cite[\citeSMMetropolisHastingsNoBrackets{}]{ElliottRepoMetropolisSims2025}.

To evaluate the level of selective signal amplification by the particle interactions, we compute the mutual information between the random vectors $\{\rho_j\}$ and $\{\epe_k\}$, given by 
\begin{equation}\label{eq:MutualInfo}
    I = \sum_{\{\epe_k\}}\sum_{\{\rho_j\}} P(\{\rho_j\},\{\epe_k\})\ln{\frac{P(\{\rho_j\},\{\epe_k\})}{P(\{\rho_j\})P(\{\epe_k\})}},
\end{equation}
in which the sums run over all possible realizations of the input and output sequences~\cite[Chapter 2]{TM_Cover1991-pa}[\citeSMMutualInformationNoBrackets{}]. The mutual information quantifies how well an output density field distinguishes different input fields. As such, rather than specifying one particular $\{\epe_k\}$, we must consider the probability distribution over all possible input fields, $P(\{\epe_k\})$, which takes a system-specific form depending on the statistics of binding energy variations across different environments. We consider in the following a uniform distribution over all realizations $\{\epe_k\}$ from a finite input range $\epe_{\rm min}<\epe_{\rm th}<\epe_{\rm max}$, where the minimum binding energy $\epe_{\rm min}$ corresponds to the most favorable energy level, and the maximum binding energy $\epe_{\rm max}$ corresponds to the least favorable input state, such that $P(\{\epe_k\})=\prod_kP(\epe_k) = N_\epe^{-B}$, where $N_\epe$ is the number of discrete energy values from which the elements of $\{\epe_k\}$ are sampled. The joint probability $P(\{\rho_j\},\{\epe_k\})$ is the product of $P(\{\epe_k\})$ with the conditional probability $P(\{\rho_j\}|\{\epe_k\})=\prod_j P(\rho_j|\{\epe_k\})$---where $P(\rho_j|\{\epe_k\})$ is approximated as discussed above. Numerical evaluation of Eq.\eqref{eq:MutualInfo} for a range of parameter values reveals an optimal regime of particle-mediated physical thresholding [Fig.~\ref{fig:SimAndContinuousResults}(b),~\citeSMMutualInformationNoBrackets{}]. Filters within this regime have gains large enough to overcome the channel noise and threshold energies within the input range.

In addition to the system parameters $d_0$ and $\bar{d}$, this optimal regime also depends on the spatial organization of interaction sites, which in biological systems are often irregular. While for uniform binding-site densities below the maximum particle density, the maximum transmitted information is reduced~\citeSMFigTwo{}, some nonuniform binding-site distributions can increase information transmission, and extend the optimal regime to a broader range of parameters. To identify, in particular, how information transmission is influenced when particles bind to structures such as cytoskeletal filaments, we consider binding sites confined to parallel lines that cover a fraction $\phi$ of the total membrane surface~\citeSMOneD{}~\cite{Hohmann2019}.  We find that lowering $\phi$ allows effective information transfer with fewer particles in the 2D membrane compared to the case of uniform binding sites [Fig.~\ref{fig:SimAndContinuousResults}(b)], because~--~so long as the readout mechanism preserves the reduced dimensionality of the binding site geometry~--~the non-binding region acts as a particle bath that buffers the particles used in the binding regions. As such, decreasing $\phi$ has a similar impact as decreasing $\bar{d}$, i.e. increasing the number of particles in the membrane [Fig.~\ref{fig:SimAndContinuousResults}(b)]. In cellular contexts, protein \emph{line}-densities associated with such quasi-1D structures indeed influence many subcellular processes such as the generation of tension by motor proteins along filaments, or molecular events at cell-cell contact lines~\cite{Cho2022,ResnikDocampo2016,Howard2009,Sun2023,Bosveld2016}. 

These optimal parameter regions for information transmission raise the question of where real surface-associated protein complexes and other macromolecular structures in biological cells fall. Bioimaging technologies are starting to overcome the challenges associated with visualizing \si{\nano\meter}-scale spatial arrangements of proteins along interfaces \cite{Norman2024,Radmacher2025}, and, in some cases, effective particle sizes and typical cellular concentrations have been reported or can be inferred from other measurements~[\citeSMTabbioExamplesNoBracket{}]. 
Indeed, we find that proteins known to associate with actin fibers or bundles (AChR~\cite{Dai2000,Chen2014,Xing2019}, E-Cadherin~\cite{Troyanovsky2025}, TCR \cite{Dustin2000}, ICAM-1 \cite{Celli2006,Carpn1992})
fall within the region in which information transmission from filamentous networks with an actin-like area fraction is optimal, whereas some proteins known to bind between neighboring cell membranes (connexons~\cite{Kirichenko2021,Alberts2002CellJunctions}) or subcellular membraneous structures (ERMES~\cite{Wozny2023,Cheema2021})
have sizes and surface densities that position them where information transmission is highest for uniform interaction sites~[Fig.~\ref{fig:SimAndContinuousResults}(c), dark orange and dark purple regions respectively, \citeSMSelectBioExamplesNoBrackets{}]. Precisely characterizing the information transmission capabilities of a particular system requires experimental measurements of the binding site geometry for a given readout mechanism. 

\paragraph*{Physical thresholding at the nuclear envelope.}---For large macromolecular structures, only a relatively small number of particles is required to form extended maximum-density regions. As a specific experimental example, we focus on nuclear pore complexes (NPCs)\cite{dultzNuclearPoreComplex2022}. These protein complexes, approximately \SI{100}{\nano\meter} in diameter, are embedded within the nuclear envelope---an intracellular membrane that forms the physical and regulatory interface between the cytoplasm and the nuclear interior \cite{Bahmanyar2020,Balaji2022}. In addition to controlling nucleocytoplasmic transport, NPCs interact with the extranuclear cytoskeleton to regulate intranuclear states, including gene expression ~\cite{Soheilypour2016,Biedzinski2020,Geng2023,Maizels2015}. Yet how the spatial distribution of NPCs influences information transfer between cellular and nuclear states remains unclear~\cite{Nair2025,Goldberg2017,Huang2023,Guo2015,Hoffman2020,Donnaloja2019}. 

\begin{figure}
\includegraphics{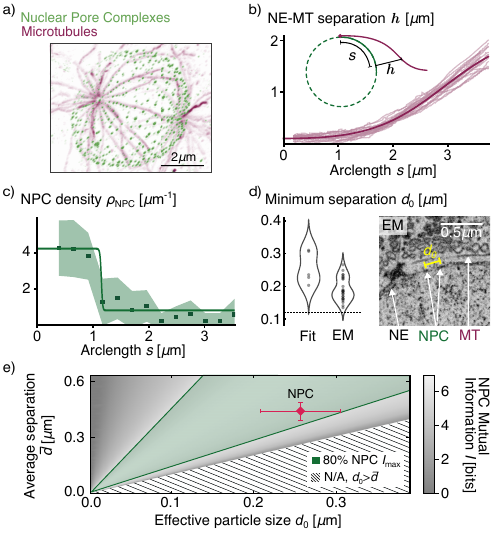}
\caption{\label{fig:NPCExample} Nuclear pore complexes (NPCs) form sigmoidal line-densities along extranuclear microtubule filaments in \textit{S. arctica} nuclei. 
a) Ultrastructure expansion microscopy shows the spatial organisation of immunofluorescently labelled microtubules (purple) and NPCs (green) on an \emph{S. arctica} nucleus (maximum intensity projection)~\citeSMUxEM{}. 
b) The shortest separation distance between the nuclear envelope (NE) and the microtubule (MT) filaments increases as a function of the arc length $s$ away from the microtubule organizing center at the pole. Light purple: individual tracks for one of five clusters. Dark purple: rational-exponential
reparameterization of these tracks~\citeSMParamFitting{}. 
c) Fitting (solid line) the measured NPC line-density (dots) for the MT cluster in (b) with Eq.~\eqref{eq:EquilibriumFermiDiracSolution} yields parameter estimates for the effective spring constant $\lambda$ and for the minimum NPC separation distance $d_0$~\citeSMTabFitVals{}. Shaded area: 95\% confidence interval. Other fitted profiles shown in~\citeSMFigNPC{}. 
d) Left: Fit values for the minimum NPC separation $d_0$ compare well with independent measurements of $d_0$ from electron  microscopy (EM) imaging, dashed line denotes the diameter of the complex \SI{120}{\nano\meter}. Right: an example EM image with labeled features.  
e) The fitted $d_0$ value, and the average NPC separation distance $\bar{d}$ measured from the data, indicate that NPC distributions can efficiently threshold microtubule proximity, given the measured filament density in this system $\phi=0.21\pm0.03$. 
}
\end{figure}

We measure how nuclear pore complex distributions respond to different configurations of the extranuclear microtubule cytoskeleton using expansion microscopy and immunostaining of \emph{S. arctica} nuclei [Fig.~\ref{fig:NPCExample}(a),~\citeSMMTNPCInteractionsandFigThreeNoBrackets{},\cite{ImageDataset}]~\cite{Shah2024,ElliottRepoImageAnalysis2025}. Quantifying the separation distance $h$ between individual microtubule filaments and the nuclear surface reveals increasing height profiles from their anchor point at the pole towards the nucleus equator [Fig.~\ref{fig:NPCExample}(b)].
We estimate the NPC density profiles along the filament arc length by taking averages across clusters of similar filaments and fitting Eq.~\eqref{eq:EquilibriumFermiDiracSolution} to this data, assuming that the effective binding energy profiles arise from the combination of a fixed interaction energy, $\epsilon_{\rm NPC} = 25 k_{\rm B}T$, and an effective Hookean elastic contribution, such that the binding energy field is given by  $\epe(s) = \epsilon_{\rm NPC}+ \lambda_{\rm{NPC}} (h-h_0)^2/2$, with effective spring constant $\lambda_{\rm{NPC}}$ and resting spring length $h_0 = \SI{84(15)}{\nano\meter}$ (measured independently from electron microscopy images) [Fig.~\ref{fig:NPCExample}(c),~\citeSMFigNPCnobrackets{}]\cite{Li2016}. This assumption implies that heterogeneities in the interactions between NPCs and microtubules arise in this system primarily due to differences in the proximity of filaments to the nuclear envelope. 
We thereby obtain estimates for the effective NPC size $d_{0,\text{NPC}} = \SI{260(50)}{\nano\meter}$, and the spring constant characterizing the elastic interaction between NPCs and microtubule filaments $\lambda_{\rm{NPC}} = \SI{0.04 (0.03)}{\pico\newton\per\nano\meter}$\citeSMTabFitVals{}. Our estimates of the minimal NPC separation distance---which we further corroborated using focused ion beam-scanning electron microscopy imaging---are larger than their diameter, suggesting additional non-steric repulsion effects, for example due to curvature-mediated interactions arising from nuclear-envelope deformations close to the NPCs [Fig.~\ref{fig:NPCExample}(d)]~\cite[ \citeSMEMestimatesAndTableNoBrackets{}]{Schindelin2012,Shah2024}. 

Computing the transmitted information given a surface area fraction corresponding to that of the MT filaments at the nuclear envelope in this system ($\phi = 0.21$) reveals that their effective size and total number allows these NPCs to efficiently threshold the MT proximity profiles [Fig.~\ref{fig:NPCExample}(e),~\citeSMSecAreaFrac].   

Together, these results suggest that NPCs form sigmoidal density profiles along microtubule filaments due to their large effective size and fixed number~\cite{Daigle2001,Rabut2004,Toyama2012,Varberg2022,Shah2024}. 
We propose that the thresholded readout of filament proximity could coordinate intranuclear functions, consistent with observations that microtubule–nucleus interactions influence e.g. chromatin organisation, gene expression, and mitotic remodelling~\cite{Roth2007,roth2011,Kelley2024,Gerlitz2012,VelasquezCarvajal2024,Maiato2004,Shah2024,Dudin2019}.

\paragraph*{Conclusions and outlook}---In summary, we find that simple physical interactions at the interfaces of membrane-enclosed cellular compartments lead to the selective transmission of information, permitting the binary classification of surface regions. 
In particular, we report that short-range repulsion between membrane-embedded particles, such as arising from shielded electrostatic repulsion~\cite{Johannes2018} or membrane-mediated interactions~\cite{Shrestha2022,Barakat2022,Yolcu2014}, produces a sigmoidal mapping between non-uniform external energies and the particle densities that form in response. We show how the nonlinear amplification of environmental heterogeneities is controlled by the effective interaction range of the particles and their total number relative to the surface size, and we identify a regime of optimal information transmission for this physical thresholding, as determined by the gain and threshold of the sigmoid. Indeed, many surface-associated subcellular structures fall within the optimal regime based on their sizes and typical cellular concentrations, and our own measurements of \emph{S. arctica} nuclear pore complex distributions reveal sigmoidal density profiles associated with extranuclear microtubule filaments. Interestingly, membrane-associated proteins often interact with irregularly shaped structures such as filamentous networks leading to quasi-1D interactions that facilitate the capacity for physical thresholding~\cite{perez-salaEditorialIntermediateFilaments2022,albertsSelfAssemblyDynamicStructure2002}. How coupling between structures of different effective dimensions influences biological processes and functions has also been investigated in the context of chemical reaction networks~\cite{braunsBulksurfaceCouplingIdentifies2021,Würthner2022} and macromolecular assembly~\cite{banterleKineticStructuralRoles2021}. 

Thresholding filters are used in control circuits, bandpass filters, and neural networks, e.g. to introduce nonlinearities, and compress outputs~\cite{Dubey2022,Astrom2008}. It will be interesting to investigate how sigmoidal particle distributions interact with downstream cellular processes, aiding pattern recognition-like functions of compartment surfaces. The selective transmission of information at the cell surface could guide for example directed movements in response to external gradients in mechanical or chemical properties.

Analyzing other types of particle interactions could reveal whether more complex transformations are possible that would permit operations such as edge detection or object recognition. We anticipate that our increasing technical ability to measure molecular patterns in cells at the sub-\si{\micro\metre} scale will enable the discovery of new physical information processing modalities through which cells and subcellular structures perceive their complex surroundings.

\begin{acknowledgments}
\textit{Acknowledgements}.---We thank Tim Dullweber, Ian Estabrook, Isabella Graf, Pamela Guruciaga, Patrick Jentsch, Johannes Jung, Thomas Quail, Jan Rombouts, Sacha Sokoloski, and Laeschkir W\"urthner for useful discussions and feedback on the manuscript, and Omaya Dudin for his advice and expertise. All authors acknowledge funding from the EMBL. H.S. was supported by the EMBL Interdisciplinary Postdoctoral Fellowship (EIPOD4) programme under Marie Sklodowska-Curie Actions Cofund (grant agreement no. 847543). G.D. and H.S. are supported by the European Union (ERC, KaryodynEvo, 101078291). 
\end{acknowledgments}

\bibliography{references}

\providecommand{\noopsort}[1]{}\providecommand{\singleletter}[1]{#1}%
\begin{thebibliography}{167}%
\makeatletter
\providecommand \@ifxundefined [1]{%
 \@ifx{#1\undefined}
}%
\providecommand \@ifnum [1]{%
 \ifnum #1\expandafter \@firstoftwo
 \else \expandafter \@secondoftwo
 \fi
}%
\providecommand \@ifx [1]{%
 \ifx #1\expandafter \@firstoftwo
 \else \expandafter \@secondoftwo
 \fi
}%
\providecommand \natexlab [1]{#1}%
\providecommand \enquote  [1]{``#1''}%
\providecommand \bibnamefont  [1]{#1}%
\providecommand \bibfnamefont [1]{#1}%
\providecommand \citenamefont [1]{#1}%
\providecommand \href@noop [0]{\@secondoftwo}%
\providecommand \href [0]{\begingroup \@sanitize@url \@href}%
\providecommand \@href[1]{\@@startlink{#1}\@@href}%
\providecommand \@@href[1]{\endgroup#1\@@endlink}%
\providecommand \@sanitize@url [0]{\catcode `\\12\catcode `\$12\catcode
  `\&12\catcode `\#12\catcode `\^12\catcode `\_12\catcode `\%12\relax}%
\providecommand \@@startlink[1]{}%
\providecommand \@@endlink[0]{}%
\providecommand \url  [0]{\begingroup\@sanitize@url \@url }%
\providecommand \@url [1]{\endgroup\@href {#1}{\urlprefix }}%
\providecommand \urlprefix  [0]{URL }%
\providecommand \Eprint [0]{\href }%
\providecommand \doibase [0]{https://doi.org/}%
\providecommand \selectlanguage [0]{\@gobble}%
\providecommand \bibinfo  [0]{\@secondoftwo}%
\providecommand \bibfield  [0]{\@secondoftwo}%
\providecommand \translation [1]{[#1]}%
\providecommand \BibitemOpen [0]{}%
\providecommand \bibitemStop [0]{}%
\providecommand \bibitemNoStop [0]{.\EOS\space}%
\providecommand \EOS [0]{\spacefactor3000\relax}%
\providecommand \BibitemShut  [1]{\csname bibitem#1\endcsname}%
\let\auto@bib@innerbib\@empty
\bibitem [{\citenamefont {Adamala}\ \emph {et~al.}(2024)\citenamefont
  {Adamala}, \citenamefont {Dogterom}, \citenamefont {Elani}, \citenamefont
  {Schwille}, \citenamefont {Takinoue},\ and\ \citenamefont
  {Tang}}]{adamalaPresentFutureSynthetic2024}%
  \BibitemOpen
  \bibfield  {author} {\bibinfo {author} {\bibfnamefont {K.~P.}\ \bibnamefont
  {Adamala}}, \bibinfo {author} {\bibfnamefont {M.}~\bibnamefont {Dogterom}},
  \bibinfo {author} {\bibfnamefont {Y.}~\bibnamefont {Elani}}, \bibinfo
  {author} {\bibfnamefont {P.}~\bibnamefont {Schwille}}, \bibinfo {author}
  {\bibfnamefont {M.}~\bibnamefont {Takinoue}},\ and\ \bibinfo {author}
  {\bibfnamefont {T.-Y.~D.}\ \bibnamefont {Tang}},\ }\bibfield  {title}
  {\bibinfo {title} {Present and future of synthetic cell development},\ }\href
  {https://doi.org/10.1038/s41580-023-00686-9} {\bibfield  {journal} {\bibinfo
  {journal} {Nature Reviews Molecular Cell Biology}\ }\textbf {\bibinfo
  {volume} {25}},\ \bibinfo {pages} {162} (\bibinfo {year} {2024})}\BibitemShut
  {NoStop}%
\bibitem [{\citenamefont
  {Shannon}(1958)}]{shannonNeumannsContributionsAutomata1958}%
  \BibitemOpen
  \bibfield  {author} {\bibinfo {author} {\bibfnamefont {C.~E.}\ \bibnamefont
  {Shannon}},\ }\bibfield  {title} {\bibinfo {title} {Von {Neumann}’s
  contributions to automata theory},\ }\href
  {https://doi.org/10.1090/S0002-9904-1958-10214-1} {\bibfield  {journal}
  {\bibinfo  {journal} {Bulletin of the American Mathematical Society}\
  }\textbf {\bibinfo {volume} {64}},\ \bibinfo {pages} {123} (\bibinfo {year}
  {1958})}\BibitemShut {NoStop}%
\bibitem [{\citenamefont {Farnsworth}\ \emph {et~al.}(2013)\citenamefont
  {Farnsworth}, \citenamefont {Nelson},\ and\ \citenamefont
  {Gershenson}}]{farnsworthLivingInformationProcessing2013}%
  \BibitemOpen
  \bibfield  {author} {\bibinfo {author} {\bibfnamefont {K.~D.}\ \bibnamefont
  {Farnsworth}}, \bibinfo {author} {\bibfnamefont {J.}~\bibnamefont {Nelson}},\
  and\ \bibinfo {author} {\bibfnamefont {C.}~\bibnamefont {Gershenson}},\
  }\bibfield  {title} {\bibinfo {title} {Living is {Information} {Processing}:
  {From} {Molecules} to {Global} {Systems}},\ }\href
  {https://doi.org/10.1007/s10441-013-9179-3} {\bibfield  {journal} {\bibinfo
  {journal} {Acta Biotheoretica}\ }\textbf {\bibinfo {volume} {61}},\ \bibinfo
  {pages} {203} (\bibinfo {year} {2013})}\BibitemShut {NoStop}%
\bibitem [{\citenamefont {Graf}\ and\ \citenamefont {Machta}(2024)}]{Graf2024}%
  \BibitemOpen
  \bibfield  {author} {\bibinfo {author} {\bibfnamefont {I.~R.}\ \bibnamefont
  {Graf}}\ and\ \bibinfo {author} {\bibfnamefont {B.~B.}\ \bibnamefont
  {Machta}},\ }\bibfield  {title} {\bibinfo {title} {A bifurcation integrates
  information from many noisy ion channels and allows for milli-kelvin thermal
  sensitivity in the snake pit organ},\ }\href
  {https://doi.org/10.1073/pnas.2308215121} {\bibfield  {journal} {\bibinfo
  {journal} {Proceedings of the National Academy of Sciences}\ }\textbf
  {\bibinfo {volume} {121}},\ \bibinfo {eid} {e2308215121} (\bibinfo {year}
  {2024})}\BibitemShut {NoStop}%
\bibitem [{\citenamefont {Dullweber}\ and\ \citenamefont
  {Erzberger}(2023)}]{Dullweber2023}%
  \BibitemOpen
  \bibfield  {author} {\bibinfo {author} {\bibfnamefont {T.}~\bibnamefont
  {Dullweber}}\ and\ \bibinfo {author} {\bibfnamefont {A.}~\bibnamefont
  {Erzberger}},\ }\bibfield  {title} {\bibinfo {title} {Mechanochemical
  feedback loops in contact-dependent fate patterning},\ }\href
  {https://doi.org/https://doi.org/10.1016/j.coisb.2023.100445} {\bibfield
  {journal} {\bibinfo  {journal} {Current Opinion in Systems Biology}\ }\textbf
  {\bibinfo {volume} {32-33}},\ \bibinfo {pages} {100445} (\bibinfo {year}
  {2023})}\BibitemShut {NoStop}%
\bibitem [{\citenamefont {Uda}(2020)}]{udaApplicationInformationTheory2020}%
  \BibitemOpen
  \bibfield  {author} {\bibinfo {author} {\bibfnamefont {S.}~\bibnamefont
  {Uda}},\ }\bibfield  {title} {\bibinfo {title} {Application of information
  theory in systems biology},\ }\href
  {https://doi.org/10.1007/s12551-020-00665-w} {\bibfield  {journal} {\bibinfo
  {journal} {Biophysical Reviews}\ }\textbf {\bibinfo {volume} {12}},\ \bibinfo
  {pages} {377} (\bibinfo {year} {2020})}\BibitemShut {NoStop}%
\bibitem [{\citenamefont {Estrada}\ \emph {et~al.}(2016)\citenamefont
  {Estrada}, \citenamefont {Wong}, \citenamefont {DePace},\ and\ \citenamefont
  {Gunawardena}}]{estradaInformationIntegrationEnergy2016}%
  \BibitemOpen
  \bibfield  {author} {\bibinfo {author} {\bibfnamefont {J.}~\bibnamefont
  {Estrada}}, \bibinfo {author} {\bibfnamefont {F.}~\bibnamefont {Wong}},
  \bibinfo {author} {\bibfnamefont {A.}~\bibnamefont {DePace}},\ and\ \bibinfo
  {author} {\bibfnamefont {J.}~\bibnamefont {Gunawardena}},\ }\bibfield
  {title} {\bibinfo {title} {Information {Integration} and {Energy}
  {Expenditure} in {Gene} {Regulation}},\ }\href
  {https://doi.org/10.1016/j.cell.2016.06.012} {\bibfield  {journal} {\bibinfo
  {journal} {Cell}\ }\textbf {\bibinfo {volume} {166}},\ \bibinfo {pages} {234}
  (\bibinfo {year} {2016})}\BibitemShut {NoStop}%
\bibitem [{\citenamefont {Collinet}\ and\ \citenamefont
  {Lecuit}(2021)}]{collinetProgrammedSelforganizedFlow2021}%
  \BibitemOpen
  \bibfield  {author} {\bibinfo {author} {\bibfnamefont {C.}~\bibnamefont
  {Collinet}}\ and\ \bibinfo {author} {\bibfnamefont {T.}~\bibnamefont
  {Lecuit}},\ }\bibfield  {title} {\bibinfo {title} {Programmed and
  self-organized flow of information during morphogenesis},\ }\href
  {https://doi.org/10.1038/s41580-020-00318-6} {\bibfield  {journal} {\bibinfo
  {journal} {Nature Reviews Molecular Cell Biology}\ }\textbf {\bibinfo
  {volume} {22}},\ \bibinfo {pages} {245} (\bibinfo {year} {2021})}\BibitemShut
  {NoStop}%
\bibitem [{\citenamefont {Ganesan}\ and\ \citenamefont
  {Zhang}(2012)}]{ganesanHowCellsProcess2012}%
  \BibitemOpen
  \bibfield  {author} {\bibinfo {author} {\bibfnamefont {A.}~\bibnamefont
  {Ganesan}}\ and\ \bibinfo {author} {\bibfnamefont {J.}~\bibnamefont
  {Zhang}},\ }\bibfield  {title} {\bibinfo {title} {How cells process
  information: {Quantification} of spatiotemporal signaling dynamics},\ }\href
  {https://doi.org/10.1002/pro.2089} {\bibfield  {journal} {\bibinfo  {journal}
  {Protein Science : A Publication of the Protein Society}\ }\textbf {\bibinfo
  {volume} {21}},\ \bibinfo {pages} {918} (\bibinfo {year} {2012})}\BibitemShut
  {NoStop}%
\bibitem [{\citenamefont {Cheong}\ \emph {et~al.}(2011)\citenamefont {Cheong},
  \citenamefont {Rhee}, \citenamefont {Wang}, \citenamefont {Nemenman},\ and\
  \citenamefont {Levchenko}}]{cheongInformationTransductionCapacity2011}%
  \BibitemOpen
  \bibfield  {author} {\bibinfo {author} {\bibfnamefont {R.}~\bibnamefont
  {Cheong}}, \bibinfo {author} {\bibfnamefont {A.}~\bibnamefont {Rhee}},
  \bibinfo {author} {\bibfnamefont {C.~J.}\ \bibnamefont {Wang}}, \bibinfo
  {author} {\bibfnamefont {I.}~\bibnamefont {Nemenman}},\ and\ \bibinfo
  {author} {\bibfnamefont {A.}~\bibnamefont {Levchenko}},\ }\bibfield  {title}
  {\bibinfo {title} {Information {Transduction} {Capacity} of {Noisy}
  {Biochemical} {Signaling} {Networks}},\ }\href
  {https://doi.org/10.1126/science.1204553} {\bibfield  {journal} {\bibinfo
  {journal} {Science}\ }\textbf {\bibinfo {volume} {334}},\ \bibinfo {pages}
  {354} (\bibinfo {year} {2011})}\BibitemShut {NoStop}%
\bibitem [{\citenamefont {Brennan}\ \emph {et~al.}(2012)\citenamefont
  {Brennan}, \citenamefont {Cheong},\ and\ \citenamefont
  {Levchenko}}]{brennanHowInformationTheory2012}%
  \BibitemOpen
  \bibfield  {author} {\bibinfo {author} {\bibfnamefont {M.~D.}\ \bibnamefont
  {Brennan}}, \bibinfo {author} {\bibfnamefont {R.}~\bibnamefont {Cheong}},\
  and\ \bibinfo {author} {\bibfnamefont {A.}~\bibnamefont {Levchenko}},\
  }\bibfield  {title} {\bibinfo {title} {How {Information} {Theory} {Handles}
  {Cell} {Signaling} and {Uncertainty}},\ }\href
  {https://doi.org/10.1126/science.1227946} {\bibfield  {journal} {\bibinfo
  {journal} {Science}\ }\textbf {\bibinfo {volume} {338}},\ \bibinfo {pages}
  {334} (\bibinfo {year} {2012})}\BibitemShut {NoStop}%
\bibitem [{\citenamefont {Kramar}\ \emph {et~al.}()\citenamefont {Kramar},
  \citenamefont {Hahn}, \citenamefont {Walczak}, \citenamefont {Mora},\ and\
  \citenamefont {Coppey}}]{kramar2024singlecellsresolvegraded}%
  \BibitemOpen
  \bibfield  {author} {\bibinfo {author} {\bibfnamefont {M.}~\bibnamefont
  {Kramar}}, \bibinfo {author} {\bibfnamefont {L.}~\bibnamefont {Hahn}},
  \bibinfo {author} {\bibfnamefont {A.~M.}\ \bibnamefont {Walczak}}, \bibinfo
  {author} {\bibfnamefont {T.}~\bibnamefont {Mora}},\ and\ \bibinfo {author}
  {\bibfnamefont {M.}~\bibnamefont {Coppey}},\ }\href
  {https://arxiv.org/abs/2410.22571} {}\Eprint
  {https://arxiv.org/abs/2410.22571} {arXiv:2410.22571 [q-bio.CB]} \BibitemShut
  {NoStop}%
\bibitem [{\citenamefont {Nandan}\ \emph {et~al.}(2022)\citenamefont {Nandan},
  \citenamefont {Das}, \citenamefont {Lott},\ and\ \citenamefont
  {Koseska}}]{Nandan2022}%
  \BibitemOpen
  \bibfield  {author} {\bibinfo {author} {\bibfnamefont {A.}~\bibnamefont
  {Nandan}}, \bibinfo {author} {\bibfnamefont {A.}~\bibnamefont {Das}},
  \bibinfo {author} {\bibfnamefont {R.}~\bibnamefont {Lott}},\ and\ \bibinfo
  {author} {\bibfnamefont {A.}~\bibnamefont {Koseska}},\ }\bibfield  {title}
  {\bibinfo {title} {Cells use molecular working memory to navigate in changing
  chemoattractant fields},\ }\href {https://doi.org/10.7554/elife.76825}
  {\bibfield  {journal} {\bibinfo  {journal} {eLife}\ }\textbf {\bibinfo
  {volume} {11}},\ \bibinfo {eid} {e76825} (\bibinfo {year}
  {2022})}\BibitemShut {NoStop}%
\bibitem [{\citenamefont {Falasco}\ \emph {et~al.}(2018)\citenamefont
  {Falasco}, \citenamefont {Rao},\ and\ \citenamefont
  {Esposito}}]{falascoInformationThermodynamicsTuring2018}%
  \BibitemOpen
  \bibfield  {author} {\bibinfo {author} {\bibfnamefont {G.}~\bibnamefont
  {Falasco}}, \bibinfo {author} {\bibfnamefont {R.}~\bibnamefont {Rao}},\ and\
  \bibinfo {author} {\bibfnamefont {M.}~\bibnamefont {Esposito}},\ }\bibfield
  {title} {\bibinfo {title} {Information {Thermodynamics} of {Turing}
  {Patterns}},\ }\href {https://doi.org/10.1103/PhysRevLett.121.108301}
  {\bibfield  {journal} {\bibinfo  {journal} {Physical Review Letters}\
  }\textbf {\bibinfo {volume} {121}},\ \bibinfo {pages} {108301} (\bibinfo
  {year} {2018})}\BibitemShut {NoStop}%
\bibitem [{\citenamefont {Avanzini}\ \emph {et~al.}(2024)\citenamefont
  {Avanzini}, \citenamefont {Aslyamov}, \citenamefont {Fodor},\ and\
  \citenamefont {Esposito}}]{avanziniNonequilibriumThermodynamicsNonIdeal2024}%
  \BibitemOpen
  \bibfield  {author} {\bibinfo {author} {\bibfnamefont {F.}~\bibnamefont
  {Avanzini}}, \bibinfo {author} {\bibfnamefont {T.}~\bibnamefont {Aslyamov}},
  \bibinfo {author} {\bibfnamefont {E.}~\bibnamefont {Fodor}},\ and\ \bibinfo
  {author} {\bibfnamefont {M.}~\bibnamefont {Esposito}},\ }\bibfield  {title}
  {\bibinfo {title} {Nonequilibrium thermodynamics of non-ideal
  reaction–diffusion systems: Implications for active self-organization},\
  }\href {https://doi.org/10.1063/5.0231520} {\bibfield  {journal} {\bibinfo
  {journal} {The Journal of Chemical Physics}\ }\textbf {\bibinfo {volume}
  {161}},\ \bibinfo {pages} {174108} (\bibinfo {year} {2024})}\BibitemShut
  {NoStop}%
\bibitem [{\citenamefont {Brückner}\ and\ \citenamefont
  {Tkačik}(2024)}]{brucknerInformationContentOptimization2023}%
  \BibitemOpen
  \bibfield  {author} {\bibinfo {author} {\bibfnamefont {D.~B.}\ \bibnamefont
  {Brückner}}\ and\ \bibinfo {author} {\bibfnamefont {G.}~\bibnamefont
  {Tkačik}},\ }\bibfield  {title} {\bibinfo {title} {Information content and
  optimization of self-organized developmental systems},\ }\href
  {https://doi.org/10.1073/pnas.2322326121} {\bibfield  {journal} {\bibinfo
  {journal} {Proceedings of the National Academy of Sciences}\ }\textbf
  {\bibinfo {volume} {121}},\ \bibinfo {pages} {e2322326121} (\bibinfo {year}
  {2024})}\BibitemShut {NoStop}%
\bibitem [{\citenamefont
  {Wolpert}(1994)}]{wolpertPositionalInformationPattern1994}%
  \BibitemOpen
  \bibfield  {author} {\bibinfo {author} {\bibfnamefont {L.}~\bibnamefont
  {Wolpert}},\ }\bibfield  {title} {\bibinfo {title} {Positional information
  and pattern formation in development},\ }\href
  {https://doi.org/10.1002/dvg.1020150607} {\bibfield  {journal} {\bibinfo
  {journal} {Developmental Genetics}\ }\textbf {\bibinfo {volume} {15}},\
  \bibinfo {pages} {485} (\bibinfo {year} {1994})}\BibitemShut {NoStop}%
\bibitem [{\citenamefont {Dubuis}\ \emph {et~al.}(2013)\citenamefont {Dubuis},
  \citenamefont {Tkačik}, \citenamefont {Wieschaus}, \citenamefont {Gregor},\
  and\ \citenamefont {Bialek}}]{dubuisPositionalInformationBits2013}%
  \BibitemOpen
  \bibfield  {author} {\bibinfo {author} {\bibfnamefont {J.~O.}\ \bibnamefont
  {Dubuis}}, \bibinfo {author} {\bibfnamefont {G.}~\bibnamefont {Tkačik}},
  \bibinfo {author} {\bibfnamefont {E.~F.}\ \bibnamefont {Wieschaus}}, \bibinfo
  {author} {\bibfnamefont {T.}~\bibnamefont {Gregor}},\ and\ \bibinfo {author}
  {\bibfnamefont {W.}~\bibnamefont {Bialek}},\ }\bibfield  {title} {\bibinfo
  {title} {Positional information, in bits},\ }\href
  {https://doi.org/10.1073/pnas.1315642110} {\bibfield  {journal} {\bibinfo
  {journal} {Proceedings of the National Academy of Sciences}\ }\textbf
  {\bibinfo {volume} {110}},\ \bibinfo {pages} {16301} (\bibinfo {year}
  {2013})}\BibitemShut {NoStop}%
\bibitem [{\citenamefont {Tkačik}\ and\ \citenamefont
  {Gregor}(2021)}]{tkacikManyBitsPositional2021}%
  \BibitemOpen
  \bibfield  {author} {\bibinfo {author} {\bibfnamefont {G.}~\bibnamefont
  {Tkačik}}\ and\ \bibinfo {author} {\bibfnamefont {T.}~\bibnamefont
  {Gregor}},\ }\bibfield  {title} {\bibinfo {title} {The many bits of
  positional information},\ }\href {https://doi.org/10.1242/dev.176065}
  {\bibfield  {journal} {\bibinfo  {journal} {Development}\ }\textbf {\bibinfo
  {volume} {148}},\ \bibinfo {pages} {dev176065} (\bibinfo {year}
  {2021})}\BibitemShut {NoStop}%
\bibitem [{\citenamefont {McGough}\ \emph {et~al.}(2024)\citenamefont
  {McGough}, \citenamefont {Casademunt}, \citenamefont {Nikolić},
  \citenamefont {Aridor}, \citenamefont {Petkova}, \citenamefont {Gregor},\
  and\ \citenamefont {Bialek}}]{mcgoughFindingLastBits2024}%
  \BibitemOpen
  \bibfield  {author} {\bibinfo {author} {\bibfnamefont {L.}~\bibnamefont
  {McGough}}, \bibinfo {author} {\bibfnamefont {H.}~\bibnamefont {Casademunt}},
  \bibinfo {author} {\bibfnamefont {M.}~\bibnamefont {Nikolić}}, \bibinfo
  {author} {\bibfnamefont {Z.}~\bibnamefont {Aridor}}, \bibinfo {author}
  {\bibfnamefont {M.~D.}\ \bibnamefont {Petkova}}, \bibinfo {author}
  {\bibfnamefont {T.}~\bibnamefont {Gregor}},\ and\ \bibinfo {author}
  {\bibfnamefont {W.}~\bibnamefont {Bialek}},\ }\bibfield  {title} {\bibinfo
  {title} {Finding the {Last} {Bits} of {Positional} {Information}},\ }\href
  {https://doi.org/10.1103/PRXLife.2.013016} {\bibfield  {journal} {\bibinfo
  {journal} {PRX Life}\ }\textbf {\bibinfo {volume} {2}},\ \bibinfo {pages}
  {013016} (\bibinfo {year} {2024})}\BibitemShut {NoStop}%
\bibitem [{\citenamefont {Flowers}\ and\ \citenamefont
  {Crews}(2020)}]{flowersRememberingWhereWe2020}%
  \BibitemOpen
  \bibfield  {author} {\bibinfo {author} {\bibfnamefont {G.~P.}\ \bibnamefont
  {Flowers}}\ and\ \bibinfo {author} {\bibfnamefont {C.~M.}\ \bibnamefont
  {Crews}},\ }\bibfield  {title} {\bibinfo {title} {Remembering where we are:
  {Positional} information in salamander limb regeneration},\ }\href
  {https://doi.org/10.1002/dvdy.167} {\bibfield  {journal} {\bibinfo  {journal}
  {Developmental Dynamics}\ }\textbf {\bibinfo {volume} {249}},\ \bibinfo
  {pages} {465} (\bibinfo {year} {2020})}\BibitemShut {NoStop}%
\bibitem [{\citenamefont {Zieske}\ and\ \citenamefont
  {Schwille}(2014)}]{zieskeReconstitutionSelforganizingProtein2014}%
  \BibitemOpen
  \bibfield  {author} {\bibinfo {author} {\bibfnamefont {K.}~\bibnamefont
  {Zieske}}\ and\ \bibinfo {author} {\bibfnamefont {P.}~\bibnamefont
  {Schwille}},\ }\bibfield  {title} {\bibinfo {title} {Reconstitution of
  self-organizing protein gradients as spatial cues in cell-free systems},\
  }\href {https://doi.org/10.7554/eLife.03949} {\bibfield  {journal} {\bibinfo
  {journal} {eLife}\ }\textbf {\bibinfo {volume} {3}},\ \bibinfo {pages}
  {e03949} (\bibinfo {year} {2014})}\BibitemShut {NoStop}%
\bibitem [{\citenamefont {Wigbers}\ \emph {et~al.}(2021)\citenamefont
  {Wigbers}, \citenamefont {Tan}, \citenamefont {Brauns}, \citenamefont {Liu},
  \citenamefont {Swartz}, \citenamefont {Frey},\ and\ \citenamefont
  {Fakhri}}]{wigbersHierarchyProteinPatterns2021}%
  \BibitemOpen
  \bibfield  {author} {\bibinfo {author} {\bibfnamefont {M.~C.}\ \bibnamefont
  {Wigbers}}, \bibinfo {author} {\bibfnamefont {T.~H.}\ \bibnamefont {Tan}},
  \bibinfo {author} {\bibfnamefont {F.}~\bibnamefont {Brauns}}, \bibinfo
  {author} {\bibfnamefont {J.}~\bibnamefont {Liu}}, \bibinfo {author}
  {\bibfnamefont {S.~Z.}\ \bibnamefont {Swartz}}, \bibinfo {author}
  {\bibfnamefont {E.}~\bibnamefont {Frey}},\ and\ \bibinfo {author}
  {\bibfnamefont {N.}~\bibnamefont {Fakhri}},\ }\bibfield  {title} {\bibinfo
  {title} {A hierarchy of protein patterns robustly decodes cell shape
  information},\ }\href {https://doi.org/10.1038/s41567-021-01164-9} {\bibfield
   {journal} {\bibinfo  {journal} {Nature Physics}\ }\textbf {\bibinfo {volume}
  {17}},\ \bibinfo {pages} {578} (\bibinfo {year} {2021})}\BibitemShut
  {NoStop}%
\bibitem [{\citenamefont {Wettmann}\ and\ \citenamefont
  {Kruse}(2018)}]{wettmannMinproteinOscillationsEscherichia2018}%
  \BibitemOpen
  \bibfield  {author} {\bibinfo {author} {\bibfnamefont {L.}~\bibnamefont
  {Wettmann}}\ and\ \bibinfo {author} {\bibfnamefont {K.}~\bibnamefont
  {Kruse}},\ }\bibfield  {title} {\bibinfo {title} {The {Min}-protein
  oscillations in {Escherichia} coli: an example of self-organized cellular
  protein waves},\ }\href {https://doi.org/10.1098/rstb.2017.0111} {\bibfield
  {journal} {\bibinfo  {journal} {Philosophical Transactions of the Royal
  Society B: Biological Sciences}\ }\textbf {\bibinfo {volume} {373}},\
  \bibinfo {pages} {20170111} (\bibinfo {year} {2018})}\BibitemShut {NoStop}%
\bibitem [{\citenamefont {Haupt}\ and\ \citenamefont
  {Minc}(2018)}]{hauptHowCellsSense2018}%
  \BibitemOpen
  \bibfield  {author} {\bibinfo {author} {\bibfnamefont {A.}~\bibnamefont
  {Haupt}}\ and\ \bibinfo {author} {\bibfnamefont {N.}~\bibnamefont {Minc}},\
  }\bibfield  {title} {\bibinfo {title} {How cells sense their own shape –
  mechanisms to probe cell geometry and their implications in cellular
  organization and function},\ }\href {https://doi.org/10.1242/jcs.214015}
  {\bibfield  {journal} {\bibinfo  {journal} {Journal of Cell Science}\
  }\textbf {\bibinfo {volume} {131}},\ \bibinfo {pages} {jcs214015} (\bibinfo
  {year} {2018})}\BibitemShut {NoStop}%
\bibitem [{\citenamefont {Rangamani}\ \emph {et~al.}(2013)\citenamefont
  {Rangamani}, \citenamefont {Lipshtat}, \citenamefont {Azeloglu},
  \citenamefont {Calizo}, \citenamefont {Hu}, \citenamefont {Ghassemi},
  \citenamefont {Hone}, \citenamefont {Scarlata}, \citenamefont {Neves},\ and\
  \citenamefont {Iyengar}}]{Rangamani2013}%
  \BibitemOpen
  \bibfield  {author} {\bibinfo {author} {\bibfnamefont {P.}~\bibnamefont
  {Rangamani}}, \bibinfo {author} {\bibfnamefont {A.}~\bibnamefont {Lipshtat}},
  \bibinfo {author} {\bibfnamefont {E.~U.}\ \bibnamefont {Azeloglu}}, \bibinfo
  {author} {\bibfnamefont {R.~C.}\ \bibnamefont {Calizo}}, \bibinfo {author}
  {\bibfnamefont {M.}~\bibnamefont {Hu}}, \bibinfo {author} {\bibfnamefont
  {S.}~\bibnamefont {Ghassemi}}, \bibinfo {author} {\bibfnamefont
  {J.}~\bibnamefont {Hone}}, \bibinfo {author} {\bibfnamefont {S.}~\bibnamefont
  {Scarlata}}, \bibinfo {author} {\bibfnamefont {S.~R.}\ \bibnamefont
  {Neves}},\ and\ \bibinfo {author} {\bibfnamefont {R.}~\bibnamefont
  {Iyengar}},\ }\bibfield  {title} {\bibinfo {title} {Decoding information in
  cell shape},\ }\href {https://doi.org/10.1016/j.cell.2013.08.026} {\bibfield
  {journal} {\bibinfo  {journal} {Cell}\ }\textbf {\bibinfo {volume} {154}},\
  \bibinfo {pages} {1356} (\bibinfo {year} {2013})}\BibitemShut {NoStop}%
\bibitem [{\citenamefont {Evans}\ \emph {et~al.}(2024)\citenamefont {Evans},
  \citenamefont {O’Brien}, \citenamefont {Winfree},\ and\ \citenamefont
  {Murugan}}]{Evans2024}%
  \BibitemOpen
  \bibfield  {author} {\bibinfo {author} {\bibfnamefont {C.~G.}\ \bibnamefont
  {Evans}}, \bibinfo {author} {\bibfnamefont {J.}~\bibnamefont {O’Brien}},
  \bibinfo {author} {\bibfnamefont {E.}~\bibnamefont {Winfree}},\ and\ \bibinfo
  {author} {\bibfnamefont {A.}~\bibnamefont {Murugan}},\ }\bibfield  {title}
  {\bibinfo {title} {Pattern recognition in the nucleation kinetics of
  non-equilibrium self-assembly},\ }\href
  {https://doi.org/10.1038/s41586-023-06890-z} {\bibfield  {journal} {\bibinfo
  {journal} {Nature}\ }\textbf {\bibinfo {volume} {625}},\ \bibinfo {pages}
  {500} (\bibinfo {year} {2024})}\BibitemShut {NoStop}%
\bibitem [{\citenamefont {Mijatović}\ \emph {et~al.}()\citenamefont
  {Mijatović}, \citenamefont {Kok}, \citenamefont {Zwanikken},\ and\
  \citenamefont {Bauer}}]{Mijatovic2025}%
  \BibitemOpen
  \bibfield  {author} {\bibinfo {author} {\bibfnamefont {T.}~\bibnamefont
  {Mijatović}}, \bibinfo {author} {\bibfnamefont {A.~R.}\ \bibnamefont {Kok}},
  \bibinfo {author} {\bibfnamefont {J.~W.}\ \bibnamefont {Zwanikken}},\ and\
  \bibinfo {author} {\bibfnamefont {M.}~\bibnamefont {Bauer}},\ }\href
  {https://arxiv.org/abs/2505.07641} {}\Eprint
  {https://arxiv.org/abs/2505.07641} {arXiv:2505.07641 [physics.bio-ph]}
  \BibitemShut {NoStop}%
\bibitem [{\citenamefont {Braz~Teixeira}\ \emph {et~al.}(2024)\citenamefont
  {Braz~Teixeira}, \citenamefont {Carugno}, \citenamefont {Neri},\ and\
  \citenamefont {Sartori}}]{BrazTeixeira2024}%
  \BibitemOpen
  \bibfield  {author} {\bibinfo {author} {\bibfnamefont {R.}~\bibnamefont
  {Braz~Teixeira}}, \bibinfo {author} {\bibfnamefont {G.}~\bibnamefont
  {Carugno}}, \bibinfo {author} {\bibfnamefont {I.}~\bibnamefont {Neri}},\ and\
  \bibinfo {author} {\bibfnamefont {P.}~\bibnamefont {Sartori}},\ }\bibfield
  {title} {\bibinfo {title} {Liquid hopfield model: Retrieval and localization
  in multicomponent liquid mixtures},\ }\href
  {https://doi.org/10.1073/pnas.2320504121} {\bibfield  {journal} {\bibinfo
  {journal} {Proceedings of the National Academy of Sciences}\ }\textbf
  {\bibinfo {volume} {121}},\ \bibinfo {eid} {e2320504121} (\bibinfo {year}
  {2024})}\BibitemShut {NoStop}%
\bibitem [{\citenamefont {Banerjee}\ \emph {et~al.}()\citenamefont {Banerjee},
  \citenamefont {Falk}, \citenamefont {Gardel}, \citenamefont {Walczak},
  \citenamefont {Mora},\ and\ \citenamefont {Vaikuntanathan}}]{Banerjee2025}%
  \BibitemOpen
  \bibfield  {author} {\bibinfo {author} {\bibfnamefont {D.~S.}\ \bibnamefont
  {Banerjee}}, \bibinfo {author} {\bibfnamefont {M.~J.}\ \bibnamefont {Falk}},
  \bibinfo {author} {\bibfnamefont {M.~L.}\ \bibnamefont {Gardel}}, \bibinfo
  {author} {\bibfnamefont {A.~M.}\ \bibnamefont {Walczak}}, \bibinfo {author}
  {\bibfnamefont {T.}~\bibnamefont {Mora}},\ and\ \bibinfo {author}
  {\bibfnamefont {S.}~\bibnamefont {Vaikuntanathan}},\ }\href
  {https://arxiv.org/abs/2504.15107} {}\Eprint
  {https://arxiv.org/abs/2504.15107} {arXiv:2504.15107 [cond-mat.soft]}
  \BibitemShut {NoStop}%
\bibitem [{\citenamefont {Rombouts}\ \emph {et~al.}(2023)\citenamefont
  {Rombouts}, \citenamefont {Elliott},\ and\ \citenamefont
  {Erzberger}}]{Rombouts2023}%
  \BibitemOpen
  \bibfield  {author} {\bibinfo {author} {\bibfnamefont {J.}~\bibnamefont
  {Rombouts}}, \bibinfo {author} {\bibfnamefont {J.}~\bibnamefont {Elliott}},\
  and\ \bibinfo {author} {\bibfnamefont {A.}~\bibnamefont {Erzberger}},\
  }\bibfield  {title} {\bibinfo {title} {Forceful patterning: theoretical
  principles of mechanochemical pattern formation},\ }\href
  {https://doi.org/https://doi.org/10.15252/embr.202357739} {\bibfield
  {journal} {\bibinfo  {journal} {EMBO reports}\ }\textbf {\bibinfo {volume}
  {24}},\ \bibinfo {pages} {e57739} (\bibinfo {year} {2023})}\BibitemShut
  {NoStop}%
\bibitem [{\citenamefont {Sitarska}\ \emph {et~al.}(2023)\citenamefont
  {Sitarska}, \citenamefont {Almeida}, \citenamefont {Beckwith}, \citenamefont
  {Stopp}, \citenamefont {Czuchnowski}, \citenamefont {Siggel}, \citenamefont
  {Roessner}, \citenamefont {Tschanz}, \citenamefont {Ejsing}, \citenamefont
  {Schwab}, \citenamefont {Kosinski}, \citenamefont {Sixt}, \citenamefont
  {Kreshuk}, \citenamefont {Erzberger},\ and\ \citenamefont
  {Diz-Muñoz}}]{Sitarska2023}%
  \BibitemOpen
  \bibfield  {author} {\bibinfo {author} {\bibfnamefont {E.}~\bibnamefont
  {Sitarska}}, \bibinfo {author} {\bibfnamefont {S.~D.}\ \bibnamefont
  {Almeida}}, \bibinfo {author} {\bibfnamefont {M.~S.}\ \bibnamefont
  {Beckwith}}, \bibinfo {author} {\bibfnamefont {J.}~\bibnamefont {Stopp}},
  \bibinfo {author} {\bibfnamefont {J.}~\bibnamefont {Czuchnowski}}, \bibinfo
  {author} {\bibfnamefont {M.}~\bibnamefont {Siggel}}, \bibinfo {author}
  {\bibfnamefont {R.}~\bibnamefont {Roessner}}, \bibinfo {author}
  {\bibfnamefont {A.}~\bibnamefont {Tschanz}}, \bibinfo {author} {\bibfnamefont
  {C.}~\bibnamefont {Ejsing}}, \bibinfo {author} {\bibfnamefont
  {Y.}~\bibnamefont {Schwab}}, \bibinfo {author} {\bibfnamefont
  {J.}~\bibnamefont {Kosinski}}, \bibinfo {author} {\bibfnamefont
  {M.}~\bibnamefont {Sixt}}, \bibinfo {author} {\bibfnamefont {A.}~\bibnamefont
  {Kreshuk}}, \bibinfo {author} {\bibfnamefont {A.}~\bibnamefont {Erzberger}},\
  and\ \bibinfo {author} {\bibfnamefont {A.}~\bibnamefont {Diz-Muñoz}},\
  }\bibfield  {title} {\bibinfo {title} {Sensing their plasma membrane
  curvature allows migrating cells to circumvent obstacles},\ }\href
  {https://doi.org/10.1038/s41467-023-41173-1} {\bibfield  {journal} {\bibinfo
  {journal} {Nature Communications}\ }\textbf {\bibinfo {volume} {14}},\
  \bibinfo {pages} {5644} (\bibinfo {year} {2023})}\BibitemShut {NoStop}%
\bibitem [{\citenamefont {Gomes}\ \emph {et~al.}(2005)\citenamefont {Gomes},
  \citenamefont {Jani},\ and\ \citenamefont {Gundersen}}]{Gomes2005}%
  \BibitemOpen
  \bibfield  {author} {\bibinfo {author} {\bibfnamefont {E.~R.}\ \bibnamefont
  {Gomes}}, \bibinfo {author} {\bibfnamefont {S.}~\bibnamefont {Jani}},\ and\
  \bibinfo {author} {\bibfnamefont {G.~G.}\ \bibnamefont {Gundersen}},\
  }\bibfield  {title} {\bibinfo {title} {Nuclear movement regulated by cdc42,
  mrck, myosin, and actin flow establishes mtoc polarization in migrating
  cells},\ }\href {https://doi.org/10.1016/j.cell.2005.02.022} {\bibfield
  {journal} {\bibinfo  {journal} {Cell}\ }\textbf {\bibinfo {volume} {121}},\
  \bibinfo {pages} {451} (\bibinfo {year} {2005})}\BibitemShut {NoStop}%
\bibitem [{\citenamefont {Vercruysse}\ \emph {et~al.}(2024)\citenamefont
  {Vercruysse}, \citenamefont {Br\"{u}ckner}, \citenamefont {Gómez-González},
  \citenamefont {Remson}, \citenamefont {Luciano}, \citenamefont {Kalukula},
  \citenamefont {Rossetti}, \citenamefont {Trepat}, \citenamefont {Hannezo},\
  and\ \citenamefont {Gabriele}}]{Vercruysse2024}%
  \BibitemOpen
  \bibfield  {author} {\bibinfo {author} {\bibfnamefont {E.}~\bibnamefont
  {Vercruysse}}, \bibinfo {author} {\bibfnamefont {D.~B.}\ \bibnamefont
  {Br\"{u}ckner}}, \bibinfo {author} {\bibfnamefont {M.}~\bibnamefont
  {Gómez-González}}, \bibinfo {author} {\bibfnamefont {A.}~\bibnamefont
  {Remson}}, \bibinfo {author} {\bibfnamefont {M.}~\bibnamefont {Luciano}},
  \bibinfo {author} {\bibfnamefont {Y.}~\bibnamefont {Kalukula}}, \bibinfo
  {author} {\bibfnamefont {L.}~\bibnamefont {Rossetti}}, \bibinfo {author}
  {\bibfnamefont {X.}~\bibnamefont {Trepat}}, \bibinfo {author} {\bibfnamefont
  {E.}~\bibnamefont {Hannezo}},\ and\ \bibinfo {author} {\bibfnamefont
  {S.}~\bibnamefont {Gabriele}},\ }\bibfield  {title} {\bibinfo {title}
  {Geometry-driven migration efficiency of autonomous epithelial cell
  clusters},\ }\href {https://doi.org/10.1038/s41567-024-02532-x} {\bibfield
  {journal} {\bibinfo  {journal} {Nature Physics}\ }\textbf {\bibinfo {volume}
  {20}},\ \bibinfo {pages} {1492} (\bibinfo {year} {2024})}\BibitemShut
  {NoStop}%
\bibitem [{\citenamefont {d’Alessandro}\ \emph {et~al.}(2021)\citenamefont
  {d’Alessandro}, \citenamefont {Barbier-Chebbah}, \citenamefont {Cellerin},
  \citenamefont {Benichou}, \citenamefont {Mège}, \citenamefont {Voituriez},\
  and\ \citenamefont {Ladoux}}]{dAlessandro2021}%
  \BibitemOpen
  \bibfield  {author} {\bibinfo {author} {\bibfnamefont {J.}~\bibnamefont
  {d’Alessandro}}, \bibinfo {author} {\bibfnamefont {A.}~\bibnamefont
  {Barbier-Chebbah}}, \bibinfo {author} {\bibfnamefont {V.}~\bibnamefont
  {Cellerin}}, \bibinfo {author} {\bibfnamefont {O.}~\bibnamefont {Benichou}},
  \bibinfo {author} {\bibfnamefont {R.~M.}\ \bibnamefont {Mège}}, \bibinfo
  {author} {\bibfnamefont {R.}~\bibnamefont {Voituriez}},\ and\ \bibinfo
  {author} {\bibfnamefont {B.}~\bibnamefont {Ladoux}},\ }\bibfield  {title}
  {\bibinfo {title} {Cell migration guided by long-lived spatial memory},\
  }\href {https://doi.org/10.1038/s41467-021-24249-8} {\bibfield  {journal}
  {\bibinfo  {journal} {Nature Communications}\ }\textbf {\bibinfo {volume}
  {12}},\ \bibinfo {pages} {4118} (\bibinfo {year} {2021})}\BibitemShut
  {NoStop}%
\bibitem [{\citenamefont {Ron}\ \emph {et~al.}(2024)\citenamefont {Ron},
  \citenamefont {Crestani}, \citenamefont {Kux}, \citenamefont {Liu},
  \citenamefont {Al-Dam}, \citenamefont {Monzo}, \citenamefont {Gauthier},
  \citenamefont {Sáez},\ and\ \citenamefont {Gov}}]{Ron2024}%
  \BibitemOpen
  \bibfield  {author} {\bibinfo {author} {\bibfnamefont {J.~E.}\ \bibnamefont
  {Ron}}, \bibinfo {author} {\bibfnamefont {M.}~\bibnamefont {Crestani}},
  \bibinfo {author} {\bibfnamefont {J.~M.}\ \bibnamefont {Kux}}, \bibinfo
  {author} {\bibfnamefont {J.}~\bibnamefont {Liu}}, \bibinfo {author}
  {\bibfnamefont {N.}~\bibnamefont {Al-Dam}}, \bibinfo {author} {\bibfnamefont
  {P.}~\bibnamefont {Monzo}}, \bibinfo {author} {\bibfnamefont {N.~C.}\
  \bibnamefont {Gauthier}}, \bibinfo {author} {\bibfnamefont {P.~J.}\
  \bibnamefont {Sáez}},\ and\ \bibinfo {author} {\bibfnamefont {N.~S.}\
  \bibnamefont {Gov}},\ }\bibfield  {title} {\bibinfo {title} {Emergent seesaw
  oscillations during cellular directional decision-making},\ }\href
  {https://doi.org/10.1038/s41567-023-02335-6} {\bibfield  {journal} {\bibinfo
  {journal} {Nature Physics}\ }\textbf {\bibinfo {volume} {20}},\ \bibinfo
  {pages} {501} (\bibinfo {year} {2024})}\BibitemShut {NoStop}%
\bibitem [{\citenamefont {Han}\ \emph {et~al.}(2025)\citenamefont {Han},
  \citenamefont {Liu}, \citenamefont {Qu}, \citenamefont {Duan}, \citenamefont
  {Xiang}, \citenamefont {Jiang}, \citenamefont {Yang}, \citenamefont {Fang},
  \citenamefont {Xu}, \citenamefont {Wen}, \citenamefont {Yu}, \citenamefont
  {Huang}, \citenamefont {Huang},\ and\ \citenamefont {Zhu}}]{Han2025}%
  \BibitemOpen
  \bibfield  {author} {\bibinfo {author} {\bibfnamefont {Y.}~\bibnamefont
  {Han}}, \bibinfo {author} {\bibfnamefont {X.}~\bibnamefont {Liu}}, \bibinfo
  {author} {\bibfnamefont {S.}~\bibnamefont {Qu}}, \bibinfo {author}
  {\bibfnamefont {X.}~\bibnamefont {Duan}}, \bibinfo {author} {\bibfnamefont
  {Y.}~\bibnamefont {Xiang}}, \bibinfo {author} {\bibfnamefont
  {N.}~\bibnamefont {Jiang}}, \bibinfo {author} {\bibfnamefont
  {S.}~\bibnamefont {Yang}}, \bibinfo {author} {\bibfnamefont {X.}~\bibnamefont
  {Fang}}, \bibinfo {author} {\bibfnamefont {L.}~\bibnamefont {Xu}}, \bibinfo
  {author} {\bibfnamefont {H.}~\bibnamefont {Wen}}, \bibinfo {author}
  {\bibfnamefont {Y.}~\bibnamefont {Yu}}, \bibinfo {author} {\bibfnamefont
  {S.}~\bibnamefont {Huang}}, \bibinfo {author} {\bibfnamefont
  {J.}~\bibnamefont {Huang}},\ and\ \bibinfo {author} {\bibfnamefont
  {K.}~\bibnamefont {Zhu}},\ }\bibfield  {title} {\bibinfo {title} {Tissue
  geometry spatiotemporally drives bacterial infections},\ }\href@noop {}
  {\bibfield  {journal} {\bibinfo  {journal} {Cell}\ } (\bibinfo {year}
  {2025})},\ \bibinfo {note} {(to be published)}\BibitemShut {NoStop}%
\bibitem [{\citenamefont {Baranov}\ \emph {et~al.}(2021)\citenamefont
  {Baranov}, \citenamefont {Kumar}, \citenamefont {Sacanna}, \citenamefont
  {Thutupalli},\ and\ \citenamefont {van~den Bogaart}}]{Baranov2021}%
  \BibitemOpen
  \bibfield  {author} {\bibinfo {author} {\bibfnamefont {M.~V.}\ \bibnamefont
  {Baranov}}, \bibinfo {author} {\bibfnamefont {M.}~\bibnamefont {Kumar}},
  \bibinfo {author} {\bibfnamefont {S.}~\bibnamefont {Sacanna}}, \bibinfo
  {author} {\bibfnamefont {S.}~\bibnamefont {Thutupalli}},\ and\ \bibinfo
  {author} {\bibfnamefont {G.}~\bibnamefont {van~den Bogaart}},\ }\bibfield
  {title} {\bibinfo {title} {Modulation of immune responses by particle size
  and shape},\ }\href {https://doi.org/10.3389/fimmu.2020.607945} {\bibfield
  {journal} {\bibinfo  {journal} {Frontiers in Immunology}\ }\textbf {\bibinfo
  {volume} {11}},\ \bibinfo {eid} {607945} (\bibinfo {year}
  {2021})}\BibitemShut {NoStop}%
\bibitem [{\citenamefont {Stow}\ and\ \citenamefont {Condon}(2016)}]{Stow2016}%
  \BibitemOpen
  \bibfield  {author} {\bibinfo {author} {\bibfnamefont {J.~L.}\ \bibnamefont
  {Stow}}\ and\ \bibinfo {author} {\bibfnamefont {N.~D.}\ \bibnamefont
  {Condon}},\ }\bibfield  {title} {\bibinfo {title} {The cell surface
  environment for pathogen recognition and entry},\ }\href
  {https://doi.org/https://doi.org/10.1038/cti.2016.15} {\bibfield  {journal}
  {\bibinfo  {journal} {Clinical \& Translational Immunology}\ }\textbf
  {\bibinfo {volume} {5}},\ \bibinfo {pages} {e71} (\bibinfo {year}
  {2016})}\BibitemShut {NoStop}%
\bibitem [{\citenamefont {Cho}\ \emph {et~al.}(2017)\citenamefont {Cho},
  \citenamefont {Irianto},\ and\ \citenamefont {Discher}}]{Cho2017}%
  \BibitemOpen
  \bibfield  {author} {\bibinfo {author} {\bibfnamefont {S.}~\bibnamefont
  {Cho}}, \bibinfo {author} {\bibfnamefont {J.}~\bibnamefont {Irianto}},\ and\
  \bibinfo {author} {\bibfnamefont {D.~E.}\ \bibnamefont {Discher}},\
  }\bibfield  {title} {\bibinfo {title} {Mechanosensing by the nucleus: From
  pathways to scaling relationships},\ }\href
  {https://doi.org/10.1083/jcb.201610042} {\bibfield  {journal} {\bibinfo
  {journal} {Journal of Cell Biology}\ }\textbf {\bibinfo {volume} {216}},\
  \bibinfo {pages} {305} (\bibinfo {year} {2017})}\BibitemShut {NoStop}%
\bibitem [{\citenamefont {Burkart}\ \emph {et~al.}(2022)\citenamefont
  {Burkart}, \citenamefont {Wigbers}, \citenamefont {Würthner},\ and\
  \citenamefont {Frey}}]{burkartControlProteinbasedPattern2022}%
  \BibitemOpen
  \bibfield  {author} {\bibinfo {author} {\bibfnamefont {T.}~\bibnamefont
  {Burkart}}, \bibinfo {author} {\bibfnamefont {M.~C.}\ \bibnamefont
  {Wigbers}}, \bibinfo {author} {\bibfnamefont {L.}~\bibnamefont {Würthner}},\
  and\ \bibinfo {author} {\bibfnamefont {E.}~\bibnamefont {Frey}},\ }\bibfield
  {title} {\bibinfo {title} {Control of protein-based pattern formation via
  guiding cues},\ }\href {https://doi.org/10.1038/s42254-022-00461-3}
  {\bibfield  {journal} {\bibinfo  {journal} {Nature Reviews Physics}\ }\textbf
  {\bibinfo {volume} {4}},\ \bibinfo {pages} {511} (\bibinfo {year}
  {2022})}\BibitemShut {NoStop}%
\bibitem [{\citenamefont {Shellard}\ and\ \citenamefont
  {Mayor}(2020)}]{Shellard2020}%
  \BibitemOpen
  \bibfield  {author} {\bibinfo {author} {\bibfnamefont {A.}~\bibnamefont
  {Shellard}}\ and\ \bibinfo {author} {\bibfnamefont {R.}~\bibnamefont
  {Mayor}},\ }\bibfield  {title} {\bibinfo {title} {All roads lead to
  directional cell migration},\ }\href
  {https://doi.org/10.1016/j.tcb.2020.08.002} {\bibfield  {journal} {\bibinfo
  {journal} {Trends in Cell Biology}\ }\textbf {\bibinfo {volume} {30}},\
  \bibinfo {pages} {852} (\bibinfo {year} {2020})}\BibitemShut {NoStop}%
\bibitem [{\citenamefont {Clark}\ \emph {et~al.}(2022)\citenamefont {Clark},
  \citenamefont {Maitra}, \citenamefont {Jacques}, \citenamefont {Bergert},
  \citenamefont {Pérez-González}, \citenamefont {Simon}, \citenamefont
  {Lederer}, \citenamefont {Diz-Muñoz}, \citenamefont {Trepat}, \citenamefont
  {Voituriez},\ and\ \citenamefont {Vignjevic}}]{Clark2022}%
  \BibitemOpen
  \bibfield  {author} {\bibinfo {author} {\bibfnamefont {A.~G.}\ \bibnamefont
  {Clark}}, \bibinfo {author} {\bibfnamefont {A.}~\bibnamefont {Maitra}},
  \bibinfo {author} {\bibfnamefont {C.}~\bibnamefont {Jacques}}, \bibinfo
  {author} {\bibfnamefont {M.}~\bibnamefont {Bergert}}, \bibinfo {author}
  {\bibfnamefont {C.}~\bibnamefont {Pérez-González}}, \bibinfo {author}
  {\bibfnamefont {A.}~\bibnamefont {Simon}}, \bibinfo {author} {\bibfnamefont
  {L.}~\bibnamefont {Lederer}}, \bibinfo {author} {\bibfnamefont
  {A.}~\bibnamefont {Diz-Muñoz}}, \bibinfo {author} {\bibfnamefont
  {X.}~\bibnamefont {Trepat}}, \bibinfo {author} {\bibfnamefont
  {R.}~\bibnamefont {Voituriez}},\ and\ \bibinfo {author} {\bibfnamefont
  {D.~M.}\ \bibnamefont {Vignjevic}},\ }\bibfield  {title} {\bibinfo {title}
  {Self-generated gradients steer collective migration on viscoelastic collagen
  networks},\ }\href {https://doi.org/10.1038/s41563-022-01259-5} {\bibfield
  {journal} {\bibinfo  {journal} {Nature Materials}\ }\textbf {\bibinfo
  {volume} {21}},\ \bibinfo {pages} {1200} (\bibinfo {year}
  {2022})}\BibitemShut {NoStop}%
\bibitem [{\citenamefont {Carter}(1967)}]{CARTER1967}%
  \BibitemOpen
  \bibfield  {author} {\bibinfo {author} {\bibfnamefont {S.~B.}\ \bibnamefont
  {Carter}},\ }\bibfield  {title} {\bibinfo {title} {Haptotaxis and the
  mechanism of cell motility},\ }\href {https://doi.org/10.1038/213256a0}
  {\bibfield  {journal} {\bibinfo  {journal} {Nature}\ }\textbf {\bibinfo
  {volume} {213}},\ \bibinfo {pages} {256} (\bibinfo {year}
  {1967})}\BibitemShut {NoStop}%
\bibitem [{\citenamefont {Luo}\ \emph {et~al.}(2020)\citenamefont {Luo},
  \citenamefont {Seveau~de Noray}, \citenamefont {Aoun}, \citenamefont
  {Biarnes-Pelicot}, \citenamefont {Strale}, \citenamefont {Studer},
  \citenamefont {Valignat},\ and\ \citenamefont {Theodoly}}]{Luo2020}%
  \BibitemOpen
  \bibfield  {author} {\bibinfo {author} {\bibfnamefont {X.}~\bibnamefont
  {Luo}}, \bibinfo {author} {\bibfnamefont {V.}~\bibnamefont {Seveau~de
  Noray}}, \bibinfo {author} {\bibfnamefont {L.}~\bibnamefont {Aoun}}, \bibinfo
  {author} {\bibfnamefont {M.}~\bibnamefont {Biarnes-Pelicot}}, \bibinfo
  {author} {\bibfnamefont {P.-O.}\ \bibnamefont {Strale}}, \bibinfo {author}
  {\bibfnamefont {V.}~\bibnamefont {Studer}}, \bibinfo {author} {\bibfnamefont
  {M.-P.}\ \bibnamefont {Valignat}},\ and\ \bibinfo {author} {\bibfnamefont
  {O.}~\bibnamefont {Theodoly}},\ }\bibfield  {title} {\bibinfo {title}
  {Lymphocytes perform reverse adhesive haptotaxis mediated by lfa-1
  integrins},\ }\href {https://doi.org/10.1242/jcs.242883} {\bibfield
  {journal} {\bibinfo  {journal} {Journal of Cell Science}\ }\textbf {\bibinfo
  {volume} {133}},\ \bibinfo {eid} {jcs242883} (\bibinfo {year}
  {2020})}\BibitemShut {NoStop}%
\bibitem [{\citenamefont {Fortunato}\ \emph {et~al.}(2024)\citenamefont
  {Fortunato}, \citenamefont {Br{\"u}ckner}, \citenamefont {Grosser},
  \citenamefont {Rossetti}, \citenamefont {Bosch-Padr{\'o}s}, \citenamefont
  {Trebicka}, \citenamefont {Roca-Cusachs}, \citenamefont {Sunyer},
  \citenamefont {Hannezo},\ and\ \citenamefont {Trepat}}]{Fortunato2024}%
  \BibitemOpen
  \bibfield  {author} {\bibinfo {author} {\bibfnamefont {I.~C.}\ \bibnamefont
  {Fortunato}}, \bibinfo {author} {\bibfnamefont {D.~B.}\ \bibnamefont
  {Br{\"u}ckner}}, \bibinfo {author} {\bibfnamefont {S.}~\bibnamefont
  {Grosser}}, \bibinfo {author} {\bibfnamefont {L.}~\bibnamefont {Rossetti}},
  \bibinfo {author} {\bibfnamefont {M.}~\bibnamefont {Bosch-Padr{\'o}s}},
  \bibinfo {author} {\bibfnamefont {J.}~\bibnamefont {Trebicka}}, \bibinfo
  {author} {\bibfnamefont {P.}~\bibnamefont {Roca-Cusachs}}, \bibinfo {author}
  {\bibfnamefont {R.}~\bibnamefont {Sunyer}}, \bibinfo {author} {\bibfnamefont
  {E.}~\bibnamefont {Hannezo}},\ and\ \bibinfo {author} {\bibfnamefont
  {X.}~\bibnamefont {Trepat}},\ }\bibfield  {title} {\bibinfo {title} {Single
  cell migration along and against confined haptotactic gradients},\ }\bibfield
   {journal} {\bibinfo  {journal} {bioRxiv}\ }\href
  {https://doi.org/10.1101/2024.12.02.626413} {10.1101/2024.12.02.626413}
  (\bibinfo {year} {2024})\BibitemShut {NoStop}%
\bibitem [{\citenamefont {Leiphart}\ \emph {et~al.}(2019)\citenamefont
  {Leiphart}, \citenamefont {Chen}, \citenamefont {Peredo}, \citenamefont
  {Loneker},\ and\ \citenamefont {Janmey}}]{Leiphart2018}%
  \BibitemOpen
  \bibfield  {author} {\bibinfo {author} {\bibfnamefont {R.~J.}\ \bibnamefont
  {Leiphart}}, \bibinfo {author} {\bibfnamefont {D.}~\bibnamefont {Chen}},
  \bibinfo {author} {\bibfnamefont {A.~P.}\ \bibnamefont {Peredo}}, \bibinfo
  {author} {\bibfnamefont {A.~E.}\ \bibnamefont {Loneker}},\ and\ \bibinfo
  {author} {\bibfnamefont {P.~A.}\ \bibnamefont {Janmey}},\ }\bibfield  {title}
  {\bibinfo {title} {Mechanosensing at cellular interfaces},\ }\href
  {https://doi.org/10.1021/acs.langmuir.8b02841} {\bibfield  {journal}
  {\bibinfo  {journal} {Langmuir}\ }\textbf {\bibinfo {volume} {35}},\ \bibinfo
  {pages} {7509} (\bibinfo {year} {2019})}\BibitemShut {NoStop}%
\bibitem [{\citenamefont {Ji}\ and\ \citenamefont {Huang}(2023)}]{Ji2023}%
  \BibitemOpen
  \bibfield  {author} {\bibinfo {author} {\bibfnamefont {C.}~\bibnamefont
  {Ji}}\ and\ \bibinfo {author} {\bibfnamefont {Y.}~\bibnamefont {Huang}},\
  }\bibfield  {title} {\bibinfo {title} {Durotaxis and negative durotaxis:
  where should cells go?},\ }\href {https://doi.org/10.1038/s42003-023-05554-y}
  {\bibfield  {journal} {\bibinfo  {journal} {Communications Biology}\ }\textbf
  {\bibinfo {volume} {6}},\ \bibinfo {eid} {1169} (\bibinfo {year}
  {2023})}\BibitemShut {NoStop}%
\bibitem [{\citenamefont {Rappel}\ and\ \citenamefont
  {Edelstein-Keshet}(2017)}]{Rappel2017}%
  \BibitemOpen
  \bibfield  {author} {\bibinfo {author} {\bibfnamefont {W.-J.}\ \bibnamefont
  {Rappel}}\ and\ \bibinfo {author} {\bibfnamefont {L.}~\bibnamefont
  {Edelstein-Keshet}},\ }\bibfield  {title} {\bibinfo {title} {Mechanisms of
  cell polarization},\ }\href {https://doi.org/10.1016/j.coisb.2017.03.005}
  {\bibfield  {journal} {\bibinfo  {journal} {Current Opinion in Systems
  Biology}\ }\textbf {\bibinfo {volume} {3}},\ \bibinfo {pages} {43} (\bibinfo
  {year} {2017})}\BibitemShut {NoStop}%
\bibitem [{\citenamefont {Yolcu}\ \emph {et~al.}(2014)\citenamefont {Yolcu},
  \citenamefont {Haussman},\ and\ \citenamefont {Deserno}}]{Yolcu2014}%
  \BibitemOpen
  \bibfield  {author} {\bibinfo {author} {\bibfnamefont {C.}~\bibnamefont
  {Yolcu}}, \bibinfo {author} {\bibfnamefont {R.~C.}\ \bibnamefont
  {Haussman}},\ and\ \bibinfo {author} {\bibfnamefont {M.}~\bibnamefont
  {Deserno}},\ }\bibfield  {title} {\bibinfo {title} {The effective field
  theory approach towards membrane-mediated interactions between particles},\
  }\href {https://doi.org/10.1016/j.cis.2014.02.017} {\bibfield  {journal}
  {\bibinfo  {journal} {Advances in Colloid and Interface Science}\ }\textbf
  {\bibinfo {volume} {208}},\ \bibinfo {pages} {89} (\bibinfo {year}
  {2014})}\BibitemShut {NoStop}%
\bibitem [{\citenamefont {Karal}\ \emph {et~al.}(2023)\citenamefont {Karal},
  \citenamefont {Billah}, \citenamefont {Ahmed},\ and\ \citenamefont
  {Ahamed}}]{Karal2023}%
  \BibitemOpen
  \bibfield  {author} {\bibinfo {author} {\bibfnamefont {M.~A.~S.}\
  \bibnamefont {Karal}}, \bibinfo {author} {\bibfnamefont {M.~M.}\ \bibnamefont
  {Billah}}, \bibinfo {author} {\bibfnamefont {M.}~\bibnamefont {Ahmed}},\ and\
  \bibinfo {author} {\bibfnamefont {M.~K.}\ \bibnamefont {Ahamed}},\ }\bibfield
   {title} {\bibinfo {title} {A review on the measurement of the bending
  rigidity of lipid membranes},\ }\href {https://doi.org/10.1039/d3sm00882g}
  {\bibfield  {journal} {\bibinfo  {journal} {Soft Matter}\ }\textbf {\bibinfo
  {volume} {19}},\ \bibinfo {pages} {8285} (\bibinfo {year}
  {2023})}\BibitemShut {NoStop}%
\bibitem [{\citenamefont {Liboff}(1959)}]{Liboff1959}%
  \BibitemOpen
  \bibfield  {author} {\bibinfo {author} {\bibfnamefont {R.~L.}\ \bibnamefont
  {Liboff}},\ }\bibfield  {title} {\bibinfo {title} {Transport coefficients
  determined using the shielded coulomb potential},\ }\href
  {https://doi.org/10.1063/1.1724389} {\bibfield  {journal} {\bibinfo
  {journal} {The Physics of Fluids}\ }\textbf {\bibinfo {volume} {2}},\
  \bibinfo {pages} {40} (\bibinfo {year} {1959})}\BibitemShut {NoStop}%
\bibitem [{\citenamefont {Rowlinson}(1989)}]{Rowlinson1989}%
  \BibitemOpen
  \bibfield  {author} {\bibinfo {author} {\bibfnamefont {J.}~\bibnamefont
  {Rowlinson}},\ }\bibfield  {title} {\bibinfo {title} {The yukawa potential},\
  }\href {https://doi.org/10.1016/0378-4371(89)90108-8} {\bibfield  {journal}
  {\bibinfo  {journal} {Physica A: Statistical Mechanics and its Applications}\
  }\textbf {\bibinfo {volume} {156}},\ \bibinfo {pages} {15} (\bibinfo {year}
  {1989})}\BibitemShut {NoStop}%
\bibitem [{\citenamefont {Milo}\ and\ \citenamefont
  {Phillips}(2015)}]{Milo2015Avesizeprotein}%
  \BibitemOpen
  \bibfield  {author} {\bibinfo {author} {\bibfnamefont {R.}~\bibnamefont
  {Milo}}\ and\ \bibinfo {author} {\bibfnamefont {R.}~\bibnamefont
  {Phillips}},\ }\bibinfo {title} {Cell biology by the numbers}\ (\bibinfo
  {publisher} {Garland Science},\ \bibinfo {address} {New York},\ \bibinfo
  {year} {2015})\ Chap.~\bibinfo {chapter} {1}\BibitemShut {NoStop}%
\bibitem [{\citenamefont {Zhdanov}(2009)}]{Zhdanov2009}%
  \BibitemOpen
  \bibfield  {author} {\bibinfo {author} {\bibfnamefont {V.~P.}\ \bibnamefont
  {Zhdanov}},\ }\bibfield  {title} {\bibinfo {title} {Conditions of appreciable
  influence of microrna on a large number of target mrnas},\ }\href
  {https://doi.org/10.1039/b808095j} {\bibfield  {journal} {\bibinfo  {journal}
  {Molecular BioSystems}\ }\textbf {\bibinfo {volume} {5}},\ \bibinfo {pages}
  {638} (\bibinfo {year} {2009})}\BibitemShut {NoStop}%
\bibitem [{\citenamefont {Hu}\ \emph {et~al.}(2012)\citenamefont {Hu},
  \citenamefont {Briguglio},\ and\ \citenamefont {Deserno}}]{Hu2012}%
  \BibitemOpen
  \bibfield  {author} {\bibinfo {author} {\bibfnamefont {M.}~\bibnamefont
  {Hu}}, \bibinfo {author} {\bibfnamefont {J.~J.}\ \bibnamefont {Briguglio}},\
  and\ \bibinfo {author} {\bibfnamefont {M.}~\bibnamefont {Deserno}},\
  }\bibfield  {title} {\bibinfo {title} {Determining the gaussian curvature
  modulus of lipid membranes in simulations},\ }\href
  {https://doi.org/10.1016/j.bpj.2012.02.013} {\bibfield  {journal} {\bibinfo
  {journal} {Biophysical Journal}\ }\textbf {\bibinfo {volume} {102}},\
  \bibinfo {pages} {1403} (\bibinfo {year} {2012})}\BibitemShut {NoStop}%
\bibitem [{\citenamefont {Wennerstr\"{o}m}\ \emph {et~al.}(2020)\citenamefont
  {Wennerstr\"{o}m}, \citenamefont {Vallina~Estrada}, \citenamefont
  {Danielsson},\ and\ \citenamefont {Oliveberg}}]{Wennerstrm2020}%
  \BibitemOpen
  \bibfield  {author} {\bibinfo {author} {\bibfnamefont {H.}~\bibnamefont
  {Wennerstr\"{o}m}}, \bibinfo {author} {\bibfnamefont {E.}~\bibnamefont
  {Vallina~Estrada}}, \bibinfo {author} {\bibfnamefont {J.}~\bibnamefont
  {Danielsson}},\ and\ \bibinfo {author} {\bibfnamefont {M.}~\bibnamefont
  {Oliveberg}},\ }\bibfield  {title} {\bibinfo {title} {Colloidal stability of
  the living cell},\ }\href {https://doi.org/10.1073/pnas.1914599117}
  {\bibfield  {journal} {\bibinfo  {journal} {Proceedings of the National
  Academy of Sciences}\ }\textbf {\bibinfo {volume} {117}},\ \bibinfo {pages}
  {10113} (\bibinfo {year} {2020})}\BibitemShut {NoStop}%
\bibitem [{\citenamefont {Seelaboyina}\ and\ \citenamefont
  {Vishwakarma}(2023)}]{Seelaboyina2023}%
  \BibitemOpen
  \bibfield  {author} {\bibinfo {author} {\bibfnamefont {R.}~\bibnamefont
  {Seelaboyina}}\ and\ \bibinfo {author} {\bibfnamefont {R.}~\bibnamefont
  {Vishwakarma}},\ }\bibfield  {title} {\bibinfo {title} {Different
  thresholding techniques in image processing : A review},\ }in\ \href@noop {}
  {\emph {\bibinfo {booktitle} {ICDSMLA 2021}}},\ \bibinfo {editor} {edited by\
  \bibinfo {editor} {\bibfnamefont {A.}~\bibnamefont {Kumar}}, \bibinfo
  {editor} {\bibfnamefont {S.}~\bibnamefont {Senatore}},\ and\ \bibinfo
  {editor} {\bibfnamefont {V.~K.}\ \bibnamefont {Gunjan}}}\ (\bibinfo
  {publisher} {Springer Nature Singapore},\ \bibinfo {address} {Singapore},\
  \bibinfo {year} {2023})\ pp.\ \bibinfo {pages} {23--29}\BibitemShut {NoStop}%
\bibitem [{\citenamefont {Dudin}\ \emph {et~al.}(2019)\citenamefont {Dudin},
  \citenamefont {Ondracka}, \citenamefont {Grau-Bové}, \citenamefont
  {Haraldsen}, \citenamefont {Toyoda}, \citenamefont {Suga}, \citenamefont
  {Bråte},\ and\ \citenamefont {Ruiz-Trillo}}]{Dudin2019}%
  \BibitemOpen
  \bibfield  {author} {\bibinfo {author} {\bibfnamefont {O.}~\bibnamefont
  {Dudin}}, \bibinfo {author} {\bibfnamefont {A.}~\bibnamefont {Ondracka}},
  \bibinfo {author} {\bibfnamefont {X.}~\bibnamefont {Grau-Bové}}, \bibinfo
  {author} {\bibfnamefont {A.~A.}\ \bibnamefont {Haraldsen}}, \bibinfo {author}
  {\bibfnamefont {A.}~\bibnamefont {Toyoda}}, \bibinfo {author} {\bibfnamefont
  {H.}~\bibnamefont {Suga}}, \bibinfo {author} {\bibfnamefont {J.}~\bibnamefont
  {Bråte}},\ and\ \bibinfo {author} {\bibfnamefont {I.}~\bibnamefont
  {Ruiz-Trillo}},\ }\bibfield  {title} {\bibinfo {title} {A unicellular
  relative of animals generates a layer of polarized cells by
  actomyosin-dependent cellularization},\ }\href
  {https://doi.org/10.7554/elife.49801} {\bibfield  {journal} {\bibinfo
  {journal} {eLife}\ }\textbf {\bibinfo {volume} {8}},\ \bibinfo {eid} {e49801}
  (\bibinfo {year} {2019})}\BibitemShut {NoStop}%
\bibitem [{\citenamefont {Shah}\ \emph {et~al.}(2024)\citenamefont {Shah},
  \citenamefont {Olivetta}, \citenamefont {Bhickta}, \citenamefont {Ronchi},
  \citenamefont {Trupinić}, \citenamefont {Tromer}, \citenamefont {Tolić},
  \citenamefont {Schwab}, \citenamefont {Dudin},\ and\ \citenamefont
  {Dey}}]{Shah2024}%
  \BibitemOpen
  \bibfield  {author} {\bibinfo {author} {\bibfnamefont {H.}~\bibnamefont
  {Shah}}, \bibinfo {author} {\bibfnamefont {M.}~\bibnamefont {Olivetta}},
  \bibinfo {author} {\bibfnamefont {C.}~\bibnamefont {Bhickta}}, \bibinfo
  {author} {\bibfnamefont {P.}~\bibnamefont {Ronchi}}, \bibinfo {author}
  {\bibfnamefont {M.}~\bibnamefont {Trupinić}}, \bibinfo {author}
  {\bibfnamefont {E.~C.}\ \bibnamefont {Tromer}}, \bibinfo {author}
  {\bibfnamefont {I.~M.}\ \bibnamefont {Tolić}}, \bibinfo {author}
  {\bibfnamefont {Y.}~\bibnamefont {Schwab}}, \bibinfo {author} {\bibfnamefont
  {O.}~\bibnamefont {Dudin}},\ and\ \bibinfo {author} {\bibfnamefont
  {G.}~\bibnamefont {Dey}},\ }\bibfield  {title} {\bibinfo {title}
  {Life-cycle-coupled evolution of mitosis in close relatives of animals},\
  }\href {https://doi.org/10.1038/s41586-024-07430-z} {\bibfield  {journal}
  {\bibinfo  {journal} {Nature}\ }\textbf {\bibinfo {volume} {630}},\ \bibinfo
  {pages} {116} (\bibinfo {year} {2024})}\BibitemShut {NoStop}%
\bibitem [{\citenamefont {Halatek}\ and\ \citenamefont
  {Frey}(2018)}]{Halatek2018}%
  \BibitemOpen
  \bibfield  {author} {\bibinfo {author} {\bibfnamefont {J.}~\bibnamefont
  {Halatek}}\ and\ \bibinfo {author} {\bibfnamefont {E.}~\bibnamefont {Frey}},\
  }\bibfield  {title} {\bibinfo {title} {Rethinking pattern formation in
  reaction–diffusion systems},\ }\href
  {https://doi.org/10.1038/s41567-017-0040-5} {\bibfield  {journal} {\bibinfo
  {journal} {Nature Physics}\ }\textbf {\bibinfo {volume} {14}},\ \bibinfo
  {pages} {507} (\bibinfo {year} {2018})}\BibitemShut {NoStop}%
\bibitem [{\citenamefont {Brauns}\ \emph {et~al.}(2020)\citenamefont {Brauns},
  \citenamefont {Halatek},\ and\ \citenamefont {Frey}}]{Brauns2020}%
  \BibitemOpen
  \bibfield  {author} {\bibinfo {author} {\bibfnamefont {F.}~\bibnamefont
  {Brauns}}, \bibinfo {author} {\bibfnamefont {J.}~\bibnamefont {Halatek}},\
  and\ \bibinfo {author} {\bibfnamefont {E.}~\bibnamefont {Frey}},\ }\bibfield
  {title} {\bibinfo {title} {Phase-space geometry of mass-conserving
  reaction-diffusion dynamics},\ }\href
  {https://doi.org/10.1103/physrevx.10.041036} {\bibfield  {journal} {\bibinfo
  {journal} {Physical Review X}\ }\textbf {\bibinfo {volume} {10}},\ \bibinfo
  {eid} {041036} (\bibinfo {year} {2020})}\BibitemShut {NoStop}%
\bibitem [{\citenamefont {Gardiner}(2010)}]{Gardiner2010-zc}%
  \BibitemOpen
  \bibfield  {author} {\bibinfo {author} {\bibfnamefont {C.~W.}\ \bibnamefont
  {Gardiner}},\ }\bibinfo {title} {Stochastic methods: A handbook for the
  natural and social sciences}\ (\bibinfo  {publisher} {Springer},\ \bibinfo
  {address} {Berlin},\ \bibinfo {year} {2010})\ Chap.~\bibinfo {chapter}
  {13}\BibitemShut {NoStop}%
\bibitem [{\citenamefont {Belousov}\ \emph {et~al.}(2022)\citenamefont
  {Belousov}, \citenamefont {Hassanali},\ and\ \citenamefont
  {Rold{\'{a}}n}}]{Belousov2022}%
  \BibitemOpen
  \bibfield  {author} {\bibinfo {author} {\bibfnamefont {R.}~\bibnamefont
  {Belousov}}, \bibinfo {author} {\bibfnamefont {A.}~\bibnamefont
  {Hassanali}},\ and\ \bibinfo {author} {\bibfnamefont {{\'{E}}.}~\bibnamefont
  {Rold{\'{a}}n}},\ }\bibfield  {title} {\bibinfo {title} {Statistical physics
  of inhomogeneous transport: Unification of diffusion laws and inference from
  first-passage statistics},\ }\href
  {https://doi.org/10.1103/physreve.106.014103} {\bibfield  {journal} {\bibinfo
   {journal} {Physical Review E}\ }\textbf {\bibinfo {volume} {106}},\ \bibinfo
  {eid} {014103} (\bibinfo {year} {2022})}\BibitemShut {NoStop}%
\bibitem [{\citenamefont {Carter}(2000)}]{Carter2000}%
  \BibitemOpen
  \bibfield  {author} {\bibinfo {author} {\bibfnamefont {A.~H.}\ \bibnamefont
  {Carter}},\ }\href@noop {} {\emph {\bibinfo {title} {Classical and
  statistical thermodynamics}}}\ (\bibinfo  {publisher} {Pearson Education},\
  \bibinfo {address} {Philadelphia},\ \bibinfo {year} {2000})\BibitemShut
  {NoStop}%
\bibitem [{\citenamefont {Landau}\ and\ \citenamefont
  {Lifshitz}(1996)}]{Landau1996-qq}%
  \BibitemOpen
  \bibfield  {author} {\bibinfo {author} {\bibfnamefont {L.~D.}\ \bibnamefont
  {Landau}}\ and\ \bibinfo {author} {\bibfnamefont {E.~M.}\ \bibnamefont
  {Lifshitz}},\ }\bibinfo {title} {Statistical physics}\ (\bibinfo  {publisher}
  {Butterworth-Heinemann},\ \bibinfo {address} {Oxford},\ \bibinfo {year}
  {1996})\ p.\ \bibinfo {pages} {134},\ \bibinfo {edition} {3rd}\
  ed.\BibitemShut {Stop}%
\bibitem [{\citenamefont {Hill}(1987)}]{Hill1987-fu}%
  \BibitemOpen
  \bibfield  {author} {\bibinfo {author} {\bibfnamefont {T.~L.}\ \bibnamefont
  {Hill}},\ }\bibinfo {title} {An introduction to statistical thermodynamics}\
  (\bibinfo  {publisher} {Dover Publications},\ \bibinfo {address} {MineolaY},\
  \bibinfo {year} {1987})\ p.\ \bibinfo {pages} {128}\BibitemShut {NoStop}%
\bibitem [{\citenamefont {Tishby}\ \emph {et~al.}()\citenamefont {Tishby},
  \citenamefont {Pereira},\ and\ \citenamefont {Bialek}}]{Tishby2000}%
  \BibitemOpen
  \bibfield  {author} {\bibinfo {author} {\bibfnamefont {N.}~\bibnamefont
  {Tishby}}, \bibinfo {author} {\bibfnamefont {F.~C.}\ \bibnamefont
  {Pereira}},\ and\ \bibinfo {author} {\bibfnamefont {W.}~\bibnamefont
  {Bialek}},\ }\href {https://arxiv.org/abs/physics/0004057} {}\Eprint
  {https://arxiv.org/abs/physics/0004057} {arXiv:physics/0004057
  [physics.data-an]} \BibitemShut {NoStop}%
\bibitem [{\citenamefont {Bauer}\ and\ \citenamefont
  {Bialek}(2023)}]{Bauer2023}%
  \BibitemOpen
  \bibfield  {author} {\bibinfo {author} {\bibfnamefont {M.}~\bibnamefont
  {Bauer}}\ and\ \bibinfo {author} {\bibfnamefont {W.}~\bibnamefont {Bialek}},\
  }\bibfield  {title} {\bibinfo {title} {Information bottleneck in molecular
  sensing},\ }\href {https://doi.org/10.1103/prxlife.1.023005} {\bibfield
  {journal} {\bibinfo  {journal} {PRX Life}\ }\textbf {\bibinfo {volume} {1}},\
  \bibinfo {eid} {023005} (\bibinfo {year} {2023})}\BibitemShut {NoStop}%
\bibitem [{\citenamefont {Bauer}(2022)}]{Bauer2022}%
  \BibitemOpen
  \bibfield  {author} {\bibinfo {author} {\bibfnamefont {M.}~\bibnamefont
  {Bauer}},\ }\bibfield  {title} {\bibinfo {title} {How does an organism
  extract relevant information from transcription factor concentrations?},\
  }\href {https://doi.org/10.1042/bst20220333} {\bibfield  {journal} {\bibinfo
  {journal} {Biochemical Society Transactions}\ }\textbf {\bibinfo {volume}
  {50}},\ \bibinfo {pages} {1365} (\bibinfo {year} {2022})}\BibitemShut
  {NoStop}%
\bibitem [{\citenamefont {Kleinman}\ \emph {et~al.}(2025)\citenamefont
  {Kleinman}, \citenamefont {Wang}, \citenamefont {Xiao}, \citenamefont
  {Feghhi}, \citenamefont {Lee}, \citenamefont {Carr}, \citenamefont {Li},
  \citenamefont {Hadidi}, \citenamefont {Chandrasekaran},\ and\ \citenamefont
  {Kao}}]{Kleinman2023}%
  \BibitemOpen
  \bibfield  {author} {\bibinfo {author} {\bibfnamefont {M.}~\bibnamefont
  {Kleinman}}, \bibinfo {author} {\bibfnamefont {T.}~\bibnamefont {Wang}},
  \bibinfo {author} {\bibfnamefont {D.}~\bibnamefont {Xiao}}, \bibinfo {author}
  {\bibfnamefont {E.}~\bibnamefont {Feghhi}}, \bibinfo {author} {\bibfnamefont
  {K.}~\bibnamefont {Lee}}, \bibinfo {author} {\bibfnamefont {N.}~\bibnamefont
  {Carr}}, \bibinfo {author} {\bibfnamefont {Y.}~\bibnamefont {Li}}, \bibinfo
  {author} {\bibfnamefont {N.}~\bibnamefont {Hadidi}}, \bibinfo {author}
  {\bibfnamefont {C.}~\bibnamefont {Chandrasekaran}},\ and\ \bibinfo {author}
  {\bibfnamefont {J.~C.}\ \bibnamefont {Kao}},\ }\bibfield  {title} {\bibinfo
  {title} {The information bottleneck as a principle underlying multi-area
  cortical representations during decision-making},\ }\bibfield  {journal}
  {\bibinfo  {journal} {bioRxiv}\ }\href
  {https://doi.org/10.1101/2023.07.12.548742} {10.1101/2023.07.12.548742}
  (\bibinfo {year} {2025})\BibitemShut {NoStop}%
\bibitem [{\citenamefont {Mancini}(2003)}]{Mancini2003}%
  \BibitemOpen
  \bibfield  {author} {\bibinfo {author} {\bibfnamefont {R.}~\bibnamefont
  {Mancini}},\ }\bibfield  {title} {\bibinfo {title} {Chapter 6 - development
  of the non ideal op amp equations},\ }in\ \href
  {https://doi.org/https://doi.org/10.1016/B978-075067701-1/50009-2} {\emph
  {\bibinfo {booktitle} {Op Amps for Everyone (Second Edition)}}},\ \bibinfo
  {editor} {edited by\ \bibinfo {editor} {\bibfnamefont {R.}~\bibnamefont
  {Mancini}}}\ (\bibinfo  {publisher} {Newnes},\ \bibinfo {address}
  {Burlington},\ \bibinfo {year} {2003})\ \bibinfo {edition} {second edition}\
  ed.,\ pp.\ \bibinfo {pages} {67--75}\BibitemShut {NoStop}%
\bibitem [{\citenamefont {Touchette}(2009)}]{Touchette2009}%
  \BibitemOpen
  \bibfield  {author} {\bibinfo {author} {\bibfnamefont {H.}~\bibnamefont
  {Touchette}},\ }\bibfield  {title} {\bibinfo {title} {The large deviation
  approach to statistical mechanics},\ }\href
  {https://doi.org/10.1016/j.physrep.2009.05.002} {\bibfield  {journal}
  {\bibinfo  {journal} {Physics Reports}\ }\textbf {\bibinfo {volume} {478}},\
  \bibinfo {pages} {1} (\bibinfo {year} {2009})}\BibitemShut {NoStop}%
\bibitem [{\citenamefont
  {Elliott}(2025{\natexlab{a}})}]{ElliottRepoMetropolisSims2025}%
  \BibitemOpen
  \bibfield  {author} {\bibinfo {author} {\bibfnamefont {J.}~\bibnamefont
  {Elliott}},\ }\href@noop {} {\bibinfo {title} {Elliott2025 metropolis
  simulations}},\ \bibinfo {howpublished}
  {\url{https://git.embl.de/elliot/Elliott2025-Metropolis-Simulations}}
  (\bibinfo {year} {2025}{\natexlab{a}})\BibitemShut {NoStop}%
\bibitem [{\citenamefont {Lembo}\ \emph {et~al.}(2023)\citenamefont {Lembo},
  \citenamefont {Strauss}, \citenamefont {Cheng}, \citenamefont {Vermeil},
  \citenamefont {Siggel}, \citenamefont {Toro-Nahuelpan}, \citenamefont {Chan},
  \citenamefont {Kosinski}, \citenamefont {Piel}, \citenamefont {Du~Roure},
  \citenamefont {Heuvingh}, \citenamefont {Mahamid},\ and\ \citenamefont
  {Diz-Mu{\~n}oz}}]{Lembo2023}%
  \BibitemOpen
  \bibfield  {author} {\bibinfo {author} {\bibfnamefont {S.}~\bibnamefont
  {Lembo}}, \bibinfo {author} {\bibfnamefont {L.}~\bibnamefont {Strauss}},
  \bibinfo {author} {\bibfnamefont {D.}~\bibnamefont {Cheng}}, \bibinfo
  {author} {\bibfnamefont {J.}~\bibnamefont {Vermeil}}, \bibinfo {author}
  {\bibfnamefont {M.}~\bibnamefont {Siggel}}, \bibinfo {author} {\bibfnamefont
  {M.}~\bibnamefont {Toro-Nahuelpan}}, \bibinfo {author} {\bibfnamefont
  {C.~J.}\ \bibnamefont {Chan}}, \bibinfo {author} {\bibfnamefont
  {J.}~\bibnamefont {Kosinski}}, \bibinfo {author} {\bibfnamefont
  {M.}~\bibnamefont {Piel}}, \bibinfo {author} {\bibfnamefont {O.}~\bibnamefont
  {Du~Roure}}, \bibinfo {author} {\bibfnamefont {J.}~\bibnamefont {Heuvingh}},
  \bibinfo {author} {\bibfnamefont {J.}~\bibnamefont {Mahamid}},\ and\ \bibinfo
  {author} {\bibfnamefont {A.}~\bibnamefont {Diz-Mu{\~n}oz}},\ }\bibfield
  {title} {\bibinfo {title} {The distance between the plasma membrane and the
  actomyosin cortex acts as a nanogate to control cell surface mechanics},\
  }\bibfield  {journal} {\bibinfo  {journal} {bioRxiv}\ }\href
  {https://doi.org/10.1101/2023.01.31.526409} {10.1101/2023.01.31.526409}
  (\bibinfo {year} {2023})\BibitemShut {NoStop}%
\bibitem [{\citenamefont {Robert}\ \emph {et~al.}(2021)\citenamefont {Robert},
  \citenamefont {Biarnes-Pelicot}, \citenamefont {Garcia-Seyda}, \citenamefont
  {Hatoum}, \citenamefont {Touchard}, \citenamefont {Brustlein}, \citenamefont
  {Nicolas}, \citenamefont {Malissen}, \citenamefont {Valignat},\ and\
  \citenamefont {Theodoly}}]{Robert2021}%
  \BibitemOpen
  \bibfield  {author} {\bibinfo {author} {\bibfnamefont {P.}~\bibnamefont
  {Robert}}, \bibinfo {author} {\bibfnamefont {M.}~\bibnamefont
  {Biarnes-Pelicot}}, \bibinfo {author} {\bibfnamefont {N.}~\bibnamefont
  {Garcia-Seyda}}, \bibinfo {author} {\bibfnamefont {P.}~\bibnamefont
  {Hatoum}}, \bibinfo {author} {\bibfnamefont {D.}~\bibnamefont {Touchard}},
  \bibinfo {author} {\bibfnamefont {S.}~\bibnamefont {Brustlein}}, \bibinfo
  {author} {\bibfnamefont {P.}~\bibnamefont {Nicolas}}, \bibinfo {author}
  {\bibfnamefont {B.}~\bibnamefont {Malissen}}, \bibinfo {author}
  {\bibfnamefont {M.-P.}\ \bibnamefont {Valignat}},\ and\ \bibinfo {author}
  {\bibfnamefont {O.}~\bibnamefont {Theodoly}},\ }\bibfield  {title} {\bibinfo
  {title} {Functional mapping of adhesiveness on live cells reveals how
  guidance phenotypes can emerge from complex spatiotemporal integrin
  regulation},\ }\href {https://doi.org/10.3389/fbioe.2021.625366} {\bibfield
  {journal} {\bibinfo  {journal} {Frontiers in Bioengineering and
  Biotechnology}\ }\textbf {\bibinfo {volume} {9}},\ \bibinfo {eid} {625366}
  (\bibinfo {year} {2021})}\BibitemShut {NoStop}%
\bibitem [{\citenamefont {Ekerdt}\ \emph {et~al.}(2013)\citenamefont {Ekerdt},
  \citenamefont {Segalman},\ and\ \citenamefont {Schaffer}}]{Ekerdt2013}%
  \BibitemOpen
  \bibfield  {author} {\bibinfo {author} {\bibfnamefont {B.~L.}\ \bibnamefont
  {Ekerdt}}, \bibinfo {author} {\bibfnamefont {R.~A.}\ \bibnamefont
  {Segalman}},\ and\ \bibinfo {author} {\bibfnamefont {D.~V.}\ \bibnamefont
  {Schaffer}},\ }\bibfield  {title} {\bibinfo {title} {Spatial organization of
  cell‐adhesive ligands for advanced cell culture},\ }\href
  {https://doi.org/10.1002/biot.201300302} {\bibfield  {journal} {\bibinfo
  {journal} {Biotechnology Journal}\ }\textbf {\bibinfo {volume} {8}},\
  \bibinfo {pages} {1411} (\bibinfo {year} {2013})}\BibitemShut {NoStop}%
\bibitem [{\citenamefont {Truong~Quang}\ \emph {et~al.}(2013)\citenamefont
  {Truong~Quang}, \citenamefont {Mani}, \citenamefont {Markova}, \citenamefont
  {Lecuit},\ and\ \citenamefont {Lenne}}]{TruongQuang2013}%
  \BibitemOpen
  \bibfield  {author} {\bibinfo {author} {\bibfnamefont {B.-A.}\ \bibnamefont
  {Truong~Quang}}, \bibinfo {author} {\bibfnamefont {M.}~\bibnamefont {Mani}},
  \bibinfo {author} {\bibfnamefont {O.}~\bibnamefont {Markova}}, \bibinfo
  {author} {\bibfnamefont {T.}~\bibnamefont {Lecuit}},\ and\ \bibinfo {author}
  {\bibfnamefont {P.-F.}\ \bibnamefont {Lenne}},\ }\bibfield  {title} {\bibinfo
  {title} {Principles of e-cadherin supramolecular organization in vivo},\
  }\href {https://doi.org/10.1016/j.cub.2013.09.015} {\bibfield  {journal}
  {\bibinfo  {journal} {Current Biology}\ }\textbf {\bibinfo {volume} {23}},\
  \bibinfo {pages} {2197} (\bibinfo {year} {2013})}\BibitemShut {NoStop}%
\bibitem [{\citenamefont {{Janeway Jr.}}\ \emph {et~al.}(2001)\citenamefont
  {{Janeway Jr.}}, \citenamefont {Travers}, \citenamefont {Walport},\ and\
  \citenamefont {et~al.}}]{Janeway2001}%
  \BibitemOpen
  \bibfield  {author} {\bibinfo {author} {\bibfnamefont {C.~A.}\ \bibnamefont
  {{Janeway Jr.}}}, \bibinfo {author} {\bibfnamefont {P.}~\bibnamefont
  {Travers}}, \bibinfo {author} {\bibfnamefont {M.}~\bibnamefont {Walport}},\
  and\ \bibinfo {author} {\bibnamefont {et~al.}},\ }in\ \href
  {https://www.ncbi.nlm.nih.gov/books/NBK27098/} {\emph {\bibinfo {booktitle}
  {Immunobiology: The Immune System in Health and Disease}}}\ (\bibinfo
  {publisher} {Garland Science},\ \bibinfo {address} {New York},\ \bibinfo
  {year} {2001})\ Chap.\ \bibinfo {chapter} {Antigen recognition by T cells},\
  \bibinfo {edition} {5th}\ ed.\BibitemShut {Stop}%
\bibitem [{\citenamefont {Jiang}\ \emph {et~al.}(2020)\citenamefont {Jiang},
  \citenamefont {Dudzinski}, \citenamefont {Beckermann}, \citenamefont {Young},
  \citenamefont {McKinley}, \citenamefont {McIntyre}, \citenamefont {Rathmell},
  \citenamefont {Xu},\ and\ \citenamefont {Gore}}]{Jiang2020}%
  \BibitemOpen
  \bibfield  {author} {\bibinfo {author} {\bibfnamefont {X.}~\bibnamefont
  {Jiang}}, \bibinfo {author} {\bibfnamefont {S.}~\bibnamefont {Dudzinski}},
  \bibinfo {author} {\bibfnamefont {K.~E.}\ \bibnamefont {Beckermann}},
  \bibinfo {author} {\bibfnamefont {K.}~\bibnamefont {Young}}, \bibinfo
  {author} {\bibfnamefont {E.}~\bibnamefont {McKinley}}, \bibinfo {author}
  {\bibfnamefont {J.~O.}\ \bibnamefont {McIntyre}}, \bibinfo {author}
  {\bibfnamefont {J.~C.}\ \bibnamefont {Rathmell}}, \bibinfo {author}
  {\bibfnamefont {J.}~\bibnamefont {Xu}},\ and\ \bibinfo {author}
  {\bibfnamefont {J.~C.}\ \bibnamefont {Gore}},\ }\bibfield  {title} {\bibinfo
  {title} {Mri of tumor t cell infiltration in response to checkpoint inhibitor
  therapy},\ }\href {https://doi.org/10.1136/jitc-2019-000328} {\bibfield
  {journal} {\bibinfo  {journal} {Journal for ImmunoTherapy of Cancer}\
  }\textbf {\bibinfo {volume} {8}},\ \bibinfo {pages} {e000328} (\bibinfo
  {year} {2020})}\BibitemShut {NoStop}%
\bibitem [{\citenamefont {Swamy}\ \emph {et~al.}(2008)\citenamefont {Swamy},
  \citenamefont {Dopfer}, \citenamefont {Molnar}, \citenamefont {Alarcón},\
  and\ \citenamefont {Schamel}}]{Swamy2008}%
  \BibitemOpen
  \bibfield  {author} {\bibinfo {author} {\bibfnamefont {M.}~\bibnamefont
  {Swamy}}, \bibinfo {author} {\bibfnamefont {E.~P.}\ \bibnamefont {Dopfer}},
  \bibinfo {author} {\bibfnamefont {E.}~\bibnamefont {Molnar}}, \bibinfo
  {author} {\bibfnamefont {B.}~\bibnamefont {Alarcón}},\ and\ \bibinfo
  {author} {\bibfnamefont {W.~W.~A.}\ \bibnamefont {Schamel}},\ }\bibfield
  {title} {\bibinfo {title} {The 450kda tcr complex has a stoichiometry of
  αβγεδεζζ},\ }\href {https://doi.org/10.1111/j.1365-3083.2008.02082.x}
  {\bibfield  {journal} {\bibinfo  {journal} {Scandinavian Journal of
  Immunology}\ }\textbf {\bibinfo {volume} {67}},\ \bibinfo {pages} {418}
  (\bibinfo {year} {2008})}\BibitemShut {NoStop}%
\bibitem [{\citenamefont {Ma}\ \emph {et~al.}(2022)\citenamefont {Ma},
  \citenamefont {Hu}, \citenamefont {Kellner}, \citenamefont {Brockman},
  \citenamefont {Velusamy}, \citenamefont {Blanchard}, \citenamefont {Evavold},
  \citenamefont {Alon},\ and\ \citenamefont {Salaita}}]{Ma2022}%
  \BibitemOpen
  \bibfield  {author} {\bibinfo {author} {\bibfnamefont {V.~P.-Y.}\
  \bibnamefont {Ma}}, \bibinfo {author} {\bibfnamefont {Y.}~\bibnamefont {Hu}},
  \bibinfo {author} {\bibfnamefont {A.~V.}\ \bibnamefont {Kellner}}, \bibinfo
  {author} {\bibfnamefont {J.~M.}\ \bibnamefont {Brockman}}, \bibinfo {author}
  {\bibfnamefont {A.}~\bibnamefont {Velusamy}}, \bibinfo {author}
  {\bibfnamefont {A.~T.}\ \bibnamefont {Blanchard}}, \bibinfo {author}
  {\bibfnamefont {B.~D.}\ \bibnamefont {Evavold}}, \bibinfo {author}
  {\bibfnamefont {R.}~\bibnamefont {Alon}},\ and\ \bibinfo {author}
  {\bibfnamefont {K.}~\bibnamefont {Salaita}},\ }\bibfield  {title} {\bibinfo
  {title} {The magnitude of lfa-1/icam-1 forces fine-tune tcr-triggered t cell
  activation},\ }\href {https://doi.org/10.1126/sciadv.abg4485} {\bibfield
  {journal} {\bibinfo  {journal} {Science Advances}\ }\textbf {\bibinfo
  {volume} {8}},\ \bibinfo {eid} {eabg4485} (\bibinfo {year}
  {2022})}\BibitemShut {NoStop}%
\bibitem [{\citenamefont {Bui}\ \emph {et~al.}(2020)\citenamefont {Bui},
  \citenamefont {Wiesolek},\ and\ \citenamefont {Sumagin}}]{Bui2020}%
  \BibitemOpen
  \bibfield  {author} {\bibinfo {author} {\bibfnamefont {T.~M.}\ \bibnamefont
  {Bui}}, \bibinfo {author} {\bibfnamefont {H.~L.}\ \bibnamefont {Wiesolek}},\
  and\ \bibinfo {author} {\bibfnamefont {R.}~\bibnamefont {Sumagin}},\
  }\bibfield  {title} {\bibinfo {title} {Icam-1: A master regulator of cellular
  responses in inflammation, injury resolution, and tumorigenesis},\ }\href
  {https://doi.org/10.1002/jlb.2mr0220-549r} {\bibfield  {journal} {\bibinfo
  {journal} {Journal of Leukocyte Biology}\ }\textbf {\bibinfo {volume}
  {108}},\ \bibinfo {pages} {787} (\bibinfo {year} {2020})}\BibitemShut
  {NoStop}%
\bibitem [{\citenamefont {Kurz-Isler}\ \emph {et~al.}(1992)\citenamefont
  {Kurz-Isler}, \citenamefont {Voigt},\ and\ \citenamefont
  {Wolburg}}]{KurzIsler1992}%
  \BibitemOpen
  \bibfield  {author} {\bibinfo {author} {\bibfnamefont {G.}~\bibnamefont
  {Kurz-Isler}}, \bibinfo {author} {\bibfnamefont {T.}~\bibnamefont {Voigt}},\
  and\ \bibinfo {author} {\bibfnamefont {H.}~\bibnamefont {Wolburg}},\
  }\bibfield  {title} {\bibinfo {title} {Modulation of connexon densities in
  gap junctions of horizontal cell perikarya and axon terminals in fish retina:
  effects of light/dark cycles, interruption of the optic nerve and application
  of dopamine},\ }\href {https://doi.org/10.1007/bf00318795} {\bibfield
  {journal} {\bibinfo  {journal} {Cell \& Tissue Research}\ }\textbf {\bibinfo
  {volume} {268}},\ \bibinfo {pages} {267} (\bibinfo {year}
  {1992})}\BibitemShut {NoStop}%
\bibitem [{\citenamefont {Goodenough}\ and\ \citenamefont
  {Revel}(1970)}]{Goodenough1970}%
  \BibitemOpen
  \bibfield  {author} {\bibinfo {author} {\bibfnamefont {D.~A.}\ \bibnamefont
  {Goodenough}}\ and\ \bibinfo {author} {\bibfnamefont {J.~P.}\ \bibnamefont
  {Revel}},\ }\bibfield  {title} {\bibinfo {title} {A fine structural analysis
  of intercellular junctions in the mouse liver},\ }\href
  {https://doi.org/10.1083/jcb.45.2.272} {\bibfield  {journal} {\bibinfo
  {journal} {The Journal of Cell Biology}\ }\textbf {\bibinfo {volume} {45}},\
  \bibinfo {pages} {272} (\bibinfo {year} {1970})}\BibitemShut {NoStop}%
\bibitem [{\citenamefont {Kuntze}\ \emph {et~al.}(2020)\citenamefont {Kuntze},
  \citenamefont {Goetsch}, \citenamefont {Fels}, \citenamefont {Najder},
  \citenamefont {Unger}, \citenamefont {Wilhelmi}, \citenamefont {Sargin},
  \citenamefont {Schimmelpfennig}, \citenamefont {Neumann}, \citenamefont
  {Schwab},\ and\ \citenamefont {Pethő}}]{Kuntze2020}%
  \BibitemOpen
  \bibfield  {author} {\bibinfo {author} {\bibfnamefont {A.}~\bibnamefont
  {Kuntze}}, \bibinfo {author} {\bibfnamefont {O.}~\bibnamefont {Goetsch}},
  \bibinfo {author} {\bibfnamefont {B.}~\bibnamefont {Fels}}, \bibinfo {author}
  {\bibfnamefont {K.}~\bibnamefont {Najder}}, \bibinfo {author} {\bibfnamefont
  {A.}~\bibnamefont {Unger}}, \bibinfo {author} {\bibfnamefont
  {M.}~\bibnamefont {Wilhelmi}}, \bibinfo {author} {\bibfnamefont
  {S.}~\bibnamefont {Sargin}}, \bibinfo {author} {\bibfnamefont
  {S.}~\bibnamefont {Schimmelpfennig}}, \bibinfo {author} {\bibfnamefont
  {I.}~\bibnamefont {Neumann}}, \bibinfo {author} {\bibfnamefont
  {A.}~\bibnamefont {Schwab}},\ and\ \bibinfo {author} {\bibfnamefont
  {Z.}~\bibnamefont {Pethő}},\ }\bibfield  {title} {\bibinfo {title}
  {Protonation of piezo1 impairs cell-matrix interactions of pancreatic
  stellate cells},\ }\href {https://doi.org/10.3389/fphys.2020.00089}
  {\bibfield  {journal} {\bibinfo  {journal} {Frontiers in Physiology}\
  }\textbf {\bibinfo {volume} {11}},\ \bibinfo {eid} {89} (\bibinfo {year}
  {2020})}\BibitemShut {NoStop}%
\bibitem [{\citenamefont {Mulhall}\ \emph {et~al.}(2023)\citenamefont
  {Mulhall}, \citenamefont {Gharpure}, \citenamefont {Lee}, \citenamefont
  {Dubin}, \citenamefont {Aaron}, \citenamefont {Marshall}, \citenamefont
  {Spencer}, \citenamefont {Reiche}, \citenamefont {Henderson}, \citenamefont
  {Chew},\ and\ \citenamefont {Patapoutian}}]{Mulhall2023}%
  \BibitemOpen
  \bibfield  {author} {\bibinfo {author} {\bibfnamefont {E.~M.}\ \bibnamefont
  {Mulhall}}, \bibinfo {author} {\bibfnamefont {A.}~\bibnamefont {Gharpure}},
  \bibinfo {author} {\bibfnamefont {R.~M.}\ \bibnamefont {Lee}}, \bibinfo
  {author} {\bibfnamefont {A.~E.}\ \bibnamefont {Dubin}}, \bibinfo {author}
  {\bibfnamefont {J.~S.}\ \bibnamefont {Aaron}}, \bibinfo {author}
  {\bibfnamefont {K.~L.}\ \bibnamefont {Marshall}}, \bibinfo {author}
  {\bibfnamefont {K.~R.}\ \bibnamefont {Spencer}}, \bibinfo {author}
  {\bibfnamefont {M.~A.}\ \bibnamefont {Reiche}}, \bibinfo {author}
  {\bibfnamefont {S.~C.}\ \bibnamefont {Henderson}}, \bibinfo {author}
  {\bibfnamefont {T.-L.}\ \bibnamefont {Chew}},\ and\ \bibinfo {author}
  {\bibfnamefont {A.}~\bibnamefont {Patapoutian}},\ }\bibfield  {title}
  {\bibinfo {title} {Direct observation of the conformational states of
  piezo1},\ }\href {https://doi.org/10.1038/s41586-023-06427-4} {\bibfield
  {journal} {\bibinfo  {journal} {Nature}\ }\textbf {\bibinfo {volume} {620}},\
  \bibinfo {pages} {1117} (\bibinfo {year} {2023})}\BibitemShut {NoStop}%
\bibitem [{\citenamefont {Rizzo}\ \emph {et~al.}(2003)\citenamefont {Rizzo},
  \citenamefont {Morton}, \citenamefont {DePaola}, \citenamefont {Schnitzer},\
  and\ \citenamefont {Davies}}]{Rizzo2003}%
  \BibitemOpen
  \bibfield  {author} {\bibinfo {author} {\bibfnamefont {V.}~\bibnamefont
  {Rizzo}}, \bibinfo {author} {\bibfnamefont {C.}~\bibnamefont {Morton}},
  \bibinfo {author} {\bibfnamefont {N.}~\bibnamefont {DePaola}}, \bibinfo
  {author} {\bibfnamefont {J.~E.}\ \bibnamefont {Schnitzer}},\ and\ \bibinfo
  {author} {\bibfnamefont {P.~F.}\ \bibnamefont {Davies}},\ }\bibfield  {title}
  {\bibinfo {title} {Recruitment of endothelial caveolae into
  mechanotransduction pathways by flow conditioning in vitro},\ }\href
  {https://doi.org/10.1152/ajpheart.00344.2002} {\bibfield  {journal} {\bibinfo
   {journal} {American Journal of Physiology-Heart and Circulatory Physiology}\
  }\textbf {\bibinfo {volume} {285}},\ \bibinfo {pages} {H1720} (\bibinfo
  {year} {2003})}\BibitemShut {NoStop}%
\bibitem [{\citenamefont {Thomsen}\ \emph {et~al.}(2002)\citenamefont
  {Thomsen}, \citenamefont {Roepstorff}, \citenamefont {Stahlhut},\ and\
  \citenamefont {van Deurs}}]{Thomsen2002}%
  \BibitemOpen
  \bibfield  {author} {\bibinfo {author} {\bibfnamefont {P.}~\bibnamefont
  {Thomsen}}, \bibinfo {author} {\bibfnamefont {K.}~\bibnamefont {Roepstorff}},
  \bibinfo {author} {\bibfnamefont {M.}~\bibnamefont {Stahlhut}},\ and\
  \bibinfo {author} {\bibfnamefont {B.}~\bibnamefont {van Deurs}},\ }\bibfield
  {title} {\bibinfo {title} {Caveolae are highly immobile plasma membrane
  microdomains, which are not involved in constitutive endocytic trafficking},\
  }\href {https://doi.org/10.1091/mbc.01-06-0317} {\bibfield  {journal}
  {\bibinfo  {journal} {Molecular Biology of the Cell}\ }\textbf {\bibinfo
  {volume} {13}},\ \bibinfo {pages} {238} (\bibinfo {year} {2002})}\BibitemShut
  {NoStop}%
\bibitem [{\citenamefont {Mannella}(2021)}]{Mannella2021}%
  \BibitemOpen
  \bibfield  {author} {\bibinfo {author} {\bibfnamefont {C.~A.}\ \bibnamefont
  {Mannella}},\ }\bibfield  {title} {\bibinfo {title} {Vdac—a primal
  perspective},\ }\href {https://doi.org/10.3390/ijms22041685} {\bibfield
  {journal} {\bibinfo  {journal} {International Journal of Molecular Sciences}\
  }\textbf {\bibinfo {volume} {22}},\ \bibinfo {pages} {1685} (\bibinfo {year}
  {2021})}\BibitemShut {NoStop}%
\bibitem [{\citenamefont {Hiller}\ \emph {et~al.}(2010)\citenamefont {Hiller},
  \citenamefont {Abramson}, \citenamefont {Mannella}, \citenamefont {Wagner},\
  and\ \citenamefont {Zeth}}]{Hiller2010}%
  \BibitemOpen
  \bibfield  {author} {\bibinfo {author} {\bibfnamefont {S.}~\bibnamefont
  {Hiller}}, \bibinfo {author} {\bibfnamefont {J.}~\bibnamefont {Abramson}},
  \bibinfo {author} {\bibfnamefont {C.}~\bibnamefont {Mannella}}, \bibinfo
  {author} {\bibfnamefont {G.}~\bibnamefont {Wagner}},\ and\ \bibinfo {author}
  {\bibfnamefont {K.}~\bibnamefont {Zeth}},\ }\bibfield  {title} {\bibinfo
  {title} {The 3d structures of vdac represent a native conformation},\ }\href
  {https://doi.org/10.1016/j.tibs.2010.03.005} {\bibfield  {journal} {\bibinfo
  {journal} {Trends in Biochemical Sciences}\ }\textbf {\bibinfo {volume}
  {35}},\ \bibinfo {pages} {514} (\bibinfo {year} {2010})}\BibitemShut
  {NoStop}%
\bibitem [{\citenamefont {Wozny}\ \emph {et~al.}(2023)\citenamefont {Wozny},
  \citenamefont {Di~Luca}, \citenamefont {Morado}, \citenamefont {Picco},
  \citenamefont {Khaddaj}, \citenamefont {Campomanes}, \citenamefont
  {Ivanović}, \citenamefont {Hoffmann}, \citenamefont {Miller}, \citenamefont
  {Vanni},\ and\ \citenamefont {Kukulski}}]{Wozny2023}%
  \BibitemOpen
  \bibfield  {author} {\bibinfo {author} {\bibfnamefont {M.~R.}\ \bibnamefont
  {Wozny}}, \bibinfo {author} {\bibfnamefont {A.}~\bibnamefont {Di~Luca}},
  \bibinfo {author} {\bibfnamefont {D.~R.}\ \bibnamefont {Morado}}, \bibinfo
  {author} {\bibfnamefont {A.}~\bibnamefont {Picco}}, \bibinfo {author}
  {\bibfnamefont {R.}~\bibnamefont {Khaddaj}}, \bibinfo {author} {\bibfnamefont
  {P.}~\bibnamefont {Campomanes}}, \bibinfo {author} {\bibfnamefont
  {L.}~\bibnamefont {Ivanović}}, \bibinfo {author} {\bibfnamefont {P.~C.}\
  \bibnamefont {Hoffmann}}, \bibinfo {author} {\bibfnamefont {E.~A.}\
  \bibnamefont {Miller}}, \bibinfo {author} {\bibfnamefont {S.}~\bibnamefont
  {Vanni}},\ and\ \bibinfo {author} {\bibfnamefont {W.}~\bibnamefont
  {Kukulski}},\ }\bibfield  {title} {\bibinfo {title} {In situ architecture of
  the er–mitochondria encounter structure},\ }\href
  {https://doi.org/10.1038/s41586-023-06050-3} {\bibfield  {journal} {\bibinfo
  {journal} {Nature}\ }\textbf {\bibinfo {volume} {618}},\ \bibinfo {pages}
  {188} (\bibinfo {year} {2023})}\BibitemShut {NoStop}%
\bibitem [{\citenamefont {Lyu}\ \emph {et~al.}(2024)\citenamefont {Lyu},
  \citenamefont {Chen}, \citenamefont {Zhao}, \citenamefont {Yuan},
  \citenamefont {Zhang}, \citenamefont {Zhang},\ and\ \citenamefont
  {Meng}}]{Lyu2024}%
  \BibitemOpen
  \bibfield  {author} {\bibinfo {author} {\bibfnamefont {Y.}~\bibnamefont
  {Lyu}}, \bibinfo {author} {\bibfnamefont {S.}~\bibnamefont {Chen}}, \bibinfo
  {author} {\bibfnamefont {Y.}~\bibnamefont {Zhao}}, \bibinfo {author}
  {\bibfnamefont {H.}~\bibnamefont {Yuan}}, \bibinfo {author} {\bibfnamefont
  {C.}~\bibnamefont {Zhang}}, \bibinfo {author} {\bibfnamefont
  {C.}~\bibnamefont {Zhang}},\ and\ \bibinfo {author} {\bibfnamefont
  {Q.}~\bibnamefont {Meng}},\ }\bibfield  {title} {\bibinfo {title} {Effect of
  gm1 concentration change on plasma membrane: molecular dynamics simulation
  and analysis},\ }\href {https://doi.org/10.1039/d3cp06161b} {\bibfield
  {journal} {\bibinfo  {journal} {Physical Chemistry Chemical Physics}\
  }\textbf {\bibinfo {volume} {26}},\ \bibinfo {pages} {12552} (\bibinfo {year}
  {2024})}\BibitemShut {NoStop}%
\bibitem [{\citenamefont {Mojumdar}\ \emph {et~al.}(2020)\citenamefont
  {Mojumdar}, \citenamefont {Grey},\ and\ \citenamefont
  {Sparr}}]{Mojumdar2019}%
  \BibitemOpen
  \bibfield  {author} {\bibinfo {author} {\bibfnamefont {E.~H.}\ \bibnamefont
  {Mojumdar}}, \bibinfo {author} {\bibfnamefont {C.}~\bibnamefont {Grey}},\
  and\ \bibinfo {author} {\bibfnamefont {E.}~\bibnamefont {Sparr}},\ }\bibfield
   {title} {\bibinfo {title} {Self-assembly in ganglioside‒phospholipid
  systems: The co-existence of vesicles, micelles, and discs},\ }\href
  {https://doi.org/10.3390/ijms21010056} {\bibfield  {journal} {\bibinfo
  {journal} {International Journal of Molecular Sciences}\ }\textbf {\bibinfo
  {volume} {21}},\ \bibinfo {pages} {56} (\bibinfo {year} {2020})}\BibitemShut
  {NoStop}%
\bibitem [{\citenamefont {Evans}\ \emph {et~al.}(1987)\citenamefont {Evans},
  \citenamefont {Goldman}, \citenamefont {Heinemann},\ and\ \citenamefont
  {Patrick}}]{Evans1987}%
  \BibitemOpen
  \bibfield  {author} {\bibinfo {author} {\bibfnamefont {S.}~\bibnamefont
  {Evans}}, \bibinfo {author} {\bibfnamefont {D.}~\bibnamefont {Goldman}},
  \bibinfo {author} {\bibfnamefont {S.}~\bibnamefont {Heinemann}},\ and\
  \bibinfo {author} {\bibfnamefont {J.}~\bibnamefont {Patrick}},\ }\bibfield
  {title} {\bibinfo {title} {Muscle acetylcholine receptor biosynthesis.
  regulation by transcript availability.},\ }\href
  {https://doi.org/10.1016/s0021-9258(18)61283-9} {\bibfield  {journal}
  {\bibinfo  {journal} {Journal of Biological Chemistry}\ }\textbf {\bibinfo
  {volume} {262}},\ \bibinfo {pages} {4911} (\bibinfo {year}
  {1987})}\BibitemShut {NoStop}%
\bibitem [{\citenamefont {McMahon}\ \emph {et~al.}(1994)\citenamefont
  {McMahon}, \citenamefont {Anderson}, \citenamefont {Nassar}, \citenamefont
  {Bunting}, \citenamefont {Saba}, \citenamefont {Oakeley},\ and\ \citenamefont
  {Malouf}}]{McMahon1994}%
  \BibitemOpen
  \bibfield  {author} {\bibinfo {author} {\bibfnamefont {D.~K.}\ \bibnamefont
  {McMahon}}, \bibinfo {author} {\bibfnamefont {P.~A.}\ \bibnamefont
  {Anderson}}, \bibinfo {author} {\bibfnamefont {R.}~\bibnamefont {Nassar}},
  \bibinfo {author} {\bibfnamefont {J.~B.}\ \bibnamefont {Bunting}}, \bibinfo
  {author} {\bibfnamefont {Z.}~\bibnamefont {Saba}}, \bibinfo {author}
  {\bibfnamefont {A.~E.}\ \bibnamefont {Oakeley}},\ and\ \bibinfo {author}
  {\bibfnamefont {N.~N.}\ \bibnamefont {Malouf}},\ }\bibfield  {title}
  {\bibinfo {title} {C2c12 cells: biophysical, biochemical, and
  immunocytochemical properties},\ }\href
  {https://doi.org/10.1152/ajpcell.1994.266.6.c1795} {\bibfield  {journal}
  {\bibinfo  {journal} {American Journal of Physiology-Cell Physiology}\
  }\textbf {\bibinfo {volume} {266}},\ \bibinfo {pages} {C1795} (\bibinfo
  {year} {1994})}\BibitemShut {NoStop}%
\bibitem [{\citenamefont {Lo}\ \emph {et~al.}(1982)\citenamefont {Lo},
  \citenamefont {Barnard},\ and\ \citenamefont {Dolly}}]{Lo1982}%
  \BibitemOpen
  \bibfield  {author} {\bibinfo {author} {\bibfnamefont {M.~M.~S.}\
  \bibnamefont {Lo}}, \bibinfo {author} {\bibfnamefont {E.~A.}\ \bibnamefont
  {Barnard}},\ and\ \bibinfo {author} {\bibfnamefont {J.~O.}\ \bibnamefont
  {Dolly}},\ }\bibfield  {title} {\bibinfo {title} {Size of acetylcholine
  receptors in the membrane. an improved version of the radiation inactivation
  method},\ }\href {https://doi.org/10.1021/bi00538a033} {\bibfield  {journal}
  {\bibinfo  {journal} {Biochemistry}\ }\textbf {\bibinfo {volume} {21}},\
  \bibinfo {pages} {2210} (\bibinfo {year} {1982})}\BibitemShut {NoStop}%
\bibitem [{\citenamefont {Geng}\ \emph {et~al.}(2009)\citenamefont {Geng},
  \citenamefont {Zhang},\ and\ \citenamefont {Peng}}]{Geng2009}%
  \BibitemOpen
  \bibfield  {author} {\bibinfo {author} {\bibfnamefont {L.}~\bibnamefont
  {Geng}}, \bibinfo {author} {\bibfnamefont {H.~L.}\ \bibnamefont {Zhang}},\
  and\ \bibinfo {author} {\bibfnamefont {H.~B.}\ \bibnamefont {Peng}},\
  }\bibfield  {title} {\bibinfo {title} {The formation of acetylcholine
  receptor clusters visualized with quantum dots},\ }\href
  {https://doi.org/10.1186/1471-2202-10-80} {\bibfield  {journal} {\bibinfo
  {journal} {BMC Neuroscience}\ }\textbf {\bibinfo {volume} {10}},\ \bibinfo
  {eid} {80} (\bibinfo {year} {2009})}\BibitemShut {NoStop}%
\bibitem [{\citenamefont {Kerntke}\ \emph {et~al.}(2020)\citenamefont
  {Kerntke}, \citenamefont {Nimmerjahn},\ and\ \citenamefont
  {Biburger}}]{Kerntke2020}%
  \BibitemOpen
  \bibfield  {author} {\bibinfo {author} {\bibfnamefont {C.}~\bibnamefont
  {Kerntke}}, \bibinfo {author} {\bibfnamefont {F.}~\bibnamefont
  {Nimmerjahn}},\ and\ \bibinfo {author} {\bibfnamefont {M.}~\bibnamefont
  {Biburger}},\ }\bibfield  {title} {\bibinfo {title} {There is (scientific)
  strength in numbers: A comprehensive quantitation of fc gamma receptor
  numbers on human and murine peripheral blood leukocytes},\ }\href
  {https://doi.org/10.3389/fimmu.2020.00118} {\bibfield  {journal} {\bibinfo
  {journal} {Frontiers in Immunology}\ }\textbf {\bibinfo {volume} {11}},\
  \bibinfo {eid} {118} (\bibinfo {year} {2020})}\BibitemShut {NoStop}%
\bibitem [{\citenamefont {Wacleche}\ \emph {et~al.}(2018)\citenamefont
  {Wacleche}, \citenamefont {Tremblay}, \citenamefont {Routy},\ and\
  \citenamefont {Ancuta}}]{Wacleche2018}%
  \BibitemOpen
  \bibfield  {author} {\bibinfo {author} {\bibfnamefont {V.}~\bibnamefont
  {Wacleche}}, \bibinfo {author} {\bibfnamefont {C.}~\bibnamefont {Tremblay}},
  \bibinfo {author} {\bibfnamefont {J.-P.}\ \bibnamefont {Routy}},\ and\
  \bibinfo {author} {\bibfnamefont {P.}~\bibnamefont {Ancuta}},\ }\bibfield
  {title} {\bibinfo {title} {The biology of monocytes and dendritic cells:
  Contribution to hiv pathogenesis},\ }\href
  {https://doi.org/10.3390/v10020065} {\bibfield  {journal} {\bibinfo
  {journal} {Viruses}\ }\textbf {\bibinfo {volume} {10}},\ \bibinfo {pages}
  {65} (\bibinfo {year} {2018})}\BibitemShut {NoStop}%
\bibitem [{\citenamefont {Saji}(1999)}]{Saji1999}%
  \BibitemOpen
  \bibfield  {author} {\bibinfo {author} {\bibfnamefont {F.}~\bibnamefont
  {Saji}},\ }\bibfield  {title} {\bibinfo {title} {Dynamics of immunoglobulins
  at the feto-maternal interface},\ }\href
  {https://doi.org/10.1530/ror.0.0040081} {\bibfield  {journal} {\bibinfo
  {journal} {Reviews of Reproduction}\ }\textbf {\bibinfo {volume} {4}},\
  \bibinfo {pages} {81} (\bibinfo {year} {1999})},\ \bibinfo {note} {bNID
  117058}\BibitemShut {NoStop}%
\bibitem [{\citenamefont {Zhang}\ \emph {et~al.}(2015)\citenamefont {Zhang},
  \citenamefont {Wang}, \citenamefont {Yin}, \citenamefont {Yang},
  \citenamefont {Guan}, \citenamefont {Wang}, \citenamefont {Xu},\ and\
  \citenamefont {Tao}}]{Zhang2015}%
  \BibitemOpen
  \bibfield  {author} {\bibinfo {author} {\bibfnamefont {F.}~\bibnamefont
  {Zhang}}, \bibinfo {author} {\bibfnamefont {S.}~\bibnamefont {Wang}},
  \bibinfo {author} {\bibfnamefont {L.}~\bibnamefont {Yin}}, \bibinfo {author}
  {\bibfnamefont {Y.}~\bibnamefont {Yang}}, \bibinfo {author} {\bibfnamefont
  {Y.}~\bibnamefont {Guan}}, \bibinfo {author} {\bibfnamefont {W.}~\bibnamefont
  {Wang}}, \bibinfo {author} {\bibfnamefont {H.}~\bibnamefont {Xu}},\ and\
  \bibinfo {author} {\bibfnamefont {N.}~\bibnamefont {Tao}},\ }\bibfield
  {title} {\bibinfo {title} {Quantification of epidermal growth factor receptor
  expression level and binding kinetics on cell surfaces by surface plasmon
  resonance imaging},\ }\href {https://doi.org/10.1021/acs.analchem.5b02572}
  {\bibfield  {journal} {\bibinfo  {journal} {Analytical Chemistry}\ }\textbf
  {\bibinfo {volume} {87}},\ \bibinfo {pages} {9960} (\bibinfo {year}
  {2015})}\BibitemShut {NoStop}%
\bibitem [{\citenamefont {Abulrob}\ \emph {et~al.}(2010)\citenamefont
  {Abulrob}, \citenamefont {Lu}, \citenamefont {Baumann}, \citenamefont
  {Vobornik}, \citenamefont {Taylor}, \citenamefont {Stanimirovic},\ and\
  \citenamefont {Johnston}}]{Abulrob2010}%
  \BibitemOpen
  \bibfield  {author} {\bibinfo {author} {\bibfnamefont {A.}~\bibnamefont
  {Abulrob}}, \bibinfo {author} {\bibfnamefont {Z.}~\bibnamefont {Lu}},
  \bibinfo {author} {\bibfnamefont {E.}~\bibnamefont {Baumann}}, \bibinfo
  {author} {\bibfnamefont {D.}~\bibnamefont {Vobornik}}, \bibinfo {author}
  {\bibfnamefont {R.}~\bibnamefont {Taylor}}, \bibinfo {author} {\bibfnamefont
  {D.}~\bibnamefont {Stanimirovic}},\ and\ \bibinfo {author} {\bibfnamefont
  {L.~J.}\ \bibnamefont {Johnston}},\ }\bibfield  {title} {\bibinfo {title}
  {Nanoscale imaging of epidermal growth factor receptor clustering},\ }\href
  {https://doi.org/10.1074/jbc.m109.073338} {\bibfield  {journal} {\bibinfo
  {journal} {Journal of Biological Chemistry}\ }\textbf {\bibinfo {volume}
  {285}},\ \bibinfo {pages} {3145} (\bibinfo {year} {2010})}\BibitemShut
  {NoStop}%
\bibitem [{\citenamefont {Horzum}\ \emph {et~al.}(2014)\citenamefont {Horzum},
  \citenamefont {Ozdil},\ and\ \citenamefont {Pesen-Okvur}}]{Horzum2014}%
  \BibitemOpen
  \bibfield  {author} {\bibinfo {author} {\bibfnamefont {U.}~\bibnamefont
  {Horzum}}, \bibinfo {author} {\bibfnamefont {B.}~\bibnamefont {Ozdil}},\ and\
  \bibinfo {author} {\bibfnamefont {D.}~\bibnamefont {Pesen-Okvur}},\
  }\bibfield  {title} {\bibinfo {title} {Step-by-step quantitative analysis of
  focal adhesions},\ }\href {https://doi.org/10.1016/j.mex.2014.06.004}
  {\bibfield  {journal} {\bibinfo  {journal} {MethodsX}\ }\textbf {\bibinfo
  {volume} {1}},\ \bibinfo {pages} {56} (\bibinfo {year} {2014})}\BibitemShut
  {NoStop}%
\bibitem [{\citenamefont {TruongVo}\ \emph {et~al.}(2017)\citenamefont
  {TruongVo}, \citenamefont {Kennedy}, \citenamefont {Chen}, \citenamefont
  {Chen}, \citenamefont {Berndt}, \citenamefont {Agarwal}, \citenamefont {Zhu},
  \citenamefont {Nakshatri}, \citenamefont {Wallace}, \citenamefont {Na},
  \citenamefont {Yokota},\ and\ \citenamefont {Ryu}}]{TruongVo2017}%
  \BibitemOpen
  \bibfield  {author} {\bibinfo {author} {\bibfnamefont {T.~N.}\ \bibnamefont
  {TruongVo}}, \bibinfo {author} {\bibfnamefont {R.~M.}\ \bibnamefont
  {Kennedy}}, \bibinfo {author} {\bibfnamefont {H.}~\bibnamefont {Chen}},
  \bibinfo {author} {\bibfnamefont {A.}~\bibnamefont {Chen}}, \bibinfo {author}
  {\bibfnamefont {A.}~\bibnamefont {Berndt}}, \bibinfo {author} {\bibfnamefont
  {M.}~\bibnamefont {Agarwal}}, \bibinfo {author} {\bibfnamefont
  {L.}~\bibnamefont {Zhu}}, \bibinfo {author} {\bibfnamefont {H.}~\bibnamefont
  {Nakshatri}}, \bibinfo {author} {\bibfnamefont {J.}~\bibnamefont {Wallace}},
  \bibinfo {author} {\bibfnamefont {S.}~\bibnamefont {Na}}, \bibinfo {author}
  {\bibfnamefont {H.}~\bibnamefont {Yokota}},\ and\ \bibinfo {author}
  {\bibfnamefont {J.~E.}\ \bibnamefont {Ryu}},\ }\bibfield  {title} {\bibinfo
  {title} {Microfluidic channel for characterizing normal and breast cancer
  cells},\ }\href {https://doi.org/10.1088/1361-6439/aa5bbb} {\bibfield
  {journal} {\bibinfo  {journal} {Journal of Micromechanics and
  Microengineering}\ }\textbf {\bibinfo {volume} {27}},\ \bibinfo {pages}
  {035017} (\bibinfo {year} {2017})}\BibitemShut {NoStop}%
\bibitem [{\citenamefont {Erickson}(2009)}]{Erickson2009}%
  \BibitemOpen
  \bibfield  {author} {\bibinfo {author} {\bibfnamefont {H.~P.}\ \bibnamefont
  {Erickson}},\ }\bibfield  {title} {\bibinfo {title} {Size and shape of
  protein molecules at the nanometer level determined by sedimentation, gel
  filtration, and electron microscopy},\ }\href
  {https://doi.org/10.1007/s12575-009-9008-x} {\bibfield  {journal} {\bibinfo
  {journal} {Biological Procedures Online}\ }\textbf {\bibinfo {volume} {11}},\
  \bibinfo {pages} {32} (\bibinfo {year} {2009})}\BibitemShut {NoStop}%
\bibitem [{\citenamefont {{T.M. Cover}}\ and\ \citenamefont
  {Thomas}(1991)}]{TM_Cover1991-pa}%
  \BibitemOpen
  \bibfield  {author} {\bibinfo {author} {\bibnamefont {{T.M. Cover}}}\ and\
  \bibinfo {author} {\bibfnamefont {J.~A.}\ \bibnamefont {Thomas}},\
  }\href@noop {} {\emph {\bibinfo {title} {Elements of Information Theory}}},\
  \bibinfo {edition} {99th}\ ed.,\ Wiley Series in Telecommunications and
  Signal Processing\ (\bibinfo  {publisher} {John Wiley \& Sons},\ \bibinfo
  {address} {Nashville},\ \bibinfo {year} {1991})\BibitemShut {NoStop}%
\bibitem [{\citenamefont {Hohmann}\ and\ \citenamefont
  {Dehghani}(2019)}]{Hohmann2019}%
  \BibitemOpen
  \bibfield  {author} {\bibinfo {author} {\bibfnamefont {T.}~\bibnamefont
  {Hohmann}}\ and\ \bibinfo {author} {\bibfnamefont {F.}~\bibnamefont
  {Dehghani}},\ }\bibfield  {title} {\bibinfo {title} {The cytoskeleton—a
  complex interacting meshwork},\ }\href {https://doi.org/10.3390/cells8040362}
  {\bibfield  {journal} {\bibinfo  {journal} {Cells}\ }\textbf {\bibinfo
  {volume} {8}},\ \bibinfo {pages} {362} (\bibinfo {year} {2019})}\BibitemShut
  {NoStop}%
\bibitem [{\citenamefont {Cho}\ \emph {et~al.}(2022)\citenamefont {Cho},
  \citenamefont {Haraguchi}, \citenamefont {Shigetomi}, \citenamefont
  {Matsuzawa}, \citenamefont {Uchida},\ and\ \citenamefont
  {Ikenouchi}}]{Cho2022}%
  \BibitemOpen
  \bibfield  {author} {\bibinfo {author} {\bibfnamefont {Y.}~\bibnamefont
  {Cho}}, \bibinfo {author} {\bibfnamefont {D.}~\bibnamefont {Haraguchi}},
  \bibinfo {author} {\bibfnamefont {K.}~\bibnamefont {Shigetomi}}, \bibinfo
  {author} {\bibfnamefont {K.}~\bibnamefont {Matsuzawa}}, \bibinfo {author}
  {\bibfnamefont {S.}~\bibnamefont {Uchida}},\ and\ \bibinfo {author}
  {\bibfnamefont {J.}~\bibnamefont {Ikenouchi}},\ }\bibfield  {title} {\bibinfo
  {title} {Tricellulin secures the epithelial barrier at tricellular junctions
  by interacting with actomyosin},\ }\href
  {https://doi.org/10.1083/jcb.202009037} {\bibfield  {journal} {\bibinfo
  {journal} {Journal of Cell Biology}\ }\textbf {\bibinfo {volume} {221}},\
  \bibinfo {pages} {e202009037} (\bibinfo {year} {2022})}\BibitemShut {NoStop}%
\bibitem [{\citenamefont {Resnik-Docampo}\ \emph {et~al.}(2017)\citenamefont
  {Resnik-Docampo}, \citenamefont {Koehler}, \citenamefont {Clark},
  \citenamefont {Schinaman}, \citenamefont {Sauer}, \citenamefont {Wong},
  \citenamefont {Lewis}, \citenamefont {D’Alterio}, \citenamefont {Walker},\
  and\ \citenamefont {Jones}}]{ResnikDocampo2016}%
  \BibitemOpen
  \bibfield  {author} {\bibinfo {author} {\bibfnamefont {M.}~\bibnamefont
  {Resnik-Docampo}}, \bibinfo {author} {\bibfnamefont {C.~L.}\ \bibnamefont
  {Koehler}}, \bibinfo {author} {\bibfnamefont {R.~I.}\ \bibnamefont {Clark}},
  \bibinfo {author} {\bibfnamefont {J.~M.}\ \bibnamefont {Schinaman}}, \bibinfo
  {author} {\bibfnamefont {V.}~\bibnamefont {Sauer}}, \bibinfo {author}
  {\bibfnamefont {D.~M.}\ \bibnamefont {Wong}}, \bibinfo {author}
  {\bibfnamefont {S.}~\bibnamefont {Lewis}}, \bibinfo {author} {\bibfnamefont
  {C.}~\bibnamefont {D’Alterio}}, \bibinfo {author} {\bibfnamefont {D.~W.}\
  \bibnamefont {Walker}},\ and\ \bibinfo {author} {\bibfnamefont {D.~L.}\
  \bibnamefont {Jones}},\ }\bibfield  {title} {\bibinfo {title} {Tricellular
  junctions regulate intestinal stem cell behaviour to maintain homeostasis},\
  }\href {https://doi.org/10.1038/ncb3454} {\bibfield  {journal} {\bibinfo
  {journal} {Nature Cell Biology}\ }\textbf {\bibinfo {volume} {19}},\ \bibinfo
  {pages} {52} (\bibinfo {year} {2017})}\BibitemShut {NoStop}%
\bibitem [{\citenamefont {Howard}(2009)}]{Howard2009}%
  \BibitemOpen
  \bibfield  {author} {\bibinfo {author} {\bibfnamefont {J.}~\bibnamefont
  {Howard}},\ }\bibfield  {title} {\bibinfo {title} {Mechanical signaling in
  networks of motor and cytoskeletal proteins},\ }\href
  {https://doi.org/https://doi.org/10.1146/annurev.biophys.050708.133732}
  {\bibfield  {journal} {\bibinfo  {journal} {Annual Review of Biophysics}\
  }\textbf {\bibinfo {volume} {38}},\ \bibinfo {pages} {217} (\bibinfo {year}
  {2009})}\BibitemShut {NoStop}%
\bibitem [{\citenamefont {Sun}\ and\ \citenamefont {Alushin}(2023)}]{Sun2023}%
  \BibitemOpen
  \bibfield  {author} {\bibinfo {author} {\bibfnamefont {X.}~\bibnamefont
  {Sun}}\ and\ \bibinfo {author} {\bibfnamefont {G.~M.}\ \bibnamefont
  {Alushin}},\ }\bibfield  {title} {\bibinfo {title} {Cellular force-sensing
  through actin filaments},\ }\href
  {https://doi.org/https://doi.org/10.1111/febs.16568} {\bibfield  {journal}
  {\bibinfo  {journal} {The FEBS Journal}\ }\textbf {\bibinfo {volume} {290}},\
  \bibinfo {pages} {2576} (\bibinfo {year} {2023})}\BibitemShut {NoStop}%
\bibitem [{\citenamefont {Bosveld}\ \emph {et~al.}(2016)\citenamefont
  {Bosveld}, \citenamefont {Markova}, \citenamefont {Guirao}, \citenamefont
  {Martin}, \citenamefont {Wang}, \citenamefont {Pierre}, \citenamefont
  {Balakireva}, \citenamefont {Gaugue}, \citenamefont {Ainslie}, \citenamefont
  {Christophorou}, \citenamefont {Lubensky}, \citenamefont {Minc},\ and\
  \citenamefont {Bellaïche}}]{Bosveld2016}%
  \BibitemOpen
  \bibfield  {author} {\bibinfo {author} {\bibfnamefont {F.}~\bibnamefont
  {Bosveld}}, \bibinfo {author} {\bibfnamefont {O.}~\bibnamefont {Markova}},
  \bibinfo {author} {\bibfnamefont {B.}~\bibnamefont {Guirao}}, \bibinfo
  {author} {\bibfnamefont {C.}~\bibnamefont {Martin}}, \bibinfo {author}
  {\bibfnamefont {Z.}~\bibnamefont {Wang}}, \bibinfo {author} {\bibfnamefont
  {A.}~\bibnamefont {Pierre}}, \bibinfo {author} {\bibfnamefont
  {M.}~\bibnamefont {Balakireva}}, \bibinfo {author} {\bibfnamefont
  {I.}~\bibnamefont {Gaugue}}, \bibinfo {author} {\bibfnamefont
  {A.}~\bibnamefont {Ainslie}}, \bibinfo {author} {\bibfnamefont
  {N.}~\bibnamefont {Christophorou}}, \bibinfo {author} {\bibfnamefont {D.~K.}\
  \bibnamefont {Lubensky}}, \bibinfo {author} {\bibfnamefont {N.}~\bibnamefont
  {Minc}},\ and\ \bibinfo {author} {\bibfnamefont {Y.}~\bibnamefont
  {Bellaïche}},\ }\bibfield  {title} {\bibinfo {title} {Epithelial tricellular
  junctions act as interphase cell shape sensors to orient mitosis},\ }\href
  {https://doi.org/10.1038/nature16970} {\bibfield  {journal} {\bibinfo
  {journal} {Nature}\ }\textbf {\bibinfo {volume} {530}},\ \bibinfo {pages}
  {495} (\bibinfo {year} {2016})}\BibitemShut {NoStop}%
\bibitem [{\citenamefont {Norman}\ \emph {et~al.}(2025)\citenamefont {Norman},
  \citenamefont {Chen}, \citenamefont {Recchia}, \citenamefont {Loi},
  \citenamefont {Rosemarie}, \citenamefont {Lesko}, \citenamefont {Patel},
  \citenamefont {Sherer}, \citenamefont {Takaku}, \citenamefont {Burkard},\
  and\ \citenamefont {Suzuki}}]{Norman2024}%
  \BibitemOpen
  \bibfield  {author} {\bibinfo {author} {\bibfnamefont {R.~X.}\ \bibnamefont
  {Norman}}, \bibinfo {author} {\bibfnamefont {Y.-C.}\ \bibnamefont {Chen}},
  \bibinfo {author} {\bibfnamefont {E.~E.}\ \bibnamefont {Recchia}}, \bibinfo
  {author} {\bibfnamefont {J.}~\bibnamefont {Loi}}, \bibinfo {author}
  {\bibfnamefont {Q.}~\bibnamefont {Rosemarie}}, \bibinfo {author}
  {\bibfnamefont {S.~L.}\ \bibnamefont {Lesko}}, \bibinfo {author}
  {\bibfnamefont {S.}~\bibnamefont {Patel}}, \bibinfo {author} {\bibfnamefont
  {N.}~\bibnamefont {Sherer}}, \bibinfo {author} {\bibfnamefont
  {M.}~\bibnamefont {Takaku}}, \bibinfo {author} {\bibfnamefont {M.~E.}\
  \bibnamefont {Burkard}},\ and\ \bibinfo {author} {\bibfnamefont
  {A.}~\bibnamefont {Suzuki}},\ }\bibfield  {title} {\bibinfo {title} {One step
  4× and 12× 3d-exm enables robust super-resolution microscopy of nanoscale
  cellular structures},\ }\href {https://doi.org/10.1083/jcb.202407116}
  {\bibfield  {journal} {\bibinfo  {journal} {Journal of Cell Biology}\
  }\textbf {\bibinfo {volume} {224}},\ \bibinfo {eid} {e202407116} (\bibinfo
  {year} {2025})}\BibitemShut {NoStop}%
\bibitem [{\citenamefont {Radmacher}\ \emph {et~al.}(2025)\citenamefont
  {Radmacher}, \citenamefont {Chizhik}, \citenamefont {Nevskyi}, \citenamefont
  {Gallea}, \citenamefont {Gregor},\ and\ \citenamefont
  {Enderlein}}]{Radmacher2025}%
  \BibitemOpen
  \bibfield  {author} {\bibinfo {author} {\bibfnamefont {N.}~\bibnamefont
  {Radmacher}}, \bibinfo {author} {\bibfnamefont {A.~I.}\ \bibnamefont
  {Chizhik}}, \bibinfo {author} {\bibfnamefont {O.}~\bibnamefont {Nevskyi}},
  \bibinfo {author} {\bibfnamefont {J.~I.}\ \bibnamefont {Gallea}}, \bibinfo
  {author} {\bibfnamefont {I.}~\bibnamefont {Gregor}},\ and\ \bibinfo {author}
  {\bibfnamefont {J.}~\bibnamefont {Enderlein}},\ }\bibfield  {title} {\bibinfo
  {title} {Molecular level super-resolution fluorescence imaging},\ }\href
  {https://doi.org/10.1146/annurev-biophys-071524-105321} {\bibfield  {journal}
  {\bibinfo  {journal} {Annual Review of Biophysics}\ }\textbf {\bibinfo
  {volume} {54}},\ \bibinfo {pages} {163} (\bibinfo {year} {2025})}\BibitemShut
  {NoStop}%
\bibitem [{\citenamefont {Dai}\ \emph {et~al.}(2000)\citenamefont {Dai},
  \citenamefont {Luo}, \citenamefont {Xie},\ and\ \citenamefont
  {Peng}}]{Dai2000}%
  \BibitemOpen
  \bibfield  {author} {\bibinfo {author} {\bibfnamefont {Z.}~\bibnamefont
  {Dai}}, \bibinfo {author} {\bibfnamefont {X.}~\bibnamefont {Luo}}, \bibinfo
  {author} {\bibfnamefont {H.}~\bibnamefont {Xie}},\ and\ \bibinfo {author}
  {\bibfnamefont {H.~B.}\ \bibnamefont {Peng}},\ }\bibfield  {title} {\bibinfo
  {title} {The actin-driven movement and formation of acetylcholine receptor
  clusters},\ }\href {https://doi.org/10.1083/jcb.150.6.1321} {\bibfield
  {journal} {\bibinfo  {journal} {The Journal of Cell Biology}\ }\textbf
  {\bibinfo {volume} {150}},\ \bibinfo {pages} {1321} (\bibinfo {year}
  {2000})}\BibitemShut {NoStop}%
\bibitem [{\citenamefont {Chen}\ \emph {et~al.}(2014)\citenamefont {Chen},
  \citenamefont {Ip}, \citenamefont {Shi}, \citenamefont {Zhang}, \citenamefont
  {Tang}, \citenamefont {Ng}, \citenamefont {Ye}, \citenamefont {Fu},\ and\
  \citenamefont {Ip}}]{Chen2014}%
  \BibitemOpen
  \bibfield  {author} {\bibinfo {author} {\bibfnamefont {Y.}~\bibnamefont
  {Chen}}, \bibinfo {author} {\bibfnamefont {F.~C.}\ \bibnamefont {Ip}},
  \bibinfo {author} {\bibfnamefont {L.}~\bibnamefont {Shi}}, \bibinfo {author}
  {\bibfnamefont {Z.}~\bibnamefont {Zhang}}, \bibinfo {author} {\bibfnamefont
  {H.}~\bibnamefont {Tang}}, \bibinfo {author} {\bibfnamefont {Y.~P.}\
  \bibnamefont {Ng}}, \bibinfo {author} {\bibfnamefont {W.-C.}\ \bibnamefont
  {Ye}}, \bibinfo {author} {\bibfnamefont {A.~K.}\ \bibnamefont {Fu}},\ and\
  \bibinfo {author} {\bibfnamefont {N.~Y.}\ \bibnamefont {Ip}},\ }\bibfield
  {title} {\bibinfo {title} {Coronin 6 regulates acetylcholine receptor
  clustering through modulating receptor anchorage to actin cytoskeleton},\
  }\href {https://doi.org/10.1523/jneurosci.3226-13.2014} {\bibfield  {journal}
  {\bibinfo  {journal} {The Journal of Neuroscience}\ }\textbf {\bibinfo
  {volume} {34}},\ \bibinfo {pages} {2413} (\bibinfo {year}
  {2014})}\BibitemShut {NoStop}%
\bibitem [{\citenamefont {Xing}\ \emph {et~al.}(2019)\citenamefont {Xing},
  \citenamefont {Jing}, \citenamefont {Zhang}, \citenamefont {Cao},
  \citenamefont {Li}, \citenamefont {Zhao}, \citenamefont {Dong}, \citenamefont
  {Chen}, \citenamefont {Wang}, \citenamefont {Cao}, \citenamefont {Xiong},\
  and\ \citenamefont {Mei}}]{Xing2019}%
  \BibitemOpen
  \bibfield  {author} {\bibinfo {author} {\bibfnamefont {G.}~\bibnamefont
  {Xing}}, \bibinfo {author} {\bibfnamefont {H.}~\bibnamefont {Jing}}, \bibinfo
  {author} {\bibfnamefont {L.}~\bibnamefont {Zhang}}, \bibinfo {author}
  {\bibfnamefont {Y.}~\bibnamefont {Cao}}, \bibinfo {author} {\bibfnamefont
  {L.}~\bibnamefont {Li}}, \bibinfo {author} {\bibfnamefont {K.}~\bibnamefont
  {Zhao}}, \bibinfo {author} {\bibfnamefont {Z.}~\bibnamefont {Dong}}, \bibinfo
  {author} {\bibfnamefont {W.}~\bibnamefont {Chen}}, \bibinfo {author}
  {\bibfnamefont {H.}~\bibnamefont {Wang}}, \bibinfo {author} {\bibfnamefont
  {R.}~\bibnamefont {Cao}}, \bibinfo {author} {\bibfnamefont {W.-C.}\
  \bibnamefont {Xiong}},\ and\ \bibinfo {author} {\bibfnamefont
  {L.}~\bibnamefont {Mei}},\ }\bibfield  {title} {\bibinfo {title} {A mechanism
  in agrin signaling revealed by a prevalent rapsyn mutation in congenital
  myasthenic syndrome},\ }\href {https://doi.org/10.7554/elife.49180}
  {\bibfield  {journal} {\bibinfo  {journal} {eLife}\ }\textbf {\bibinfo
  {volume} {8}},\ \bibinfo {eid} {e49180} (\bibinfo {year} {2019})}\BibitemShut
  {NoStop}%
\bibitem [{\citenamefont {Troyanovsky}\ \emph {et~al.}(2025)\citenamefont
  {Troyanovsky}, \citenamefont {Indra},\ and\ \citenamefont
  {Troyanovsky}}]{Troyanovsky2025}%
  \BibitemOpen
  \bibfield  {author} {\bibinfo {author} {\bibfnamefont {R.~B.}\ \bibnamefont
  {Troyanovsky}}, \bibinfo {author} {\bibfnamefont {I.}~\bibnamefont {Indra}},\
  and\ \bibinfo {author} {\bibfnamefont {S.~M.}\ \bibnamefont {Troyanovsky}},\
  }\bibfield  {title} {\bibinfo {title} {Actin-dependent α-catenin
  oligomerization contributes to adherens junction assembly},\ }\href
  {https://doi.org/10.1038/s41467-025-57079-z} {\bibfield  {journal} {\bibinfo
  {journal} {Nature Communications}\ }\textbf {\bibinfo {volume} {16}},\
  \bibinfo {eid} {1801} (\bibinfo {year} {2025})}\BibitemShut {NoStop}%
\bibitem [{\citenamefont {Dustin}\ and\ \citenamefont
  {Cooper}(2000)}]{Dustin2000}%
  \BibitemOpen
  \bibfield  {author} {\bibinfo {author} {\bibfnamefont {M.~L.}\ \bibnamefont
  {Dustin}}\ and\ \bibinfo {author} {\bibfnamefont {J.~A.}\ \bibnamefont
  {Cooper}},\ }\bibfield  {title} {\bibinfo {title} {The immunological synapse
  and the actin cytoskeleton: molecular hardware for t cell signaling},\ }\href
  {https://doi.org/10.1038/76877} {\bibfield  {journal} {\bibinfo  {journal}
  {Nature Immunology}\ }\textbf {\bibinfo {volume} {1}},\ \bibinfo {pages} {23}
  (\bibinfo {year} {2000})}\BibitemShut {NoStop}%
\bibitem [{\citenamefont {Celli}\ \emph {et~al.}(2006)\citenamefont {Celli},
  \citenamefont {Ryckewaert}, \citenamefont {Delachanal},\ and\ \citenamefont
  {Duperray}}]{Celli2006}%
  \BibitemOpen
  \bibfield  {author} {\bibinfo {author} {\bibfnamefont {L.}~\bibnamefont
  {Celli}}, \bibinfo {author} {\bibfnamefont {J.-J.}\ \bibnamefont
  {Ryckewaert}}, \bibinfo {author} {\bibfnamefont {E.}~\bibnamefont
  {Delachanal}},\ and\ \bibinfo {author} {\bibfnamefont {A.}~\bibnamefont
  {Duperray}},\ }\bibfield  {title} {\bibinfo {title} {Evidence of a functional
  role for interaction between icam-1 and nonmuscle α-actinins in leukocyte
  diapedesis},\ }\href {https://doi.org/10.4049/jimmunol.177.6.4113} {\bibfield
   {journal} {\bibinfo  {journal} {The Journal of Immunology}\ }\textbf
  {\bibinfo {volume} {177}},\ \bibinfo {pages} {4113} (\bibinfo {year}
  {2006})}\BibitemShut {NoStop}%
\bibitem [{\citenamefont {Carpén}\ \emph {et~al.}(1992)\citenamefont
  {Carpén}, \citenamefont {Pallai}, \citenamefont {Staunton},\ and\
  \citenamefont {Springer}}]{Carpn1992}%
  \BibitemOpen
  \bibfield  {author} {\bibinfo {author} {\bibfnamefont {O.}~\bibnamefont
  {Carpén}}, \bibinfo {author} {\bibfnamefont {P.}~\bibnamefont {Pallai}},
  \bibinfo {author} {\bibfnamefont {D.~E.}\ \bibnamefont {Staunton}},\ and\
  \bibinfo {author} {\bibfnamefont {T.~A.}\ \bibnamefont {Springer}},\
  }\bibfield  {title} {\bibinfo {title} {Association of intercellular adhesion
  molecule-1 (icam-1) with actin-containing cytoskeleton and alpha-actinin.},\
  }\href {https://doi.org/10.1083/jcb.118.5.1223} {\bibfield  {journal}
  {\bibinfo  {journal} {The Journal of cell biology}\ }\textbf {\bibinfo
  {volume} {118}},\ \bibinfo {pages} {1223} (\bibinfo {year}
  {1992})}\BibitemShut {NoStop}%
\bibitem [{\citenamefont {Kirichenko}\ \emph {et~al.}(2021)\citenamefont
  {Kirichenko}, \citenamefont {Skatchkov},\ and\ \citenamefont
  {Ermakov}}]{Kirichenko2021}%
  \BibitemOpen
  \bibfield  {author} {\bibinfo {author} {\bibfnamefont {E.~Y.}\ \bibnamefont
  {Kirichenko}}, \bibinfo {author} {\bibfnamefont {S.~N.}\ \bibnamefont
  {Skatchkov}},\ and\ \bibinfo {author} {\bibfnamefont {A.~M.}\ \bibnamefont
  {Ermakov}},\ }\bibfield  {title} {\bibinfo {title} {Structure and functions
  of gap junctions and their constituent connexins in the mammalian cns},\
  }\href {https://doi.org/10.1134/s1990747821020069} {\bibfield  {journal}
  {\bibinfo  {journal} {Biochemistry (Moscow), Supplement Series A: Membrane
  and Cell Biology}\ }\textbf {\bibinfo {volume} {15}},\ \bibinfo {pages} {107}
  (\bibinfo {year} {2021})}\BibitemShut {NoStop}%
\bibitem [{\citenamefont {Alberts}\ \emph
  {et~al.}(2002{\natexlab{a}})\citenamefont {Alberts}, \citenamefont {Bray},
  \citenamefont {Lewis}, \citenamefont {Raff}, \citenamefont {Roberts},\ and\
  \citenamefont {Watson}}]{Alberts2002CellJunctions}%
  \BibitemOpen
  \bibfield  {author} {\bibinfo {author} {\bibfnamefont {B.}~\bibnamefont
  {Alberts}}, \bibinfo {author} {\bibfnamefont {D.}~\bibnamefont {Bray}},
  \bibinfo {author} {\bibfnamefont {J.}~\bibnamefont {Lewis}}, \bibinfo
  {author} {\bibfnamefont {M.}~\bibnamefont {Raff}}, \bibinfo {author}
  {\bibfnamefont {K.}~\bibnamefont {Roberts}},\ and\ \bibinfo {author}
  {\bibfnamefont {J.}~\bibnamefont {Watson}},\ }\bibinfo {title} {{Molecular
  Biology of the Cell}}\ (\bibinfo  {publisher} {Garland},\ \bibinfo {address}
  {New York},\ \bibinfo {year} {2002})\ Chap.\ \bibinfo {chapter} {Cell
  Junctions},\ \bibinfo {edition} {4th}\ ed.\BibitemShut {Stop}%
\bibitem [{\citenamefont {Cheema}\ \emph {et~al.}(2021)\citenamefont {Cheema},
  \citenamefont {He}, \citenamefont {Wei},\ and\ \citenamefont
  {Fu}}]{Cheema2021}%
  \BibitemOpen
  \bibfield  {author} {\bibinfo {author} {\bibfnamefont {J.~Y.}\ \bibnamefont
  {Cheema}}, \bibinfo {author} {\bibfnamefont {J.}~\bibnamefont {He}}, \bibinfo
  {author} {\bibfnamefont {W.}~\bibnamefont {Wei}},\ and\ \bibinfo {author}
  {\bibfnamefont {C.}~\bibnamefont {Fu}},\ }\bibfield  {title} {\bibinfo
  {title} {The endoplasmic reticulum-mitochondria encounter structure and its
  regulatory proteins},\ }\href {https://doi.org/10.1177/25152564211064491}
  {\bibfield  {journal} {\bibinfo  {journal} {Contact}\ }\textbf {\bibinfo
  {volume} {4}},\ \bibinfo {pages} {25152564211064491} (\bibinfo {year}
  {2021})}\BibitemShut {NoStop}%
\bibitem [{\citenamefont {Dultz}\ \emph {et~al.}(2022)\citenamefont {Dultz},
  \citenamefont {Wojtynek}, \citenamefont {Medalia},\ and\ \citenamefont
  {Onischenko}}]{dultzNuclearPoreComplex2022}%
  \BibitemOpen
  \bibfield  {author} {\bibinfo {author} {\bibfnamefont {E.}~\bibnamefont
  {Dultz}}, \bibinfo {author} {\bibfnamefont {M.}~\bibnamefont {Wojtynek}},
  \bibinfo {author} {\bibfnamefont {O.}~\bibnamefont {Medalia}},\ and\ \bibinfo
  {author} {\bibfnamefont {E.}~\bibnamefont {Onischenko}},\ }\bibfield  {title}
  {\bibinfo {title} {The {Nuclear} {Pore} {Complex}: {Birth}, {Life}, and
  {Death} of a {Cellular} {Behemoth}},\ }\href
  {https://doi.org/10.3390/cells11091456} {\bibfield  {journal} {\bibinfo
  {journal} {Cells}\ }\textbf {\bibinfo {volume} {11}},\ \bibinfo {pages}
  {1456} (\bibinfo {year} {2022})}\BibitemShut {NoStop}%
\bibitem [{\citenamefont {Bahmanyar}\ and\ \citenamefont
  {Schlieker}(2020)}]{Bahmanyar2020}%
  \BibitemOpen
  \bibfield  {author} {\bibinfo {author} {\bibfnamefont {S.}~\bibnamefont
  {Bahmanyar}}\ and\ \bibinfo {author} {\bibfnamefont {C.}~\bibnamefont
  {Schlieker}},\ }\bibfield  {title} {\bibinfo {title} {Lipid and protein
  dynamics that shape nuclear envelope identity},\ }\href
  {https://doi.org/10.1091/mbc.e18-10-0636} {\bibfield  {journal} {\bibinfo
  {journal} {Molecular Biology of the Cell}\ }\textbf {\bibinfo {volume}
  {31}},\ \bibinfo {pages} {1315} (\bibinfo {year} {2020})}\BibitemShut
  {NoStop}%
\bibitem [{\citenamefont {Balaji}\ \emph {et~al.}(2022)\citenamefont {Balaji},
  \citenamefont {Saha}, \citenamefont {Deshpande}, \citenamefont {Poola},\ and\
  \citenamefont {Sengupta}}]{Balaji2022}%
  \BibitemOpen
  \bibfield  {author} {\bibinfo {author} {\bibfnamefont {A.~K.}\ \bibnamefont
  {Balaji}}, \bibinfo {author} {\bibfnamefont {S.}~\bibnamefont {Saha}},
  \bibinfo {author} {\bibfnamefont {S.}~\bibnamefont {Deshpande}}, \bibinfo
  {author} {\bibfnamefont {D.}~\bibnamefont {Poola}},\ and\ \bibinfo {author}
  {\bibfnamefont {K.}~\bibnamefont {Sengupta}},\ }\bibfield  {title} {\bibinfo
  {title} {Nuclear envelope, chromatin organizers, histones, and dna: The many
  achilles heels exploited across cancers},\ }\href
  {https://doi.org/10.3389/fcell.2022.1068347} {\bibfield  {journal} {\bibinfo
  {journal} {Frontiers in Cell and Developmental Biology}\ }\textbf {\bibinfo
  {volume} {10}},\ \bibinfo {eid} {1068347} (\bibinfo {year}
  {2022})}\BibitemShut {NoStop}%
\bibitem [{\citenamefont {Soheilypour}\ \emph {et~al.}(2016)\citenamefont
  {Soheilypour}, \citenamefont {Peyro}, \citenamefont {Jahed},\ and\
  \citenamefont {Mofrad}}]{Soheilypour2016}%
  \BibitemOpen
  \bibfield  {author} {\bibinfo {author} {\bibfnamefont {M.}~\bibnamefont
  {Soheilypour}}, \bibinfo {author} {\bibfnamefont {M.}~\bibnamefont {Peyro}},
  \bibinfo {author} {\bibfnamefont {Z.}~\bibnamefont {Jahed}},\ and\ \bibinfo
  {author} {\bibfnamefont {M.~R.~K.}\ \bibnamefont {Mofrad}},\ }\bibfield
  {title} {\bibinfo {title} {On the nuclear pore complex and its roles in
  nucleo-cytoskeletal coupling and mechanobiology},\ }\href
  {https://doi.org/10.1007/s12195-016-0443-x} {\bibfield  {journal} {\bibinfo
  {journal} {Cellular and Molecular Bioengineering}\ }\textbf {\bibinfo
  {volume} {9}},\ \bibinfo {pages} {217} (\bibinfo {year} {2016})}\BibitemShut
  {NoStop}%
\bibitem [{\citenamefont {Biedzinski}\ \emph {et~al.}(2020)\citenamefont
  {Biedzinski}, \citenamefont {Agsu}, \citenamefont {Vianay}, \citenamefont
  {Delord}, \citenamefont {Blanchoin}, \citenamefont {Larghero}, \citenamefont
  {Faivre}, \citenamefont {Théry},\ and\ \citenamefont
  {Brunet}}]{Biedzinski2020}%
  \BibitemOpen
  \bibfield  {author} {\bibinfo {author} {\bibfnamefont {S.}~\bibnamefont
  {Biedzinski}}, \bibinfo {author} {\bibfnamefont {G.}~\bibnamefont {Agsu}},
  \bibinfo {author} {\bibfnamefont {B.}~\bibnamefont {Vianay}}, \bibinfo
  {author} {\bibfnamefont {M.}~\bibnamefont {Delord}}, \bibinfo {author}
  {\bibfnamefont {L.}~\bibnamefont {Blanchoin}}, \bibinfo {author}
  {\bibfnamefont {J.}~\bibnamefont {Larghero}}, \bibinfo {author}
  {\bibfnamefont {L.}~\bibnamefont {Faivre}}, \bibinfo {author} {\bibfnamefont
  {M.}~\bibnamefont {Théry}},\ and\ \bibinfo {author} {\bibfnamefont
  {S.}~\bibnamefont {Brunet}},\ }\bibfield  {title} {\bibinfo {title}
  {Microtubules control nuclear shape and gene expression during early stages
  of hematopoietic differentiation},\ }\href
  {https://doi.org/10.15252/embj.2019103957} {\bibfield  {journal} {\bibinfo
  {journal} {The EMBO Journal}\ }\textbf {\bibinfo {volume} {39}},\ \bibinfo
  {eid} {e103957} (\bibinfo {year} {2020})}\BibitemShut {NoStop}%
\bibitem [{\citenamefont {Geng}\ \emph {et~al.}(2023)\citenamefont {Geng},
  \citenamefont {Kang}, \citenamefont {Sun}, \citenamefont {Zhang},
  \citenamefont {Wang}, \citenamefont {Li}, \citenamefont {Li}, \citenamefont
  {Su},\ and\ \citenamefont {Wei}}]{Geng2023}%
  \BibitemOpen
  \bibfield  {author} {\bibinfo {author} {\bibfnamefont {J.}~\bibnamefont
  {Geng}}, \bibinfo {author} {\bibfnamefont {Z.}~\bibnamefont {Kang}}, \bibinfo
  {author} {\bibfnamefont {Q.}~\bibnamefont {Sun}}, \bibinfo {author}
  {\bibfnamefont {M.}~\bibnamefont {Zhang}}, \bibinfo {author} {\bibfnamefont
  {P.}~\bibnamefont {Wang}}, \bibinfo {author} {\bibfnamefont {Y.}~\bibnamefont
  {Li}}, \bibinfo {author} {\bibfnamefont {J.}~\bibnamefont {Li}}, \bibinfo
  {author} {\bibfnamefont {B.}~\bibnamefont {Su}},\ and\ \bibinfo {author}
  {\bibfnamefont {Q.}~\bibnamefont {Wei}},\ }\bibfield  {title} {\bibinfo
  {title} {Microtubule assists actomyosin to regulate cell nuclear mechanics
  and chromatin accessibility},\ }\href
  {https://doi.org/10.34133/research.0054} {\bibfield  {journal} {\bibinfo
  {journal} {Research}\ }\textbf {\bibinfo {volume} {6}},\ \bibinfo {eid}
  {0054} (\bibinfo {year} {2023})}\BibitemShut {NoStop}%
\bibitem [{\citenamefont {Maizels}\ and\ \citenamefont
  {Gerlitz}(2015)}]{Maizels2015}%
  \BibitemOpen
  \bibfield  {author} {\bibinfo {author} {\bibfnamefont {Y.}~\bibnamefont
  {Maizels}}\ and\ \bibinfo {author} {\bibfnamefont {G.}~\bibnamefont
  {Gerlitz}},\ }\bibfield  {title} {\bibinfo {title} {Shaping of interphase
  chromosomes by the microtubule network},\ }\href
  {https://doi.org/10.1111/febs.13334} {\bibfield  {journal} {\bibinfo
  {journal} {The FEBS Journal}\ }\textbf {\bibinfo {volume} {282}},\ \bibinfo
  {pages} {3500} (\bibinfo {year} {2015})}\BibitemShut {NoStop}%
\bibitem [{\citenamefont {Nair}\ \emph {et~al.}(2025)\citenamefont {Nair},
  \citenamefont {Khanna}, \citenamefont {Kler}, \citenamefont {Ragesh},\ and\
  \citenamefont {Sengupta}}]{Nair2025}%
  \BibitemOpen
  \bibfield  {author} {\bibinfo {author} {\bibfnamefont {A.}~\bibnamefont
  {Nair}}, \bibinfo {author} {\bibfnamefont {J.}~\bibnamefont {Khanna}},
  \bibinfo {author} {\bibfnamefont {J.}~\bibnamefont {Kler}}, \bibinfo {author}
  {\bibfnamefont {R.}~\bibnamefont {Ragesh}},\ and\ \bibinfo {author}
  {\bibfnamefont {K.}~\bibnamefont {Sengupta}},\ }\bibfield  {title} {\bibinfo
  {title} {Nuclear envelope and chromatin choreography direct cellular
  differentiation},\ }\href {https://doi.org/10.1080/19491034.2024.2449520}
  {\bibfield  {journal} {\bibinfo  {journal} {Nucleus}\ }\textbf {\bibinfo
  {volume} {16}},\ \bibinfo {eid} {2449520} (\bibinfo {year}
  {2025})}\BibitemShut {NoStop}%
\bibitem [{\citenamefont {Goldberg}(2017)}]{Goldberg2017}%
  \BibitemOpen
  \bibfield  {author} {\bibinfo {author} {\bibfnamefont {M.~W.}\ \bibnamefont
  {Goldberg}},\ }\bibfield  {title} {\bibinfo {title} {Nuclear pore complex
  tethers to the cytoskeleton},\ }\href
  {https://doi.org/10.1016/j.semcdb.2017.06.017} {\bibfield  {journal}
  {\bibinfo  {journal} {Seminars in Cell \&; Developmental Biology}\ }\textbf
  {\bibinfo {volume} {68}},\ \bibinfo {pages} {52} (\bibinfo {year}
  {2017})}\BibitemShut {NoStop}%
\bibitem [{\citenamefont {Huang}\ \emph {et~al.}(2024)\citenamefont {Huang},
  \citenamefont {Zhang}, \citenamefont {Cheng}, \citenamefont {Wang},
  \citenamefont {Miao}, \citenamefont {Huang}, \citenamefont {Fu},\ and\
  \citenamefont {Feng}}]{Huang2023}%
  \BibitemOpen
  \bibfield  {author} {\bibinfo {author} {\bibfnamefont {P.}~\bibnamefont
  {Huang}}, \bibinfo {author} {\bibfnamefont {X.}~\bibnamefont {Zhang}},
  \bibinfo {author} {\bibfnamefont {Z.}~\bibnamefont {Cheng}}, \bibinfo
  {author} {\bibfnamefont {X.}~\bibnamefont {Wang}}, \bibinfo {author}
  {\bibfnamefont {Y.}~\bibnamefont {Miao}}, \bibinfo {author} {\bibfnamefont
  {G.}~\bibnamefont {Huang}}, \bibinfo {author} {\bibfnamefont {Y.-F.}\
  \bibnamefont {Fu}},\ and\ \bibinfo {author} {\bibfnamefont {X.}~\bibnamefont
  {Feng}},\ }\bibfield  {title} {\bibinfo {title} {The nuclear pore y-complex
  functions as a platform for transcriptional regulation of flowering locus c
  in arabidopsis},\ }\href {https://doi.org/10.1093/plcell/koad271} {\bibfield
  {journal} {\bibinfo  {journal} {The Plant Cell}\ }\textbf {\bibinfo {volume}
  {36}},\ \bibinfo {pages} {346} (\bibinfo {year} {2024})}\BibitemShut
  {NoStop}%
\bibitem [{\citenamefont {Guo}\ and\ \citenamefont {Zheng}(2015)}]{Guo2015}%
  \BibitemOpen
  \bibfield  {author} {\bibinfo {author} {\bibfnamefont {Y.}~\bibnamefont
  {Guo}}\ and\ \bibinfo {author} {\bibfnamefont {Y.}~\bibnamefont {Zheng}},\
  }\bibfield  {title} {\bibinfo {title} {Lamins position the nuclear pores and
  centrosomes by modulating dynein},\ }\href
  {https://doi.org/10.1091/mbc.e15-07-0482} {\bibfield  {journal} {\bibinfo
  {journal} {Molecular Biology of the Cell}\ }\textbf {\bibinfo {volume}
  {26}},\ \bibinfo {pages} {3379} (\bibinfo {year} {2015})}\BibitemShut
  {NoStop}%
\bibitem [{\citenamefont {Hoffman}\ \emph {et~al.}(2020)\citenamefont
  {Hoffman}, \citenamefont {Smith}, \citenamefont {Jensen}, \citenamefont
  {Yoshigi}, \citenamefont {Blankman}, \citenamefont {Ullman},\ and\
  \citenamefont {Beckerle}}]{Hoffman2020}%
  \BibitemOpen
  \bibfield  {author} {\bibinfo {author} {\bibfnamefont {L.~M.}\ \bibnamefont
  {Hoffman}}, \bibinfo {author} {\bibfnamefont {M.~A.}\ \bibnamefont {Smith}},
  \bibinfo {author} {\bibfnamefont {C.~C.}\ \bibnamefont {Jensen}}, \bibinfo
  {author} {\bibfnamefont {M.}~\bibnamefont {Yoshigi}}, \bibinfo {author}
  {\bibfnamefont {E.}~\bibnamefont {Blankman}}, \bibinfo {author}
  {\bibfnamefont {K.~S.}\ \bibnamefont {Ullman}},\ and\ \bibinfo {author}
  {\bibfnamefont {M.~C.}\ \bibnamefont {Beckerle}},\ }\bibfield  {title}
  {\bibinfo {title} {Mechanical stress triggers nuclear remodeling and the
  formation of transmembrane actin nuclear lines with associated nuclear pore
  complexes},\ }\href {https://doi.org/10.1091/mbc.e19-01-0027} {\bibfield
  {journal} {\bibinfo  {journal} {Molecular Biology of the Cell}\ }\textbf
  {\bibinfo {volume} {31}},\ \bibinfo {pages} {1774} (\bibinfo {year}
  {2020})}\BibitemShut {NoStop}%
\bibitem [{\citenamefont {Donnaloja}\ \emph {et~al.}(2019)\citenamefont
  {Donnaloja}, \citenamefont {Jacchetti}, \citenamefont {Soncini},\ and\
  \citenamefont {Raimondi}}]{Donnaloja2019}%
  \BibitemOpen
  \bibfield  {author} {\bibinfo {author} {\bibfnamefont {F.}~\bibnamefont
  {Donnaloja}}, \bibinfo {author} {\bibfnamefont {E.}~\bibnamefont
  {Jacchetti}}, \bibinfo {author} {\bibfnamefont {M.}~\bibnamefont {Soncini}},\
  and\ \bibinfo {author} {\bibfnamefont {M.~T.}\ \bibnamefont {Raimondi}},\
  }\bibfield  {title} {\bibinfo {title} {Mechanosensing at the nuclear envelope
  by nuclear pore complex stretch activation and its effect in physiology and
  pathology},\ }\href {https://doi.org/10.3389/fphys.2019.00896} {\bibfield
  {journal} {\bibinfo  {journal} {Frontiers in Physiology}\ }\textbf {\bibinfo
  {volume} {10}},\ \bibinfo {eid} {896} (\bibinfo {year} {2019})}\BibitemShut
  {NoStop}%
\bibitem [{\citenamefont {Elliott}\ \emph {et~al.}(2025)\citenamefont
  {Elliott}, \citenamefont {Shah}, \citenamefont {Belousov}, \citenamefont
  {Dey},\ and\ \citenamefont {Erzberger}}]{ImageDataset}%
  \BibitemOpen
  \bibfield  {author} {\bibinfo {author} {\bibfnamefont {J.}~\bibnamefont
  {Elliott}}, \bibinfo {author} {\bibfnamefont {H.}~\bibnamefont {Shah}},
  \bibinfo {author} {\bibfnamefont {R.}~\bibnamefont {Belousov}}, \bibinfo
  {author} {\bibfnamefont {G.}~\bibnamefont {Dey}},\ and\ \bibinfo {author}
  {\bibfnamefont {A.}~\bibnamefont {Erzberger}},\ }\href@noop {} {\bibinfo
  {title} {{Image dataset used in 'Repulsive particle interactions enable
  selective information processing at cellular interfaces'}}},\ \bibinfo
  {howpublished}
  {\url{https://www.ebi.ac.uk/biostudies/bioimages/studies/S-BIAD2081}}
  (\bibinfo {year} {2025})\BibitemShut {NoStop}%
\bibitem [{\citenamefont
  {Elliott}(2025{\natexlab{b}})}]{ElliottRepoImageAnalysis2025}%
  \BibitemOpen
  \bibfield  {author} {\bibinfo {author} {\bibfnamefont {J.}~\bibnamefont
  {Elliott}},\ }\href@noop {} {\bibinfo {title} {Elliott2025 image analysis
  s-arctica}},\ \bibinfo {howpublished}
  {\url{https://git.embl.de/elliot/Elliott2025-Image-Analysis-S-arctica}}
  (\bibinfo {year} {2025}{\natexlab{b}})\BibitemShut {NoStop}%
\bibitem [{\citenamefont {Li}\ \emph {et~al.}(2016)\citenamefont {Li},
  \citenamefont {Alper},\ and\ \citenamefont {Alexov}}]{Li2016}%
  \BibitemOpen
  \bibfield  {author} {\bibinfo {author} {\bibfnamefont {L.}~\bibnamefont
  {Li}}, \bibinfo {author} {\bibfnamefont {J.}~\bibnamefont {Alper}},\ and\
  \bibinfo {author} {\bibfnamefont {E.}~\bibnamefont {Alexov}},\ }\bibfield
  {title} {\bibinfo {title} {Cytoplasmic dynein binding, run length, and
  velocity are guided by long-range electrostatic interactions},\ }\href
  {https://doi.org/10.1038/srep31523} {\bibfield  {journal} {\bibinfo
  {journal} {Scientific Reports}\ }\textbf {\bibinfo {volume} {6}},\ \bibinfo
  {eid} {31523} (\bibinfo {year} {2016})}\BibitemShut {NoStop}%
\bibitem [{\citenamefont {Schindelin}\ \emph {et~al.}(2012)\citenamefont
  {Schindelin}, \citenamefont {Arganda-Carreras}, \citenamefont {Frise},
  \citenamefont {Kaynig}, \citenamefont {Longair}, \citenamefont {Pietzsch},
  \citenamefont {Preibisch}, \citenamefont {Rueden}, \citenamefont {Saalfeld},
  \citenamefont {Schmid}, \citenamefont {Tinevez}, \citenamefont {White},
  \citenamefont {Hartenstein}, \citenamefont {Eliceiri}, \citenamefont
  {Tomancak},\ and\ \citenamefont {Cardona}}]{Schindelin2012}%
  \BibitemOpen
  \bibfield  {author} {\bibinfo {author} {\bibfnamefont {J.}~\bibnamefont
  {Schindelin}}, \bibinfo {author} {\bibfnamefont {I.}~\bibnamefont
  {Arganda-Carreras}}, \bibinfo {author} {\bibfnamefont {E.}~\bibnamefont
  {Frise}}, \bibinfo {author} {\bibfnamefont {V.}~\bibnamefont {Kaynig}},
  \bibinfo {author} {\bibfnamefont {M.}~\bibnamefont {Longair}}, \bibinfo
  {author} {\bibfnamefont {T.}~\bibnamefont {Pietzsch}}, \bibinfo {author}
  {\bibfnamefont {S.}~\bibnamefont {Preibisch}}, \bibinfo {author}
  {\bibfnamefont {C.}~\bibnamefont {Rueden}}, \bibinfo {author} {\bibfnamefont
  {S.}~\bibnamefont {Saalfeld}}, \bibinfo {author} {\bibfnamefont
  {B.}~\bibnamefont {Schmid}}, \bibinfo {author} {\bibfnamefont {J.-Y.}\
  \bibnamefont {Tinevez}}, \bibinfo {author} {\bibfnamefont {D.~J.}\
  \bibnamefont {White}}, \bibinfo {author} {\bibfnamefont {V.}~\bibnamefont
  {Hartenstein}}, \bibinfo {author} {\bibfnamefont {K.}~\bibnamefont
  {Eliceiri}}, \bibinfo {author} {\bibfnamefont {P.}~\bibnamefont {Tomancak}},\
  and\ \bibinfo {author} {\bibfnamefont {A.}~\bibnamefont {Cardona}},\
  }\bibfield  {title} {\bibinfo {title} {Fiji: an open-source platform for
  biological-image analysis},\ }\href {https://doi.org/10.1038/nmeth.2019}
  {\bibfield  {journal} {\bibinfo  {journal} {Nature Methods}\ }\textbf
  {\bibinfo {volume} {9}},\ \bibinfo {pages} {676} (\bibinfo {year}
  {2012})}\BibitemShut {NoStop}%
\bibitem [{\citenamefont {Daigle}\ \emph {et~al.}(2001)\citenamefont {Daigle},
  \citenamefont {Beaudouin}, \citenamefont {Hartnell}, \citenamefont {Imreh},
  \citenamefont {Hallberg}, \citenamefont {Lippincott-Schwartz},\ and\
  \citenamefont {Ellenberg}}]{Daigle2001}%
  \BibitemOpen
  \bibfield  {author} {\bibinfo {author} {\bibfnamefont {N.}~\bibnamefont
  {Daigle}}, \bibinfo {author} {\bibfnamefont {J.}~\bibnamefont {Beaudouin}},
  \bibinfo {author} {\bibfnamefont {L.}~\bibnamefont {Hartnell}}, \bibinfo
  {author} {\bibfnamefont {G.}~\bibnamefont {Imreh}}, \bibinfo {author}
  {\bibfnamefont {E.}~\bibnamefont {Hallberg}}, \bibinfo {author}
  {\bibfnamefont {J.}~\bibnamefont {Lippincott-Schwartz}},\ and\ \bibinfo
  {author} {\bibfnamefont {J.}~\bibnamefont {Ellenberg}},\ }\bibfield  {title}
  {\bibinfo {title} {Nuclear pore complexes form immobile networks and have a
  very low turnover in live mammalian cells},\ }\href
  {https://doi.org/10.1083/jcb.200101089} {\bibfield  {journal} {\bibinfo
  {journal} {The Journal of Cell Biology}\ }\textbf {\bibinfo {volume} {154}},\
  \bibinfo {pages} {71} (\bibinfo {year} {2001})}\BibitemShut {NoStop}%
\bibitem [{\citenamefont {Rabut}\ \emph {et~al.}(2004)\citenamefont {Rabut},
  \citenamefont {Doye},\ and\ \citenamefont {Ellenberg}}]{Rabut2004}%
  \BibitemOpen
  \bibfield  {author} {\bibinfo {author} {\bibfnamefont {G.}~\bibnamefont
  {Rabut}}, \bibinfo {author} {\bibfnamefont {V.}~\bibnamefont {Doye}},\ and\
  \bibinfo {author} {\bibfnamefont {J.}~\bibnamefont {Ellenberg}},\ }\bibfield
  {title} {\bibinfo {title} {Mapping the dynamic organization of the nuclear
  pore complex inside single living cells},\ }\href
  {https://doi.org/10.1038/ncb1184} {\bibfield  {journal} {\bibinfo  {journal}
  {Nature Cell Biology}\ }\textbf {\bibinfo {volume} {6}},\ \bibinfo {pages}
  {1114} (\bibinfo {year} {2004})}\BibitemShut {NoStop}%
\bibitem [{\citenamefont {Toyama}\ and\ \citenamefont
  {Hetzer}(2013)}]{Toyama2012}%
  \BibitemOpen
  \bibfield  {author} {\bibinfo {author} {\bibfnamefont {B.~H.}\ \bibnamefont
  {Toyama}}\ and\ \bibinfo {author} {\bibfnamefont {M.~W.}\ \bibnamefont
  {Hetzer}},\ }\bibfield  {title} {\bibinfo {title} {Protein homeostasis: live
  long, won’t prosper},\ }\href {https://doi.org/10.1038/nrm3496} {\bibfield
  {journal} {\bibinfo  {journal} {Nature Reviews Molecular Cell Biology}\
  }\textbf {\bibinfo {volume} {14}},\ \bibinfo {pages} {55} (\bibinfo {year}
  {2013})}\BibitemShut {NoStop}%
\bibitem [{\citenamefont {Varberg}\ \emph {et~al.}(2022)\citenamefont
  {Varberg}, \citenamefont {Unruh}, \citenamefont {Bestul}, \citenamefont
  {Khan},\ and\ \citenamefont {Jaspersen}}]{Varberg2022}%
  \BibitemOpen
  \bibfield  {author} {\bibinfo {author} {\bibfnamefont {J.~M.}\ \bibnamefont
  {Varberg}}, \bibinfo {author} {\bibfnamefont {J.~R.}\ \bibnamefont {Unruh}},
  \bibinfo {author} {\bibfnamefont {A.~J.}\ \bibnamefont {Bestul}}, \bibinfo
  {author} {\bibfnamefont {A.~A.}\ \bibnamefont {Khan}},\ and\ \bibinfo
  {author} {\bibfnamefont {S.~L.}\ \bibnamefont {Jaspersen}},\ }\bibfield
  {title} {\bibinfo {title} {Quantitative analysis of nuclear pore complex
  organization in schizosaccharomyces pombe},\ }\href
  {https://doi.org/10.26508/lsa.202201423} {\bibfield  {journal} {\bibinfo
  {journal} {Life Science Alliance}\ }\textbf {\bibinfo {volume} {5}},\
  \bibinfo {pages} {e202201423} (\bibinfo {year} {2022})}\BibitemShut {NoStop}%
\bibitem [{\citenamefont {Roth}\ \emph {et~al.}(2007)\citenamefont {Roth},
  \citenamefont {Moseley}, \citenamefont {Glover}, \citenamefont {Pouton},\
  and\ \citenamefont {Jans}}]{Roth2007}%
  \BibitemOpen
  \bibfield  {author} {\bibinfo {author} {\bibfnamefont {D.~M.}\ \bibnamefont
  {Roth}}, \bibinfo {author} {\bibfnamefont {G.~W.}\ \bibnamefont {Moseley}},
  \bibinfo {author} {\bibfnamefont {D.}~\bibnamefont {Glover}}, \bibinfo
  {author} {\bibfnamefont {C.~W.}\ \bibnamefont {Pouton}},\ and\ \bibinfo
  {author} {\bibfnamefont {D.~A.}\ \bibnamefont {Jans}},\ }\bibfield  {title}
  {\bibinfo {title} {A microtubule‐facilitated nuclear import pathway for
  cancer regulatory proteins},\ }\href
  {https://doi.org/10.1111/j.1600-0854.2007.00564.x} {\bibfield  {journal}
  {\bibinfo  {journal} {Traffic}\ }\textbf {\bibinfo {volume} {8}},\ \bibinfo
  {pages} {673} (\bibinfo {year} {2007})}\BibitemShut {NoStop}%
\bibitem [{\citenamefont {Roth}\ \emph {et~al.}(2011)\citenamefont {Roth},
  \citenamefont {Moseley}, \citenamefont {Pouton},\ and\ \citenamefont
  {Jans}}]{roth2011}%
  \BibitemOpen
  \bibfield  {author} {\bibinfo {author} {\bibfnamefont {D.~M.}\ \bibnamefont
  {Roth}}, \bibinfo {author} {\bibfnamefont {G.~W.}\ \bibnamefont {Moseley}},
  \bibinfo {author} {\bibfnamefont {C.~W.}\ \bibnamefont {Pouton}},\ and\
  \bibinfo {author} {\bibfnamefont {D.~A.}\ \bibnamefont {Jans}},\ }\bibfield
  {title} {\bibinfo {title} {Mechanism of microtubule-facilitated “fast
  track” nuclear import},\ }\href {https://doi.org/10.1074/jbc.m110.210302}
  {\bibfield  {journal} {\bibinfo  {journal} {Journal of Biological Chemistry}\
  }\textbf {\bibinfo {volume} {286}},\ \bibinfo {pages} {14335} (\bibinfo
  {year} {2011})}\BibitemShut {NoStop}%
\bibitem [{\citenamefont {Kelley}\ \emph {et~al.}(2024)\citenamefont {Kelley},
  \citenamefont {Carlini}, \citenamefont {Kornakov}, \citenamefont {Aher},
  \citenamefont {Khodjakov},\ and\ \citenamefont {Kapoor}}]{Kelley2024}%
  \BibitemOpen
  \bibfield  {author} {\bibinfo {author} {\bibfnamefont {M.~E.}\ \bibnamefont
  {Kelley}}, \bibinfo {author} {\bibfnamefont {L.}~\bibnamefont {Carlini}},
  \bibinfo {author} {\bibfnamefont {N.}~\bibnamefont {Kornakov}}, \bibinfo
  {author} {\bibfnamefont {A.}~\bibnamefont {Aher}}, \bibinfo {author}
  {\bibfnamefont {A.}~\bibnamefont {Khodjakov}},\ and\ \bibinfo {author}
  {\bibfnamefont {T.~M.}\ \bibnamefont {Kapoor}},\ }\bibfield  {title}
  {\bibinfo {title} {Spastin regulates anaphase chromosome separation distance
  and microtubule-containing nuclear tunnels},\ }\href
  {https://doi.org/10.1091/mbc.e24-01-0031-t} {\bibfield  {journal} {\bibinfo
  {journal} {Molecular Biology of the Cell}\ }\textbf {\bibinfo {volume}
  {35}},\ \bibinfo {pages} {ar48} (\bibinfo {year} {2024})}\BibitemShut
  {NoStop}%
\bibitem [{\citenamefont {Gerlitz}\ \emph {et~al.}(2013)\citenamefont
  {Gerlitz}, \citenamefont {Reiner},\ and\ \citenamefont
  {Bustin}}]{Gerlitz2012}%
  \BibitemOpen
  \bibfield  {author} {\bibinfo {author} {\bibfnamefont {G.}~\bibnamefont
  {Gerlitz}}, \bibinfo {author} {\bibfnamefont {O.}~\bibnamefont {Reiner}},\
  and\ \bibinfo {author} {\bibfnamefont {M.}~\bibnamefont {Bustin}},\
  }\bibfield  {title} {\bibinfo {title} {Microtubule dynamics alter the
  interphase nucleus},\ }\href {https://doi.org/10.1007/s00018-012-1200-5}
  {\bibfield  {journal} {\bibinfo  {journal} {Cellular and Molecular Life
  Sciences}\ }\textbf {\bibinfo {volume} {70}},\ \bibinfo {pages} {1255}
  (\bibinfo {year} {2013})}\BibitemShut {NoStop}%
\bibitem [{\citenamefont {Velasquez-Carvajal}\ \emph
  {et~al.}(2024)\citenamefont {Velasquez-Carvajal}, \citenamefont {Garampon},
  \citenamefont {Besnardeau}, \citenamefont {Lemée}, \citenamefont {Schaub},\
  and\ \citenamefont {Castagnetti}}]{VelasquezCarvajal2024}%
  \BibitemOpen
  \bibfield  {author} {\bibinfo {author} {\bibfnamefont {D.}~\bibnamefont
  {Velasquez-Carvajal}}, \bibinfo {author} {\bibfnamefont {F.}~\bibnamefont
  {Garampon}}, \bibinfo {author} {\bibfnamefont {L.}~\bibnamefont
  {Besnardeau}}, \bibinfo {author} {\bibfnamefont {R.}~\bibnamefont {Lemée}},
  \bibinfo {author} {\bibfnamefont {S.}~\bibnamefont {Schaub}},\ and\ \bibinfo
  {author} {\bibfnamefont {S.}~\bibnamefont {Castagnetti}},\ }\bibfield
  {title} {\bibinfo {title} {Microtubule reorganization during mitotic cell
  division in the dinoflagellate ostreospis cf. ovata},\ }\href
  {https://doi.org/10.1242/jcs.261733} {\bibfield  {journal} {\bibinfo
  {journal} {Journal of Cell Science}\ }\textbf {\bibinfo {volume} {137}},\
  \bibinfo {eid} {jcs261733} (\bibinfo {year} {2024})}\BibitemShut {NoStop}%
\bibitem [{\citenamefont {Maiato}\ \emph {et~al.}(2004)\citenamefont {Maiato},
  \citenamefont {Sampaio},\ and\ \citenamefont {Sunkel}}]{Maiato2004}%
  \BibitemOpen
  \bibfield  {author} {\bibinfo {author} {\bibfnamefont {H.}~\bibnamefont
  {Maiato}}, \bibinfo {author} {\bibfnamefont {P.}~\bibnamefont {Sampaio}},\
  and\ \bibinfo {author} {\bibfnamefont {C.~E.}\ \bibnamefont {Sunkel}},\
  }\bibfield  {title} {\bibinfo {title} {Microtubule-associated proteins and
  their essential roles during mitosis}\ }(\bibinfo  {publisher} {Academic
  Press},\ \bibinfo {year} {2004})\ pp.\ \bibinfo {pages} {53--153}\BibitemShut
  {NoStop}%
\bibitem [{\citenamefont {Johannes}\ \emph {et~al.}(2018)\citenamefont
  {Johannes}, \citenamefont {Pezeshkian}, \citenamefont {Ipsen},\ and\
  \citenamefont {Shillcock}}]{Johannes2018}%
  \BibitemOpen
  \bibfield  {author} {\bibinfo {author} {\bibfnamefont {L.}~\bibnamefont
  {Johannes}}, \bibinfo {author} {\bibfnamefont {W.}~\bibnamefont
  {Pezeshkian}}, \bibinfo {author} {\bibfnamefont {J.~H.}\ \bibnamefont
  {Ipsen}},\ and\ \bibinfo {author} {\bibfnamefont {J.~C.}\ \bibnamefont
  {Shillcock}},\ }\bibfield  {title} {\bibinfo {title} {Clustering on
  membranes: Fluctuations and more},\ }\href
  {https://doi.org/10.1016/j.tcb.2018.01.009} {\bibfield  {journal} {\bibinfo
  {journal} {Trends in Cell Biology}\ }\textbf {\bibinfo {volume} {28}},\
  \bibinfo {pages} {405} (\bibinfo {year} {2018})}\BibitemShut {NoStop}%
\bibitem [{\citenamefont {Shrestha}\ \emph {et~al.}(2022)\citenamefont
  {Shrestha}, \citenamefont {Kahraman},\ and\ \citenamefont
  {Haselwandter}}]{Shrestha2022}%
  \BibitemOpen
  \bibfield  {author} {\bibinfo {author} {\bibfnamefont {A.}~\bibnamefont
  {Shrestha}}, \bibinfo {author} {\bibfnamefont {O.}~\bibnamefont {Kahraman}},\
  and\ \bibinfo {author} {\bibfnamefont {C.~A.}\ \bibnamefont {Haselwandter}},\
  }\bibfield  {title} {\bibinfo {title} {Mechanochemical coupling of lipid
  organization and protein function through membrane thickness deformations},\
  }\href {https://doi.org/10.1103/physreve.105.054410} {\bibfield  {journal}
  {\bibinfo  {journal} {Physical Review E}\ }\textbf {\bibinfo {volume}
  {105}},\ \bibinfo {eid} {054410} (\bibinfo {year} {2022})}\BibitemShut
  {NoStop}%
\bibitem [{\citenamefont {Barakat}\ and\ \citenamefont
  {Squires}(2022)}]{Barakat2022}%
  \BibitemOpen
  \bibfield  {author} {\bibinfo {author} {\bibfnamefont {J.~M.}\ \bibnamefont
  {Barakat}}\ and\ \bibinfo {author} {\bibfnamefont {T.~M.}\ \bibnamefont
  {Squires}},\ }\bibfield  {title} {\bibinfo {title} {Curvature-mediated forces
  on elastic inclusions in fluid interfaces},\ }\href
  {https://doi.org/10.1021/acs.langmuir.1c02709} {\bibfield  {journal}
  {\bibinfo  {journal} {Langmuir}\ }\textbf {\bibinfo {volume} {38}},\ \bibinfo
  {pages} {1099} (\bibinfo {year} {2022})}\BibitemShut {NoStop}%
\bibitem [{\citenamefont {Pérez-Sala}\ and\ \citenamefont
  {Guo}(2022)}]{perez-salaEditorialIntermediateFilaments2022}%
  \BibitemOpen
  \bibfield  {author} {\bibinfo {author} {\bibfnamefont {D.}~\bibnamefont
  {Pérez-Sala}}\ and\ \bibinfo {author} {\bibfnamefont {M.}~\bibnamefont
  {Guo}},\ }\bibfield  {title} {\bibinfo {title} {Editorial: {Intermediate}
  filaments structure, function, and clinical significance},\ }\href
  {https://doi.org/10.3389/fcell.2022.1103110} {\bibfield  {journal} {\bibinfo
  {journal} {Frontiers in Cell and Developmental Biology}\ }\textbf {\bibinfo
  {volume} {10}},\ \bibinfo {pages} {1103110} (\bibinfo {year}
  {2022})}\BibitemShut {NoStop}%
\bibitem [{\citenamefont {Alberts}\ \emph
  {et~al.}(2002{\natexlab{b}})\citenamefont {Alberts}, \citenamefont {Johnson},
  \citenamefont {Lewis}, \citenamefont {Raff}, \citenamefont {Roberts},\ and\
  \citenamefont {Walter}}]{albertsSelfAssemblyDynamicStructure2002}%
  \BibitemOpen
  \bibfield  {author} {\bibinfo {author} {\bibfnamefont {B.}~\bibnamefont
  {Alberts}}, \bibinfo {author} {\bibfnamefont {A.}~\bibnamefont {Johnson}},
  \bibinfo {author} {\bibfnamefont {J.}~\bibnamefont {Lewis}}, \bibinfo
  {author} {\bibfnamefont {M.}~\bibnamefont {Raff}}, \bibinfo {author}
  {\bibfnamefont {K.}~\bibnamefont {Roberts}},\ and\ \bibinfo {author}
  {\bibfnamefont {P.}~\bibnamefont {Walter}},\ }\bibinfo {title} {{Molecular
  Biology of the Cell}}\ (\bibinfo  {publisher} {Garland Science},\ \bibinfo
  {address} {New York},\ \bibinfo {year} {2002})\ Chap.\ \bibinfo {chapter}
  {The {Self}-{Assembly} and {Dynamic} {Structure} of {Cytoskeletal}
  {Filaments}},\ \bibinfo {edition} {4th}\ ed.\BibitemShut {Stop}%
\bibitem [{\citenamefont {Brauns}\ \emph {et~al.}(2021)\citenamefont {Brauns},
  \citenamefont {Pawlik}, \citenamefont {Halatek}, \citenamefont
  {Kerssemakers}, \citenamefont {Frey},\ and\ \citenamefont
  {Dekker}}]{braunsBulksurfaceCouplingIdentifies2021}%
  \BibitemOpen
  \bibfield  {author} {\bibinfo {author} {\bibfnamefont {F.}~\bibnamefont
  {Brauns}}, \bibinfo {author} {\bibfnamefont {G.}~\bibnamefont {Pawlik}},
  \bibinfo {author} {\bibfnamefont {J.}~\bibnamefont {Halatek}}, \bibinfo
  {author} {\bibfnamefont {J.}~\bibnamefont {Kerssemakers}}, \bibinfo {author}
  {\bibfnamefont {E.}~\bibnamefont {Frey}},\ and\ \bibinfo {author}
  {\bibfnamefont {C.}~\bibnamefont {Dekker}},\ }\bibfield  {title} {\bibinfo
  {title} {Bulk-surface coupling identifies the mechanistic connection between
  {Min}-protein patterns in vivo and in vitro},\ }\href
  {https://doi.org/10.1038/s41467-021-23412-5} {\bibfield  {journal} {\bibinfo
  {journal} {Nature Communications}\ }\textbf {\bibinfo {volume} {12}},\
  \bibinfo {pages} {3312} (\bibinfo {year} {2021})}\BibitemShut {NoStop}%
\bibitem [{\citenamefont {Würthner}\ \emph {et~al.}(2022)\citenamefont
  {Würthner}, \citenamefont {Brauns}, \citenamefont {Pawlik}, \citenamefont
  {Halatek}, \citenamefont {Kerssemakers}, \citenamefont {Dekker},\ and\
  \citenamefont {Frey}}]{Würthner2022}%
  \BibitemOpen
  \bibfield  {author} {\bibinfo {author} {\bibfnamefont {L.}~\bibnamefont
  {Würthner}}, \bibinfo {author} {\bibfnamefont {F.}~\bibnamefont {Brauns}},
  \bibinfo {author} {\bibfnamefont {G.}~\bibnamefont {Pawlik}}, \bibinfo
  {author} {\bibfnamefont {J.}~\bibnamefont {Halatek}}, \bibinfo {author}
  {\bibfnamefont {J.}~\bibnamefont {Kerssemakers}}, \bibinfo {author}
  {\bibfnamefont {C.}~\bibnamefont {Dekker}},\ and\ \bibinfo {author}
  {\bibfnamefont {E.}~\bibnamefont {Frey}},\ }\bibfield  {title} {\bibinfo
  {title} {Bridging scales in a multiscale pattern-forming system},\ }\href
  {https://doi.org/10.1073/pnas.2206888119} {\bibfield  {journal} {\bibinfo
  {journal} {Proceedings of the National Academy of Sciences}\ }\textbf
  {\bibinfo {volume} {119}},\ \bibinfo {pages} {e2206888119} (\bibinfo {year}
  {2022})}\BibitemShut {NoStop}%
\bibitem [{\citenamefont {Banterle}\ \emph {et~al.}(2021)\citenamefont
  {Banterle}, \citenamefont {Nievergelt}, \citenamefont {de~Buhr},
  \citenamefont {Hatzopoulos}, \citenamefont {Brillard}, \citenamefont
  {Andany}, \citenamefont {Hübscher}, \citenamefont {Sorgenfrei},
  \citenamefont {Schwarz}, \citenamefont {Gräter}, \citenamefont {Fantner},\
  and\ \citenamefont {Gönczy}}]{banterleKineticStructuralRoles2021}%
  \BibitemOpen
  \bibfield  {author} {\bibinfo {author} {\bibfnamefont {N.}~\bibnamefont
  {Banterle}}, \bibinfo {author} {\bibfnamefont {A.~P.}\ \bibnamefont
  {Nievergelt}}, \bibinfo {author} {\bibfnamefont {S.}~\bibnamefont {de~Buhr}},
  \bibinfo {author} {\bibfnamefont {G.~N.}\ \bibnamefont {Hatzopoulos}},
  \bibinfo {author} {\bibfnamefont {C.}~\bibnamefont {Brillard}}, \bibinfo
  {author} {\bibfnamefont {S.}~\bibnamefont {Andany}}, \bibinfo {author}
  {\bibfnamefont {T.}~\bibnamefont {Hübscher}}, \bibinfo {author}
  {\bibfnamefont {F.~A.}\ \bibnamefont {Sorgenfrei}}, \bibinfo {author}
  {\bibfnamefont {U.~S.}\ \bibnamefont {Schwarz}}, \bibinfo {author}
  {\bibfnamefont {F.}~\bibnamefont {Gräter}}, \bibinfo {author} {\bibfnamefont
  {G.~E.}\ \bibnamefont {Fantner}},\ and\ \bibinfo {author} {\bibfnamefont
  {P.}~\bibnamefont {Gönczy}},\ }\bibfield  {title} {\bibinfo {title} {Kinetic
  and structural roles for the surface in guiding {SAS}-6 self-assembly to
  direct centriole architecture},\ }\href
  {https://doi.org/10.1038/s41467-021-26329-1} {\bibfield  {journal} {\bibinfo
  {journal} {Nature Communications}\ }\textbf {\bibinfo {volume} {12}},\
  \bibinfo {pages} {6180} (\bibinfo {year} {2021})}\BibitemShut {NoStop}%
\bibitem [{\citenamefont {Dubey}\ \emph {et~al.}(2022)\citenamefont {Dubey},
  \citenamefont {Singh},\ and\ \citenamefont {Chaudhuri}}]{Dubey2022}%
  \BibitemOpen
  \bibfield  {author} {\bibinfo {author} {\bibfnamefont {S.~R.}\ \bibnamefont
  {Dubey}}, \bibinfo {author} {\bibfnamefont {S.~K.}\ \bibnamefont {Singh}},\
  and\ \bibinfo {author} {\bibfnamefont {B.~B.}\ \bibnamefont {Chaudhuri}},\
  }\bibfield  {title} {\bibinfo {title} {Activation functions in deep learning:
  A comprehensive survey and benchmark},\ }\href
  {https://doi.org/10.1016/j.neucom.2022.06.111} {\bibfield  {journal}
  {\bibinfo  {journal} {Neurocomputing}\ }\textbf {\bibinfo {volume} {503}},\
  \bibinfo {pages} {92–108} (\bibinfo {year} {2022})}\BibitemShut {NoStop}%
\bibitem [{\citenamefont {Astrom}\ and\ \citenamefont
  {Murray}(2008)}]{Astrom2008}%
  \BibitemOpen
  \bibfield  {author} {\bibinfo {author} {\bibfnamefont {K.~J.}\ \bibnamefont
  {Astrom}}\ and\ \bibinfo {author} {\bibfnamefont {R.~M.}\ \bibnamefont
  {Murray}},\ }\href@noop {} {\emph {\bibinfo {title} {Feedback systems}}}\
  (\bibinfo  {publisher} {Princeton University Press},\ \bibinfo {address}
  {Princeton, NJ},\ \bibinfo {year} {2008})\BibitemShut {NoStop}%
\bibitem [{\citenamefont {Inc.}()}]{Mathematica14}%
  \BibitemOpen
  \bibfield  {author} {\bibinfo {author} {\bibfnamefont {W.~R.}\ \bibnamefont
  {Inc.}},\ }\href@noop {} {\bibinfo {title} {Mathematica, {V}ersion 14.0}},\
  \bibinfo {note} {champaign, IL, 2024}\BibitemShut {NoStop}%
\bibitem [{\citenamefont
  {Elliott}(2025{\natexlab{c}})}]{ElliottRepoMutualInformation2025}%
  \BibitemOpen
  \bibfield  {author} {\bibinfo {author} {\bibfnamefont {J.}~\bibnamefont
  {Elliott}},\ }\href@noop {} {\bibinfo {title} {Elliott2025 mathematica
  files}},\ \bibinfo {howpublished}
  {\url{https://git.embl.de/elliot/Elliott2025-Mathematica-Files}} (\bibinfo
  {year} {2025}{\natexlab{c}})\BibitemShut {NoStop}%
\bibitem [{\citenamefont {Olivetta}\ and\ \citenamefont
  {Dudin}(2023)}]{Olivetta2023}%
  \BibitemOpen
  \bibfield  {author} {\bibinfo {author} {\bibfnamefont {M.}~\bibnamefont
  {Olivetta}}\ and\ \bibinfo {author} {\bibfnamefont {O.}~\bibnamefont
  {Dudin}},\ }\bibfield  {title} {\bibinfo {title} {The nuclear-to-cytoplasmic
  ratio drives cellularization in the close animal relative sphaeroforma
  arctica},\ }\href {https://doi.org/https://doi.org/10.1016/j.cub.2023.03.019}
  {\bibfield  {journal} {\bibinfo  {journal} {Current Biology}\ }\textbf
  {\bibinfo {volume} {33}},\ \bibinfo {pages} {1597} (\bibinfo {year}
  {2023})}\BibitemShut {NoStop}%
\bibitem [{\citenamefont {Ondracka}\ \emph {et~al.}(2018)\citenamefont
  {Ondracka}, \citenamefont {Dudin},\ and\ \citenamefont
  {Ruiz-Trillo}}]{Ondracka2018}%
  \BibitemOpen
  \bibfield  {author} {\bibinfo {author} {\bibfnamefont {A.}~\bibnamefont
  {Ondracka}}, \bibinfo {author} {\bibfnamefont {O.}~\bibnamefont {Dudin}},\
  and\ \bibinfo {author} {\bibfnamefont {I.}~\bibnamefont {Ruiz-Trillo}},\
  }\bibfield  {title} {\bibinfo {title} {Decoupling of nuclear division cycles
  and cell size during the coenocytic growth of the ichthyosporean sphaeroforma
  arctica},\ }\href {https://doi.org/https://doi.org/10.1016/j.cub.2018.04.074}
  {\bibfield  {journal} {\bibinfo  {journal} {Current Biology}\ }\textbf
  {\bibinfo {volume} {28}},\ \bibinfo {pages} {1964} (\bibinfo {year}
  {2018})}\BibitemShut {NoStop}%
\bibitem [{\citenamefont {Gambarotto}\ \emph {et~al.}(2018)\citenamefont
  {Gambarotto}, \citenamefont {Zwettler}, \citenamefont {Le~Guennec},
  \citenamefont {Schmidt-Cernohorska}, \citenamefont {Fortun}, \citenamefont
  {Borgers}, \citenamefont {Heine}, \citenamefont {Schloetel}, \citenamefont
  {Reuss}, \citenamefont {Unser}, \citenamefont {Boyden}, \citenamefont
  {Sauer}, \citenamefont {Hamel},\ and\ \citenamefont
  {Guichard}}]{Gambarotto2018}%
  \BibitemOpen
  \bibfield  {author} {\bibinfo {author} {\bibfnamefont {D.}~\bibnamefont
  {Gambarotto}}, \bibinfo {author} {\bibfnamefont {F.~U.}\ \bibnamefont
  {Zwettler}}, \bibinfo {author} {\bibfnamefont {M.}~\bibnamefont
  {Le~Guennec}}, \bibinfo {author} {\bibfnamefont {M.}~\bibnamefont
  {Schmidt-Cernohorska}}, \bibinfo {author} {\bibfnamefont {D.}~\bibnamefont
  {Fortun}}, \bibinfo {author} {\bibfnamefont {S.}~\bibnamefont {Borgers}},
  \bibinfo {author} {\bibfnamefont {J.}~\bibnamefont {Heine}}, \bibinfo
  {author} {\bibfnamefont {J.-G.}\ \bibnamefont {Schloetel}}, \bibinfo {author}
  {\bibfnamefont {M.}~\bibnamefont {Reuss}}, \bibinfo {author} {\bibfnamefont
  {M.}~\bibnamefont {Unser}}, \bibinfo {author} {\bibfnamefont {E.~S.}\
  \bibnamefont {Boyden}}, \bibinfo {author} {\bibfnamefont {M.}~\bibnamefont
  {Sauer}}, \bibinfo {author} {\bibfnamefont {V.}~\bibnamefont {Hamel}},\ and\
  \bibinfo {author} {\bibfnamefont {P.}~\bibnamefont {Guichard}},\ }\bibfield
  {title} {\bibinfo {title} {Imaging cellular ultrastructures using expansion
  microscopy (u-exm)},\ }\href {https://doi.org/10.1038/s41592-018-0238-1}
  {\bibfield  {journal} {\bibinfo  {journal} {Nature Methods}\ }\textbf
  {\bibinfo {volume} {16}},\ \bibinfo {pages} {71–74} (\bibinfo {year}
  {2018})}\BibitemShut {NoStop}%
\end{thebibliography}%


\pagebreak
\onecolumngrid
\setcounter{equation}{0}
\setcounter{figure}{0}
\setcounter{table}{0}

\renewcommand{\thesection}{\arabic{section}}
\renewcommand{\thefigure}{S\arabic{figure}}
\renewcommand{\theequation}{S\arabic{equation}}
\renewcommand{\thetable}{S\arabic{table}}

\newcommand{\citeMTFigOne}[1]{Fig.~\ref{fig:ModelIntro}(#1)} 
\newcommand{\citeMTFigTwo}[1]{Fig.~\ref{fig:SimAndContinuousResults}(#1)} 
\newcommand{\citeMTFigThree}[1]{Fig.~\ref{fig:NPCExample}(#1)} 
\newcommand{\citeMTEqNormalization}{Eq.~\eqref{eq:normalization}} 
\newcommand{\citeMTEqEquilibriumFermiDiracSolution}{Eq.~\eqref{eq:EquilibriumFermiDiracSolution}} 

\renewcommand{\thesection}{}
\renewcommand{\thesubsubsection}{\ifnum\value{subsubsection}>0  \arabic{subsubsection}\fi}

\newcolumntype{C}[1]{>{\centering\let\newline\\\arraybackslash\hspace{0pt}}m{#1}}

\let\paragraph\oldparagraph

\setlength{\parskip}{1\baselineskip}

\vspace{3em}
\begin{center}\large\textbf{Supplementary Material~--~
Repulsive particle interactions enable selective information processing at cellular interfaces}\end{center}
\vspace{-1em}


\subsection{\label{secSM:ContinuousInteractions}Equilibrium distribution of interacting particles}

In the following, we derive analytical equations for the equilibrium distributions of diffusive particles on a 2D fluid membrane, which undergo binding/unbinding transitions that are governed by an interaction energy field $\epsilon(\mathbf{r})$. We derive the general equilibrium distribution of particles interacting via a density-dependent interaction potential, first as a solution to the Smoluchowski equations of the system dynamics and then from entropy maximization. 

\paragraph{\label{secSM:EqDynamicalEquations}Dynamical equations}
The time evolution of particle densities in the bound, $i=b$, and unbound, $i=u$, states can be described by the reaction-diffusion equation 
\begin{equation}
    \partial_t \rho_i =\nabla\cdot\Big{[}
        D_i\nabla\left(\rho_i\right)
        +\beta D_i\rho_i\nabla(\epp_i+\epe_i)
    \Big{]}+ \mathcal{R}_i,
\end{equation}
with reaction rates 
\begin{equation}
    \mathcal{R}_{\rm{b}} = f(\rho_{\rm{u}},\rho_{\rm{b}},\epe_{\rm{u}})( \kappa_{\text{off}} e^{-\beta (\Delta\epe+\Delta\epp)}\rho_{\rm{u}} - \kappa_{\text{off}}  \rho_{\rm{b}})
\end{equation}
and 
\begin{equation}
    \mathcal{R}_{\rm{u}} = f(\rho_{\rm{u}},\rho_{\rm{b}},\epe_{\rm{u}})(  \kappa_{\text{off}} \rho_{\rm{b}} -\kappa_{\text{off}}e^{-\beta  (\Delta\epe+\Delta\epp)}\rho_{\rm{u}}),
\end{equation}
where $\Delta\epe = \epe_{\rm{b}}-\epe_{\rm{u}}$, $\epe_{\rm{u}}\equiv\text{const}$, $\Delta\epp= \epp_{\rm{b}}-\epp_{\rm{u}}$, and $f(\rho_{\rm{u}},\rho_{\rm{b}},\epe_{\rm{u}})$ is a density dependent prefactor that has no effect on the equilibrium densities and is included in $k_{\text{off}}= f(\rho_{\rm{u}},\rho_{\rm{b}},\epe_{\rm{u}})\kappa_{\text{off}}$ in the main text. Such equations can be routinely derived from Langevin equations of the particle dynamics as detailed in \cite{Gardiner2010-zc}. For interacting particles, $\epp_i$ is a particle density-dependent potential field $\epp_i(\rho_{\rm{u}},\rho_{\rm{b}})$ that takes into account particle-particle interactions. The coarse-graining of interaction potentials into such density-dependent fields is discussed in section~\ref{secSM:CoarsegrainingPotentials}. 

We solve these coupled dynamical equations for the equilibrium distribution of particles in the membrane, using no-flux boundary conditions. In equilibrium all fluxes vanish identically, including the reaction rates $\mathcal{R}_i \equiv 0$, which account for fluxes between particle states, and the term $\bm{j} = D_i\nabla\rho_i +\beta D_i\rho_i\nabla(\epp_i+\epe_i)=0$ for the flow of the matter in space. These conditions result in two equilibrium equations:
\begin{equation}\label{eq:condition1}
0= \kappa_{\text{off}} e^{-\beta(\Delta\epe+\Delta\epp)}\rho_{\rm{u}} - \kappa_{\text{off}}  \rho_{\rm{b}},
\end{equation}
and 
\begin{equation}
0=D_i\nabla\left(\rho_i\right) +\beta D_i\rho_i\nabla(\epp_i+\epe_i).
\end{equation} 

The latter condition has the formal solution 
\begin{equation}
    \rho_i(\bm{r}) = \rho_{i,0}e^{-\beta\left[\epe_i(\bm{r}) + \epp_i\bigl(\rho_{\rm{u}}(\bm{r}),\rho_{\rm{b}}(\bm{r})\bigr)\right]}
\end{equation}
where $\rho_{i,0}$ is an integration constant. Substituting this solution into the condition Eq.~\eqref{eq:condition1}, we find $\rho_{u,0} = \rho_{b,0}$. With the constants $\rho_{i,0}$ and $\epe_{\rm{u}}$ incorporated into the normalization factor $l^{-2} = \rho_{b,0}e^{-\beta\epe_{\rm{u}}}$, the equilibrium solutions take the form 
\begin{equation}\label{eq:with_ell}
    \rho_{\rm{u}} = \frac{1}{l^2}e^{-\beta\epp_{\rm{u}}},\qquad
    \rho_{\rm{b}} = \frac{1}{l^2}e^{-\beta(\Delta\epe+\Delta\epp + \epp_{\rm{u}})},
\end{equation}
which yield the total particle density
\begin{equation}\label{eqSM:SmoluchowskiEquilibriumSolutionGeneral}
    \rho = \rho_{\rm{u}}+\rho_{\rm{b}} = \frac{1}{l^2}e^{-\beta\epp_{\rm{u}}(\rho)}(1+e^{-\beta(\Delta\epe+\Delta\epp)}).
\end{equation}
As shown in the following section, the constant $l$ is related to the chemical potential and can be found from the conservation of the particle number $N = \int dA\,\rho$ over the whole domain area $A$.

In the main text, we assume that each particle can exist in either of the two states. In this case $\epp_{\rm{u}}(\rho) = \epp_{\rm{b}}(\rho)=\epp(\rho)$ and depends only on the total particle density [see sections~\ref{secSM:BoltEntropyMaximization} and~\ref{secSM:StericInteractions}]. Resulting in the equilibrium total particle density
\begin{equation}\label{eqSM:SmoluchowskiEquilibriumSolution}
    \rho =  \frac{1}{l^2}e^{-\beta\epp(\rho)}(1+e^{-\beta\epe}).
\end{equation}
where we have set $\Delta\epe= \epe$ for notation simplicity. Consequences of relaxing this binding site density assumption are discussed in section~\ref{secSM:influenceOfBindingSiteDensity}.

\paragraph{\label{secSM:BoltEntropyMaximization}Equilibrium solution from entropy maximization}

Re-deriving the equilibrium solution through entropy maximization provides insight into the origins of the interaction potentials $\epp_i(\rho_{\rm{u}},\rho_{\rm{b}})$. Here, we obtain equations \eqref{eq:with_ell} by applying the principle of maximum entropy to a system of \textit{indistinguishable} particles with continuous, pairwise-additive interactions and assuming that multiply occupied energy states are allowed. For simplicity, we analyse a one-dimensional system, as the generalization of the simplified model to higher dimensions is straightforward. 

In particular, we consider a system divided into $B$ boxes of size $dx$ indexed by $j > 0$ in contact with a reservoir with $j=0$, characterized by particle number $N_0$. Acting as a heat bath, the reservoir can exchange energy, but not particles, with the system. Within each subsystem $j$ there are $K_j$ unbound particles and $M_j$ bound particles. The boxes have degeneracy $g_K$ and $g_M$ of the unbound and bound states with energies $\epsilon_{\rm{u}}$ (constant) and $\epe_{b,j}$ (box-dependent), respectively.

Besides the single-particle contributions of $\epe_{\rm{b}}$ and $\epsilon_{\rm{u}}$, we also include an energy term $E_r(K_j, M_j)$, which models local interactions between particles inside each box, as discussed further in Sec.~\ref{secSM:ShortRangeRepInteractions}. Being local, this term depends only on the number of particles in a given box regardless of their state distributions.

With the above assumptions we can treat the system of particles as a Boltzmann gas. In the microcanonical framework we count the number of microstates $W$ which make up the macrostate defined by $\{K_j,M_j\}$. The number of ways to arrange $K_j$ and $M_j$ indistinguishable particles into $g_K$ and $g_M$ states, respectively, is given by Maxwell-Boltzmann statistics~\cite[Chapter 13]{Carter2000}
\begin{equation}
    W_{j\ge0} = \frac{g_K^{K_j}}{K_j!}\frac{g_M^{M_j}}{M_j!}.
\end{equation}
Under the imposed constraints of the total energy $U$ and total number of particles $N$ across all boxes, and their number $N_0$ in the reservoir, the equilibrium state features the maximum number of realizations $W = \prod_{j\ge0} W_j$ or, equivalently, the maximum Boltzmann entropy $S_{\rm B}/k_{\rm B} = \ln W$. We introduce the Lagrange multipliers $\alpha$, $\alpha_0$, $\beta$ to the constraints on $N$, $N_0$, and $U$, respectively, and extremize the objective function
\begin{equation}\label{eqSM:BoltzmanObjectiveFunction}
        f(\{K_j\},\{M_j\}) = \ln W + \alpha \bigg(N - N_0 -\sum_{j>0}(K_j+M_j)\bigg) + \alpha_0 (N_0 - K_0 - M_0)
    + \beta\bigg(U - \sum_{j\ge0}\big(K_j\epsilon_{\rm{u}}+M_j\epe_{b,j}+E_r(K_j+M_j)\big)\bigg),
\end{equation}
by requiring that its derivatives vanish
\begin{eqnarray}\label{eq:fN}
    \frac{\partial f}{\partial K_j} &=& \ln g_K-\ln{K_j} -\delta_{0j} \alpha_0 - (1 - \delta_{0j}) \alpha - \beta \epsilon_{\rm{u}} - \beta\frac{\partial E_r(K_j+M_j)}{\partial K_j} = 0,\\\label{eq:fM}
    \frac{\partial f}{\partial M_j} &=&\ln g_M - \ln M_j
        -\delta_{0j} \alpha_0 - (1 - \delta_{0j}) \alpha -\beta\epe_{b,j}- \beta\frac{\partial E_r(K_j+M_j)}{\partial M_j} = 0,
\end{eqnarray}
in which we apply the Stirling approximation, and use the Kronecker $\delta_{jk}$. From Eqs.~\eqref{eq:fN} and $\eqref{eq:fM}$ we obtain
\begin{alignat}{2}\label{eqSM:BoltEntropySol0} 
    K_0 =& g_K e^{\displaystyle \beta\left(\mu_0 - \epsilon_{\rm{u}} - \frac{\partial E_r(K_0+M_0)}{\partial K_0}\right)},&\qquad
    M_0 =& g_M e^{\displaystyle \beta\left(\mu_0 - \epe_{b,0} - \frac{\partial E_r(K_0+M_0)}{\partial M_0}\right)}
\\\label{eqSM:BoltEntropySoli}
    K_{j > 0} =& g_{K} e^{\displaystyle \beta \left(\mu - \epsilon_{\rm{u}} - \dfrac{\partial E_r(K_j+M_j)}{\partial K_j}\right)},&\qquad
    M_{j > 0} =& g_{M} e^{\displaystyle \beta \left(\mu - \epe_{b,j} - \dfrac{\partial E_r(K_j+M_j)}{\partial M_j}\right)}
\end{alignat}
which represent Boltzmann distribution of particles over energy levels with the chemical potentials of the bath $\mu_0 = -\alpha_0 / \beta$ and of the system $\mu = -\alpha / \beta$. The reservoir part of the solution, Eq.~\eqref{eqSM:BoltEntropySol0}---which we keep for completeness, plays no role in the following.

Now we divide Eq.~\eqref{eqSM:BoltEntropySoli} by the box size $dx$, and take the continuum limit $dx\to0$ as the number of boxes increases to infinity. This procedure yields the particle densities at $x = \lim_{dx\to0} j\, dx$:
\begin{equation}\label{eq:climN}
    \lim_{dx\to0} \frac{K_j}{dx} = \rho_{\rm{u}}(x)= g_{\rm{u}} e^{\beta\mu-\beta \epsilon_{\rm{u}} - \beta E(\rho)},
\end{equation}
\begin{equation}\label{eq:climM}
    \lim_{dx\to0} \frac{M_j}{dx} = \rho_{\rm{b}}(x)= g_{\rm{b}} e^{\beta\mu-\beta \epe_{\rm{b}}(x)- \beta E(\rho)},
\end{equation}
with $\rho = \rho_{\rm{u}} + \rho_{\rm{b}}$ and the implied limits $\lim_{dx\to0}\epsilon_{b,j} = \epsilon_{\rm{b}}(x)$,
\begin{alignat}{2}\label{eq:lim_u}
    \lim_{dx\to0} \frac{g_K}{dx} =& g_{\rm{u}},&\quad
    \lim_{dx\to0} \frac{\partial}{\partial (K_j / dx)} \frac{1}{dx} E_r\left[
        dx \left(\frac{K_i}{dx} + \frac{M_j}{dx}\right)
    \right]
    =& \frac{\partial}{\partial\rho_{\rm{u}}} E_\rho(\rho_{\rm{u}} + \rho_{\rm{b}}) =
    \frac{\partial\rho}{\partial\rho_{\rm{u}}} \frac{\partial E_\rho(\rho)}{\partial\rho}
    = E(\rho),    
    \\\label{eq:lim_b}
    \lim_{dx\to0} \frac{g_M}{dx} =& g_{\rm{b}},&\quad
    \lim_{dx\to0} \frac{\partial}{\partial (M_j / dx)} \frac{1}{dx} E_r\left[
        dx \left(\frac{K_j}{dx} + \frac{M_j}{dx}\right)
    \right]
    =& \frac{\partial}{\partial\rho_{\rm{b}}} E_\rho(\rho_{\rm{u}} + \rho_{\rm{b}})
    = \frac{\partial\rho}{\partial\rho_{\rm{b}}} \frac{\partial E_\rho(\rho)}{\partial\rho}
    = E(\rho).
\end{alignat}

The normalization constant $l$, introduced in the previous section, can be explicitly related to the chemical potential $\mu$ via the formula
$$l = \left(g_{\rm{u}} e^{\beta (\mu - \epsilon_{\rm{u}})}\right)^{-\frac{1}{2}},$$
which leads to
\begin{equation}
        \rho_{\rm{u}}(x) = \frac{1}{l^2} e^{- \beta E(\rho(x))},\qquad
        \rho_{\rm{b}}(x) = \frac{1}{l^2} e^{- \beta \big(\epsilon_{\rm{b}}(x) - \epsilon_{\rm{u}} + E(\rho(x))+ k_{\rm B} T \ln(g_{\rm{u}} / g_{\rm{b}})\big)}.
\end{equation}
The above equations recapitulate \eqref{eq:with_ell} if we identify $\epp_{\rm{u}}(\rho)=\epp(\rho)$ and $\Delta\epp(\rho) = k_{\rm B} T \ln(g_{\rm{u}} / g_{\rm{b}})$. Here we find that, for particles acting as a Boltzmann gas, $\epp_{\rm{u}}$ and $\epp_{\rm{b}}$ remain functions of only the total density $\rho$, even when relaxing the assumption on binding site density such that the ratio of the densities of states $f_{\rm{BS}} = g_{\rm{b}}/g_{\rm{u}} = g_M / g_K < 1$ . This ratio characterizes the average number of locally available binding sites, and introduces a constant $\Delta\epp$, the consequences of which are discussed in section~\ref{secSM:influenceOfBindingSiteDensity}.

\subsection{\label{secSM:CoarsegrainingPotentials} Mean-field potentials for repulsive particle interactions}

In this letter we focus on short-range repulsive particle-particle interactions, and first derive the expression for $\epp_{\rm{u}}$ and $\Delta\epp$ in the presence of steric interactions. In this case, the mean-field potential emerges as a correction to the chemical potential of the ideal gas. We then also derive expressions for $\epp(\rho)$ for other, continuous interactions.

\paragraph{\label{secSM:StericInteractions}Hard-core repulsion}
In section \ref{secSM:ContinuousInteractions}, we derived the equilibrium distribution of particles that obey the Maxwell-Boltzmann statistics. Therefore Eq.~\eqref{eqSM:SmoluchowskiEquilibriumSolutionGeneral} corresponds to the distribution of a Boltzmann gas if we set $\epp(\rho) \equiv 0$. In the presence of steric exclusion effects, we find a correction $\Delta\mu = \epp(\rho)$ to the chemical potential $\mu^* = \mu + \Delta{\mu}$. In particular, the lattice gas, where each site of the lattice can host at most one particle of effective size $d_0$, models a short-range repulsion mediated by the hard-core potential. This potential is infinite for interparticle separations shorter than $d_0$ and equals zero otherwise. Such interactions describe particles which effectively exclude the volume they occupy from that available to other particles.

Considering bound and unbound particles in boxes indexed by $j$ as before,  we introduce within each box a lattice with a total of $\Omega$ sites, of which a subset $Q$ permits the particles to be in the bound state. While the particles are treated as indistinguishable, the sites are not.
We count the number of microstates $W$ that make up the macrostate defined by ${K_j,M_j}$ by first counting the arrangements of the bound particles. These can occupy any of the $Q$ sites, but no site can have more than one particle, resulting in the number of combinations of the $M_j$ indistinguishable particles
\begin{equation}\label{eqSM:LatticeNumOfArrangementsBound}
    W_{M,j} = \frac{Q!}{(Q-M_j)!M_j!}.
\end{equation}
Next, we count the possible arrangements for the free particles. For a total of $\Omega$ sites within the lattice, $M_j$ of which are already occupied, there remain $\Omega-M_j$ available spaces. Therefore the number of arrangements for the free particles is given by
\begin{equation}\label{eqSM:LatticeNumOfArrangementsUnbound}
    W_{K,j} = \frac{(\Omega-M_j)!}{(\Omega-M_j-K_j)!K_j!}.
\end{equation}
The number of microstates for one box is then the product of equations \eqref{eqSM:LatticeNumOfArrangementsBound} and \eqref{eqSM:LatticeNumOfArrangementsUnbound}, 
\begin{equation}
    W_j = W_{K,j} W_{M,j} = \frac{Q! (\Omega-M_j)!}{(Q-M_j)!(\Omega-M_j-K_j)!K_j! M_j!}.
\end{equation}

Extremizing $\ln W = \sum_{j\ge0} \ln W_j$ subject to the same constraints of energy and particle numbers as before yields
\begin{eqnarray}
    \label{eq:Ni_fd}
    K_j &=& (\Omega-M_j-K_j) e^{\displaystyle - \delta_{0j} \alpha_0 - (1 - \delta_{0j}) \alpha  -\beta \epsilon_{\rm{u}}},\\\label{eq:Mi_fd}
    M_j &=& (\Omega-M_j-K_j)\frac{Q-M_j}{\Omega-M_j} e^{\displaystyle - \delta_{0j} \alpha_0 - (1 - \delta_{0j}) \alpha -\beta \epsilon_{\mathrm{b},j}}.
\end{eqnarray}

Although Eqs.~\eqref{eq:Ni_fd} and \eqref{eq:Mi_fd} can be explicitly solved for $K_j$ and $M_j$, leading to a Fermi-Dirac-like distribution, we keep the $K_j$- and $M_j$-dependent prefactors in the form convenient to determine the Lagrange multiplier $\alpha = -\beta \mu^*$ related to the chemical potential $\mu^*$ of the hard-core lattice gas. In the continuum limit $dx\to 0$, cf. Eqs.~\eqref{eq:climN} and \eqref{eq:climM}, we obtain thus
\begin{equation}\label{eqSM:LatticeRhoUnbound}
    \rho_{\rm{u}}(x)= \big(\rho_{\max}-\rho_{\rm{b}}(x)-\rho_{\rm{u}}(x)\big) e^{-\alpha-\beta \epsilon_{\rm{u}}}
\end{equation}
\begin{equation}\label{eqSM:LatticeRhoBound}
    \rho_{\rm{b}}(x)= \big(\rho_{\max}-\rho_{\rm{b}}(x)-\rho_{\rm{u}}(x)\big) \frac{\rho_{\text{ext}}-\rho_{\rm{b}}(x)}{\rho_{\max}-\rho_{\rm{b}}(x)} e^{-\alpha-\beta \epsilon_{\rm{b}}(x)}
\end{equation}
where $\rho_{\rm{u}}(x)$ and $\rho_{\rm{b}}(x)$ are the densities of unbound and bound particles, and $\rho_{\max}  = 1/d_0^2 = \lim_{dx\to0} \Omega / dx$ and $\rho_{\text{ext}} = \lim_{dx\to0} Q / dx$ are the maximum-packing and binding-site densities respectively. C.f. Eq.~\eqref{eq:with_ell}, we find
\begin{equation}\label{eqSM:FermiDiracE}
    \epp_{\rm{u}}(\rho)=-k_{\rm B} T \ln \left(1-\frac{\rho}{\rho_{\rm max}}\right)
\end{equation} 
and 
\begin{equation}
\Delta\epp(\rho)=-k_{\rm B} T \ln\left(\frac{\rho_{\rm ext}-\rho_{\rm{b}}}{\rho_{\rm max}-\rho_{\rm{b}}}\right),
\end{equation}
for $l^2 = e^{\alpha + \beta\epsilon_{\rm{u}}}/\rho_{\rm max}$, where the discrepancy between the number of lattice sites and the number of binding sites introduces a dependence of $\Delta\epp$ on the density of bound particles. The assumption that each particle has exactly one binding site is equivalent to the assumption that the binding site density is equal to the maximum particle density (i.e. $\rho_{\max} = \rho_{\rm ext}$). Under this assumption, $\Delta\epp=0$ and $\epp_{\rm{u}}(\rho)=\epp(\rho)$, and we recover Eq.~\ref{eqSM:SmoluchowskiEquilibriumSolution} from Eqs.~\eqref{eqSM:LatticeRhoUnbound} and~\eqref{eqSM:LatticeRhoBound}. The consequences of foregoing this simplifying assumption are addressed in Sec.~\ref{secSM:influenceOfBindingSiteDensity}.

\paragraph{\label{secSM:ShortRangeRepInteractions}Soft-core repulsion}

Next we demonstrate how generic soft-core repulsive interactions give rise to a density-dependent mean-field potential $\epp(\rho)$. In Sec. \ref{secSM:BoltEntropyMaximization}, we obtained $\epp(\rho) = \partial_{\rho} E_\rho(\rho)$ from the local many-body interaction $E_r(K_j + M_j)$ in the continuum limit of the coarse-graining box size $dx\to0$. 
Considering now two-dimensional boxes of size $a = dx^2$, we assume a pairwise-additive form of the interparticle potential, neglect non-nearest neighbor interactions, and approximate the pairwise distances between neighboring particles $m$ and $k$ by their average, i.e. $d_{mk}\approx d = \sqrt{a/N_j}$, given a uniform distribution of particles in the box. With $\xi$ denoting the number of nearest neighbors, the interparticle potential in the $j$th box and its derivative are given by
\begin{align}
    E_r(N_j) &\approx N_j \frac{\xi}{2} u(d),\\
     \frac{\partial \epp_r(N_j)}{\partial N_j} &\approx \frac{\xi}{2} \left( u(d)+  N_j \frac{\partial u(d)}{\partial d}\frac{\partial d}{\partial N_j} \right)= \frac{\xi}{2} u(d)- \frac{\xi}{4}  \frac{\partial u(d)}{\partial d} d,
\end{align}
 which in the continuum limit reads 
 \begin{equation}
 \begin{split}
     \epp(\rho) = \frac{\partial \epp_\rho(\rho)}{\partial \rho} &= \frac{\xi}{2}u\left(d\right)- \frac{\xi}{4}  \frac{\partial u\left(d\right)}{\partial d} d, 
 \end{split}
 \end{equation}
with $d = \sqrt{1/\rho}$.

Similar interaction potentials arise from other physical repulsion mechanisms, for example due to membrane-curvature mediated interactions, or shielded electrostatic interactions.
Curvature mediated repulsive interactions can be approximated as pairwise additive at low particle densities, given by~\cite{Yolcu2014} 
\begin{equation}
    u_{\text{curvature}}(d_{mk}) =  8\pi \kappa \theta^2 \left(\frac{d_0}{2d_{mk}}\right)^4,
\end{equation}
where $\theta$ is the membrane contact angle, and $\kappa$ denotes the membrane bending rigidity~\cite{Karal2023}, leading to the mean-field potential  
\begin{equation}\label{eqSM:CurveE}
    \epp_{\text{curvature}}(d) =  24\pi \kappa \theta^2 \frac{\xi}{2} \left(\frac{d_0}{2d}\right)^4.
\end{equation}

Similarly, shielded electrostatic interactions described by the Yukawa potential~\cite{Liboff1959,Rowlinson1989} lead to an interaction energy
\begin{equation}
    u_{\text{electrostatic}}(d_{mk}) = \frac{Q^2}{4\pi\epsilon_0} \dfrac{e^{\displaystyle -d_{mk}/\lambda_d}}{d_{mk}},
\end{equation}
where the vacuum permittivity is $\epsilon_0=55.2\text{e}^2\text{eV}^{-1}\mu\text{m}^{-1}$, which can be written in units of $K_{\rm{b}}T$, for $T=\SI{300}{K}$, as $\epsilon_0= 2.14\times 10^{6}
\text{e}^2\text{k}_{\rm{b}}\text{T}^{-1}\text{nm}^{-1}$. 
The corresponding mean-field potential reads 
\begin{equation}\label{eqSM:electroE}
    \epp_{\text{electrostatic}}(d) =\frac{\xi}{2} \frac{Q^2}{4\pi\epsilon_0}  \frac{e^{\displaystyle-d/\lambda_d}}{d}\left(\frac{3}{2}+\frac{d}{2\lambda_d}\right).
\end{equation}
In \citeMTFigOne{b}, we show the comparison between Eqs.~\eqref{eqSM:electroE},~\eqref{eqSM:CurveE}, and~\eqref{eqSM:FermiDiracE} using a particle size $d_0=1\text{nm}$, which is the typical order of magnitude for protein size~\cite[pg. 45]{Milo2015Avesizeprotein}~\cite[BNID 100018]{Zhdanov2009}, and assuming the number of nearest neighbors to be $\xi=6$. For the curvature-mediated interactions we use the membrane bending rigidity $\kappa=20k_{\rm{B}}T$~\cite{Hu2012} and the membrane contact angle $\theta\approx\pi/24$, and for the electrostatic interactions we use the Debye length $\lambda_d =$\SI{0.8}{\nano\meter}~\cite{Wennerstrm2020} and consider particles with a charge of four elementary charges, $Q\approx 4e$.

\subsection{\label{secSM:MutualInfo}Quantification of information transmission}

As discussed in the main text, the binding energy fields contain information on the physical properties of the environment, which is selectively encoded by the particle density distribution. We ask how \emph{fluctuating} particle densities respond to variations in the binding energy field across different environments, and characterize how much information the densities carry about the external binding energy, given the compression that results from the sigmoidal relation \citeMTEqEquilibriumFermiDiracSolution. To ease analysis and allow verification by Metropolis sampling, we compute the relevant probability distributions and the mutual information between the input energy and the output particle density by discretizing these quantities, and introducing random vectors with realizations $\{\rho_j\}$ and $\{\epe_k\}$ with box indices $j,k = 1,...,B$. 
The calculations detailed in this section were implemented in Mathematica~\cite{Mathematica14} and python, the notebooks and scripts of which are provided in the repository~\cite{ElliottRepoMutualInformation2025}.

\subsubsection{\label{secSM:particlefluctuations}Gaussian approximation of particle fluctuations}
Similar to before, we discretize our surface into boxes of area $a$ indexed by $j$, and now introduce a random vector for the particle numbers, with one entry per box, denoting a given realization by $\{N_j\}$. In the grand canonical ensemble, fluctuations in the total number of particles $\{N_j\}$ in all the boxes obey the distribution $P(\{N_{j}\}) \propto e^{S/k_{\rm B}}$, where $S$ is the entropy of the system and the reservoir. We find the entropy $S=k_{\rm{B}}\ln{W}$ from the number of microstates $W$. 
To obtain an approximate expression for the particle number fluctuations, we do not distinguish between bound and unbound particles, in contrast to section~\ref{secSM:ContinuousInteractions}, such that the number of particle arrangements is given by
\begin{equation}
    W =\prod_j \frac{\Omega!}{(\Omega-N_j)!N_j!},
\end{equation}
resulting in the probability of a given set of particle numbers following
\begin{equation}\label{eq:distOffluct1}
    P(\{N_{j}\}) \propto \prod_j e^{\Omega(\ln{\Omega}-1)-(\Omega-N_j)(\ln{(\Omega-N_j)}-1)-N_j(\ln{N_j}-1)},
\end{equation}
in which we have applied the Stirling approximation.
By applying the saddle-point approximation about the most likely values $\bar{N}_j$~\cite{Touchette2009}, we get
\begin{equation}\label{eqSM:distPN}
    \ln{P(\{N_{j}\})} \asymp \sum_j 
    \frac{1}{2}\left(
        \frac{\partial^2 \ln{P}}{\partial N_{j}^2}\bigg|_{N_{j}=\bar{N}_{j}}\, (N_{j}-\bar{N}_{j})^2 \right),
\end{equation}
in which the first-order derivatives vanish, as well as the cross-term $\partial \ln{P(\{N_{j}\})}/(\partial N_{j}\partial N_{k\neq j})$. Note that this approximation treats the particle number fluctuations in each box as independent, since Eq.~\eqref{eqSM:distPN} implies $P(\{N_j\}) = \prod_j P(N_j)$.
By substituting \eqref{eq:distOffluct1} into \eqref{eqSM:distPN} and identifying the discrete densities $\rho_j = N_j/a$ and $\bar\rho_j = \bar N_j/a$, and $\Omega = a/d_0^2$, we evaluate the derivative as
\begin{equation}\label{eq:lim_gauss}
   \sum_j \frac{\partial^2 \ln{P}}{\partial N_{j}^2}\bigg|_{N_j = \bar{N}_j} \frac{(N_j - \bar{N}_j)^2}{2}
    = -\sum_j
    \frac{\Omega}{(\Omega-N_j)N_j}\bigg|_{N_j = \bar{N}_j} \frac{(N_j - \bar{N}_j)^2}{2}  = -\sum_j \frac{a}{(1-d_0^2\bar\rho_j)\bar\rho_j}\frac{(\rho_j - \bar\rho_j)^2}{2}.
\end{equation}
From Eqs.~\ref{eqSM:distPN}-\ref{eq:lim_gauss}, and taking into account the dependence of the mean $\bar{\rho}_j$ on the binding energy, we obtain the following expression for the conditional probability of a given particle density in box $j$
\begin{equation}\label{eqSM:localConditionalProb}
    P\big(\rho_{j}|\{\epe_{k}\}\big) \propto 
    \exp\left[
        -\frac{\big(\rho_{j}-\bar{\rho}_{j}(\{\epe_{k}\})\big)^2}{2\sigma_\rho^2\big(\{\epe_{k}\}\big)}
    \right]
\end{equation}
with variance $\sigma_\rho^2\big(\{\epe_{k}\}\big) = \big[1-d_0^2\bar\rho_j(\{\epe_{k}\})]\bar\rho_j(\{\epe_{k}\}) /a$ and mean $\bar\rho_j(\{\epe_{k}\})$ [\citeMTEqEquilibriumFermiDiracSolution{}], where $\{\epe_k\}$ denotes the vector of input energy values. 
The mean depends on the full set of realizations in all boxes $\{\epe_k\}$ through the constraint of particle conservation, \citeMTEqNormalization{}, which in the discrete limit takes the form
\begin{equation}
    \sum_{j} \bar\rho_{j}(\{\epe_{k}\})= \frac{B}{\bar{d}^{2}}.
\end{equation}
Note that our approximation of the variance $\sigma_\rho^2$ neglects any explicit dependence of the particle fluctuations on the binding energy.

\paragraph{\label{secSM:MetropolisSampling}Metropolis sampling}

\begin{figure}[!t]
\includegraphics[width=\textwidth]{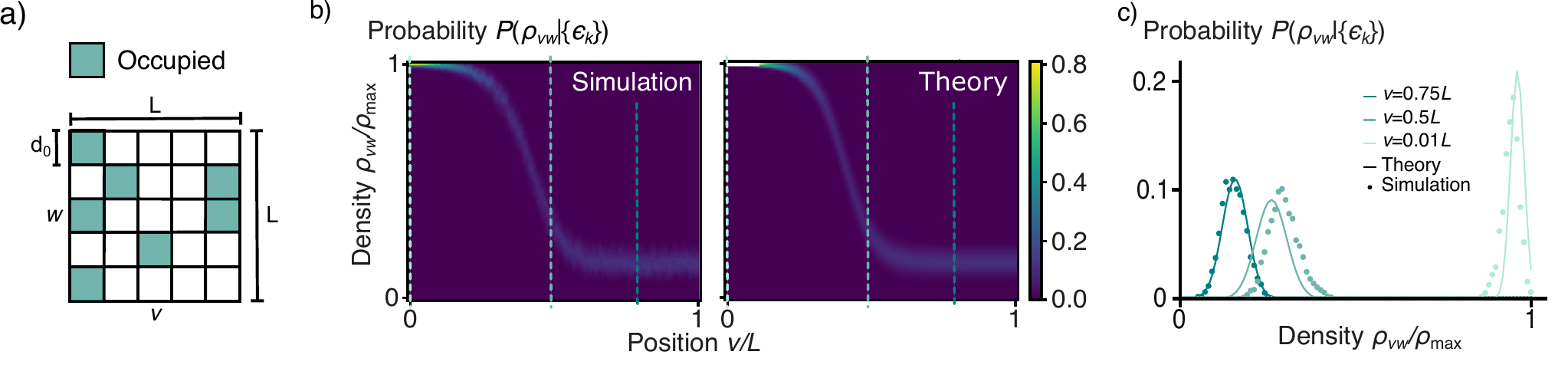}
\caption{\label{figSM:localConditionalProbability} The conditional probability $P\big(\rho_{j}|\{\epe_{k}\}\big)$ is well approximated by a Gaussian distribution Eq.~\ref{eqSM:localConditionalProb}. a) Sketch of lattice setup for Metropolis-Hastings simulations. b) Comparisons of the estimates for $P\big(\rho_{vw}|\{\epe_{vw}\}\big)$ from simulations (left) and the Gaussian approximation (right) for $(d_0/\bar{d})^2= 0.5$ and input energy $\epe_{vw} = (2 v/10 -10) k_{\rm B} T$. Colored dashed lines mark positions plotted in (c), and white shading indicates values above $0.8$. c) The conditional distribution estimates at various positions show that the Gaussian expression approximates the simulated system well for mean densities at the two plateaus, while a slight shift appears in at intermediate positions. 
}
\end{figure}

To investigate the validity of our approximation for $P\big(\rho_{j}|\{\epe_{k}\}\big)$, we numerically estimate the conditional probability Eq.~\ref{eqSM:localConditionalProb} by performing Metropolis-Hastings sampling of the distribution of particles given a fixed set of input energies 
(code provided in~\cite{ElliottRepoMetropolisSims2025}). On a 100-by-100 lattice, where the box index $j$ can be identified with a lattice coordinate $(v,w)$, we fix the input energy to a linear profile $\epsilon_{vw} = (2 v/10 -10) k_{\rm B} T$, and sample the particle distributions for 1000, 5000, and 9000 particles, which in terms of $(d_0/\bar{d})^2$ correspond to the values $0.1$, $0.5$, and $0.9$ respectively.
We discretize the possible binding energies into 241 evenly-spaced values between $-12k_{\rm{B}}T$ and $12k_{\rm{B}}T$, and add a Gaussian noise with standard deviation $1\ k_{\rm B} T$.
The sampled particle densities, resolved along the $v$-axis, are shown together with the analytical density profiles in \citeMTFigOne{d}.
We compare the sampled local conditional probability for the $(d_0/\bar{d})^2= 0.5$ case  with Eq.~\eqref{eqSM:localConditionalProb} (Fig.~\ref{figSM:localConditionalProbability}), and find that, as expected, the high-density region has a lower channel noise than the low-density region where the number of possible particle configurations over the available distinguishable energy levels is larger. We observe a good agreement between the simulations and the Gaussian approximation Eq.~\eqref{eqSM:localConditionalProb}, especially for low and high densities. Although we find that~--~in the biologically-relevant fluctuation regime~--~the channel noise is dominated by fluctuations in the particle arrangements, we expect that a large increase in the noise of the binding energies would correspondingly lower information transmission. 

\paragraph{Joint probability of particle density and binding energy}
The joint probability $P(\{\rho_j\},\{\epe_k\})=P(\{\rho_j\}|\{\epe_k\})P(\{\epe_k\})$ of observing a particular set of density values $\{\rho_j\}$ together with an input energy vector $\{\epe_k\}$, is the product of the conditional probability $P(\{\rho_j\}|\{\epe_k\})$ and the marginal probability of the input $P(\{\epe_k\})$. To investigate how well the particle patterns distinguish between different inputs, we define the support of the distribution $P(\{\epe_k\})$ to include all possible inputs that the surface could receive, i.e. to keep the framework general, we assume the elements of the  vector $\{\epe_k\}$ to be independently and uniformly distributed as $P(\epe_k) = 1/n$, where $n$ is the number of possible energy levels so that
\begin{equation}
    P(\{\epe_k\})=\prod_k^B P(\epe_k) = n^{-B}.
\end{equation}
Similarly, the conditional probability $P(\{\rho_j\}|\{\epe_k\})$ of a density profile $\{\rho_j\}$ given an input energy profile $\{\epe_k\}$ can be calculated as the product of the local conditional probability $P(\rho_j|\{\epe_k\})$ -- given by Eq.~\eqref{eqSM:localConditionalProb} -- over the whole space
\begin{equation}
    P(\{\rho_j\}|\{\epe_k\}) = \prod_j^B P(\rho_j|\{\epe_k\}),
\end{equation}
such that the joint probability distribution takes the form
\begin{equation}
P(\{\rho_j\},\{\epe_k\}) = \frac{1}{n^B} \prod_j^B P(\rho_j|\{\epe_k\}).\label{eqSM:JointProbAsProdExpanded}
\end{equation}
For completeness we remark that in the continuum limit $a\to0$ and $B\to\infty$ Eq.~\eqref{eqSM:JointProbAsProdExpanded} yields a functional
$$
    P[\rho(\cdot),\epe(\cdot)] \propto g_{\rm DoS} \exp\left\{
        \int_A dA\, \ln P\left[\rho(\mathbf{r})|\epe(\cdot)\right]
    \right\}.
$$
where $g_{\rm DoS}$ is the constant density of energy states, and the integral is taken over the area $A$ of the domain. 

\subsubsection{\label{secSM:MutualInfoDerivation}Mutual information between energy and particle densities}
The information gained about the input field by observing the output density field is given by the mutual information~\cite[Chapter 2]{TM_Cover1991-pa} 
\begin{equation}\label{eqSM:MutualInformationDefinition}
    I = \sum_{\{\epe_k\}}\sum_{\{\rho_j\}} P(\{\rho_j\},\{\epe_k\})\ln{\frac{P(\{\rho_j\},\{\epe_k\})}{P(\{\rho_j\})P(\{\epe_k\})}}.
\end{equation}
Here $\sum_{\{\epe_k\}} = \sum_{\epe_1}\sum_{\epe_2}...\sum_{\epe_{\rm{B}}}$ and $\sum_{\{\rho_j\}} = \sum_{\rho_1}\sum_{\rho_2}...\sum_{\rho_{\rm{B}}}$ denote sums over all possible realizations of the input and output random vectors respectively, and the marginal probability $P(\{\rho_j\})=\sum_{\{\epe_k\}}P(\{\rho_j\},\{\epe_k\})$, can be found by summing the joint probability over all combinations of possible input vectors. 

This equation was evaluated for a system with coarse-graining area $a = 10d_0^2$ and $B=5$. We constructed all possible input vectors given a minimum interaction energy of $-10 k_{\rm B}T$ (a typical chemical-interaction potential between biological proteins), an upper cutoff $+10 k_{\rm B} T$, and discretizing the possible levels to integer values in units of $2 k_{\rm B} T$, such that each element $\epe_j$ was sampled from the set $\bm{\mathcal{E}} = \{-10k_{\rm{B}}T,-8k_{\rm{B}}T,...,10k_{\rm{B}}T\}$. We  discretized the possible density values according to the number of particles in the coarse-graining area $a$, such that the value of each density element is sampled from the set $\{0,d_0^2/a,2 d_0^2/a,...,1\}$. Calculating the corresponding value of $P(\rho_m|\{\epe_k\})$---as given by Eq.~\eqref{eqSM:localConditionalProb}---for each possible value of $\rho_m$, and using Eq.~\eqref{eqSM:JointProbAsProdExpanded}, then permitted the evaluation of Eq.~\eqref{eqSM:MutualInformationDefinition}. 

\subsubsection{\label{secSM:influenceOfBindingSiteDensity}Influence of binding site density and heterogeneity on information transmission}

\begin{figure}[!t]
\includegraphics[width=0.5\textwidth]{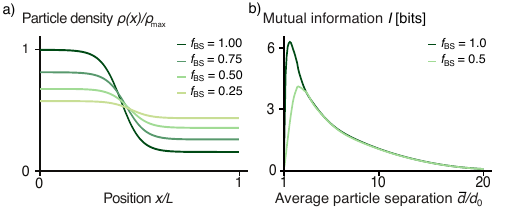}
\caption{\label{figSM:fBScomparison} Reducing the \emph{uniform} binding site density reduces the amount of information read out via the particle density. a) Equilibrium particle density profiles forming in response to a linear input energy profile $\epe(x)=(20 x/L -10)k_{\rm{b}}T$ over a domain of length $L$ show that reducing the area fraction $f_{\rm BS}$ lowers the gain of the filter. b) The reduced gain leads to a lower mutual information for uniform binding sites with $f_{\rm BS}=0.5$ compared to $f_{\rm BS}=1$.
}
\end{figure}

Equation \eqref{eqSM:SmoluchowskiEquilibriumSolution} from section \ref{secSM:ContinuousInteractions} is valid under the assumption that the interaction sites, or binding sites, are uniformly distributed with the same density as the maximum density of the particles, such that each particle can exist in either a bound or unbound state. 
Relaxing this assumption, and introducing a uniform area fraction of binding sites $f_{\rm{BS}} = \rho_{\text{ext}}/\rho_{\text{max}}$, Eqs.~\eqref{eqSM:LatticeRhoUnbound} and \eqref{eqSM:LatticeRhoBound} yield the total equilibrium particle density
 \begin{equation}
     \rho(x) = \frac{\rho_{\rm{max}}}{1+e^{\alpha + \beta\epsilon_{\rm{u}}}}\bigg(1 + f_{\rm{BS}}\frac{e^{\alpha + \beta\epsilon_{\rm{u}}}}{e^{\beta\epsilon(x)}+1+e^{\alpha + \beta\epsilon_{\rm{u}}+\beta\epsilon (x)}}\bigg),
 \end{equation}
and we find that a reduction in the uniform binding site density results in a smaller separation between the maximum and minimum densities, and a smaller gain, leading to reduced information transmission (Fig.~\ref{figSM:fBScomparison}). 
However, cellular structures at membranes have typically irregular shapes~--~membrane-bound proteins, for example, often interact with slender cytoskeletal filaments. Considering a non-uniform area-fraction $\phi = \zeta Ld_0/A = \zeta d_0/L$ for binding sites arranged along $\zeta$ parallel lines of length $L$ and with the width of a single particle $d_0$, we divide the membrane into parallel strips of width $d_0$, and consider the set of line-densities $\rho_{\rm{1D}}$ obtained by taking the integral of the 2D particle density over the width of each strip. 
The constant $l$ defined by particle conservation across the whole surface is then determined through
\begin{equation}\label{eqSM:lfrom1D}
    \frac{\phi}{L d_0} \left(\int_0^L dx \rho_{\rm{1D}}(x) \right) +  \frac{1-\phi}{l^2 + d_0^2}   = \frac{1}{\bar{d}^2},
\end{equation}
in which $x$ parameterizes the coordinates along the $\zeta$ lines, within which $\epsilon = \epsilon(x)$, and the interaction energy is infinite within the remaining region. 
We compare the information transmission capacity of such nonuniform area fraction systems to the uniform case by evaluating~\eqref{eqSM:MutualInformationDefinition} along strips discretized into $B=5$ segments of length $10d_0$, such that the coarse-graining area remains $a=10d_0^2$, and use the same discretization of energy and density values as before. We find that for filament-like input binding structures a smaller number of particles are required for effective information transfer compared to uniform surfaces, since the membrane regions without binding sites act as a particle bath for the regions with binding sites. Although we consider parallel lines for analytical convenience, our findings hold for non-overlapping lines of any orientation. 

\subsection{\label{secSM:ParticleLikeStructures}Estimating the effective size and average spacing of particle-like structures at cellular membranes}
The properties of the sigmoidal mapping that arises between binding energy and particle distribution due to repulsive interactions are determined by the gain and threshold parameters, which are in turn set by the effective particle size $d_0$ and their average spacing $\bar{d}$.
Table~\ref{tabSM:biologicalExamples} shows measured or estimated values for these parameters for a range of protein complexes and macromolecular complexes, together with the corresponding sources. We aimed to compile an exhaustive list of proteins and subcellular structures for which this information is currently available~--~no selection was made beyond restricting the search to \emph{membrane-bound} particle-like structures. 
\citeMTFigTwo{c} shows these compiled parameter values together with the optimal information-processing regimes for various binding site area fractions, including an actin-like area fraction of binding sites $\phi = 10^{-2}$ (dark orange). 
This estimate is based on measurements of cortical actin filament densities obtained from high-resolution cryo-ET tomograms in fibroblasts [Fig.~2c in~\cite{Lembo2023}].

\newcolumntype{L}{>{\centering\arraybackslash}p{0.15\linewidth}}

\begin{table}[]
    \centering
    \begin{tabular}{l |l |l |l}
         Protein structure  & Average density $1/\bar{d}^2$ [\SI{}{\per\micro\meter\squared}] & Average separation $\bar{d}$ [\SI{}{\nano\meter}]& Interaction length $d_0$  [\SI{}{\nano\meter}] \\
         \hline
         Integrin LFA-1 & 120 \cite{Robert2021} & 91 & 10 \cite{Ekerdt2013}\\
         Integrin VLA-4 & 75 \cite{Robert2021} & 115 & 10 \cite{Ekerdt2013}  \\
         E-Cadherin & 630 
         \cite{TruongQuang2013}& 40  & 7 \cite{TruongQuang2013}  \\
         TCR 
         & 42.4 ~\cite[BNID 103567]{Janeway2001,Jiang2020}\textsuperscript{**} & 154 & 10 \cite{Swamy2008} \textsuperscript{*} \\
         ICAM-1 
         & 900 \cite{Ma2022}& 33 &  3 \cite{Bui2020} \textsuperscript{*} \\
         Connexons & 5000 \cite{KurzIsler1992} & 14 & 9.5 \cite{Goodenough1970}  \\
         Piezo-1 & 0.52 \cite{Kuntze2020} & 1387 & 24 \cite{Mulhall2023}  \\
         Caveolae & 0.04  \cite{Rizzo2003} & 5000 & 75 \cite{Thomsen2002} \\
         VDAC & 5000 \cite{Mannella2021} & 14 & 3.8 \cite{Hiller2010} \\
         ERMES & 1200 \cite{Wozny2023} & 29 & 15 \cite{Wozny2023} \\
         Lipid GM1 & 17000~\cite{Lyu2024}\textsuperscript{**} & 8 & 3  \cite{Mojumdar2019}  \\
         AChR & 55 \cite{Evans1987,McMahon1994}\textsuperscript{**} & 135 & 10 \cite{Lo1982,Geng2009} \\
         Fc$\gamma$-1 & 11 \cite{Kerntke2020,Wacleche2018} \textsuperscript{**} & 302 & 2.7  \cite[BNID 117058]{Saji1999}\textsuperscript{*} \\
         EGFR & 636 \cite{Zhang2015} & 40 & 17 \cite{Abulrob2010}\\
         Focal adhesions 
         & 0.14 \cite{Horzum2014,TruongVo2017}\textsuperscript{**} & 2673 & 700 \cite{Horzum2014} \\
         NPCs & - & 440 \textsuperscript{***}  & 260  \textsuperscript{***}\\
    \end{tabular}
    \caption{Reported parameters for particle-like structures at cellular and subcellular membrane surfaces. }
    \small\textsuperscript{*} $d_0 = 2r_{\rm{min}}$ with $r_{\rm{min}}$ calculated from molecular mass value using \cite{Erickson2009}.
    \small\textsuperscript{**} Calculated as number of protein complexes per area of a spherical cell of a given diameter.
    \small\textsuperscript{***} Measured in our study.
    \label{tabSM:biologicalExamples}
\end{table}

\subsection{\label{secSM:DistNPC}Distribution of nuclear pore complexes in the nuclear envelope of the ichthyosporean \emph{S. arctica}}

To test our predictions for the mapping between interaction energy fields and particle density profiles, we measured the distribution of nuclear pore complexes (NPCs) in the nuclei of the multinucleate unicellular organism \textit{Sphaeroforma arctica}~\cite{Dudin2019,Olivetta2023}. NPCs are embedded within the double-bilayer membrane of the nuclear envelope (NE), and associate with microtubules (MTs) that emanate from microtubule organising centers (MTOCs) at the nuclear poles in a regular, radial fashion in this system (\citeMTFigThree{a},\cite{Shah2024}). 

\subsubsection{\label{secSM:ExpMethods}Experimental methods}
\paragraph{Culture conditions}
Sphaeroforma arctica was cultured at \SI{17}{\celsius} in Marine broth (Difco, 37.4 g l-1) and synchronized as previously described~\cite{Dudin2019,Ondracka2018}. Briefly, the cultures were synchronized in 1/16 Marine broth diluted in artificial seawater (Instant Ocean, 37 g l-1). Cultures were diluted 1:100 in 1/16 Marine broth and grown for 3 days to obtain synchronised cultures. The synchronised cultures were inoculated 1:50 in fresh 10ml Marine broth in 25ml cell culture flasks. To obtain the 8-32 nuclear stage, cells were fixed around 28 - 30h after inoculation. The cell culture flasks were scraped and the suspension was added to 15 ml Falcon tubes and collected by centrifugation at 500rpm for 5 min. The supernatant was removed and cells were transferred to 1.5 ml microfuge tubes and fixative was added for 30 min. The cells were fixed with 4\% formaldehyde in 250 mM sorbitol in 1× phosphate buffer saline (PBS), washed twice with PBS and resuspended in 20-30 $\mu$l of PBS.

\paragraph{\label{secSM:UExM}Ultrastructure expansion microscopy}
Ultrastructure expansion microscopy (U-ExM) was performed as previously described~\cite{Gambarotto2018, Shah2024}. Briefly, the fixed cells were anchored in 1\% acrylamide/ 0.7\% formaldehyde solution overnight at \SI{37}{\celsius}. The anchored cells were attached to 6 mm poly-l-lysine-coated coverslips for 30 mins - 1 h. The coverslips with cells were inverted on a \SI{9}{\micro\litre} drop of monomer solution (19\% (wt/wt) sodium acrylate (7446-81-3), 10\% (wt/wt) acrylamide (Sigma-Aldrich A4058), 0.1\% (wt/wt) N,N'-methylenebisacrylamide (Sigma-Aldrich M1533) in PBS). After a 5 min incubation on ice, gels were allowed to polymerize for 1 h at \SI{37}{\celsius} in a moist chamber. For denaturation, coverslips along with gels were transferred to the microfuge tubes with 1 ml pre-heated denaturation buffer (50 mM Tris pH 9.0, 200 mM NaCl, 200 mM SDS, pH to 9.0) and incubated at \SI{95}{\celsius} for \SI{1.5}{hours}. Following denaturation, the gels were expanded with 3 water exchanges as previously described. Post expansion, the gel diameter was measured and used to determine the expansion factor.

\paragraph{Immunostaining and imaging}
For staining, the gels were re-incubated in 1x PBS to shrink them. This was followed by blocking in 3\% bovine serum albumin (BSA) in PBST (1× PBS with 0.1\% Tween20) at room temperature for 30min – 1h. The gels were then incubated for 5 h at 37 °C in primary antibody [Tubulin - AA344 and AA345 (ABCD antibodies) and NPC - MAb414 (Biolegend 902901)] prepared at 1:500 in blocking solution. This was followed by three washes for 10 min at room temperature in PBST and addition of the secondary antibody [Goat anti-mouse secondary antibody, Alexa Fluor 488 (Thermo A-11001), Goat anti-guinea pig secondary antibody, Alexa Fluor 647 (Thermo A-21450)] at 1:500 final concentration in blocking solution. Incubation was done at 4 °C overnight. Next gels were stained with protein and lipid pan-labelling dyes, Dylight 405 (ThermoFischer, 46400) and BODIPY TR ceramide (ThermoFischer D7540, 2 mM stock in dimethylsulfoxide), at 1:500 dilution in 1× PBS for 2 – 5 h. The gels were then washed and re-expanded prior to imaging. Ibidi chamber slides (two-well, Ibidi 80286) were pre-coated with poly-l-lysine. Gels were cut to an appropriate size to fit the Ibidi chambers and added onto the wells. The gels were overlaid with water to prevent drying or shrinkage during imaging. The gels were imaged using the Zeiss LSM 880 with the Airy fast mode using a Plan-Apochromat 63×/1.4 Oil DIC M27. The images were processed with the default 3D airyscan processing on the ZEN software prior to further analysis.

\subsubsection{\label{secSM:ImageAnalysisMethods}Image analysis and parameter estimation}

\begin{figure}[!t]
\includegraphics[width=\textwidth]{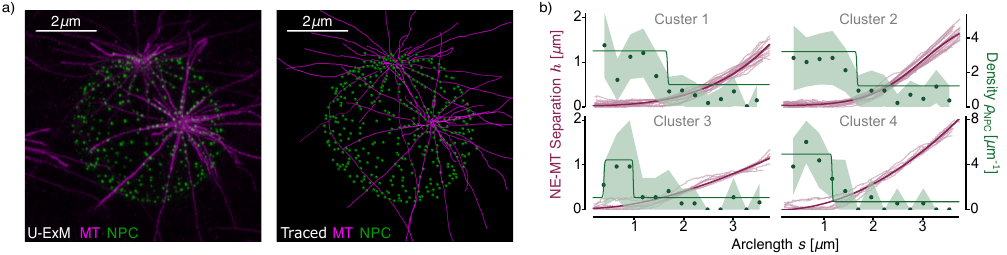}
\caption{\label{figSM:FitResultsAllClusters} Microtubules clustered according to their NE-MT separation profiles give rise to sigmoidal NPC line-densities. a) [Right] Three-dimensional rendering of microtubule traces (magenta) and NPC coordinates (green) measured from an ultrastructure expansion microscopy image [Left: 3D rendering of the original image stack] of an \emph{S. arctica} nucleus. b) Independent fits of \citeMTEqEquilibriumFermiDiracSolution{} (solid green line) to the measured NPC line-density (green dots, shaded area: 95\% confidence interval) for the four MT clusters not presented in the main text, using the differentiable representation of the separation $\hat{h}$ (dark purple) between the nuclear envelope (NE) and the microtubule (MT) filaments (individual tracks in light purple). Each fit resulted in an estimate of the separation $d_0$ and the effective spring constant $\lambda$.
}
\end{figure}

We estimated i) the distance between the MTs and the NE, and ii) the NPC coordinates in the NE as detailed in our protocol~\cite{ElliottRepoImageAnalysis2025}. 
Briefly, we traced MT filaments from their point of origin at the MTOC (identified manually) using Sato and Frangi tubeness filters, skimage.morphology's skeletonize\_3d function, and a custom algorithm for connecting branches into their most likely filament combinations, which attaches branch ends that are close to each other and aligned in the same direction, allowing for the correction of patchy MT labelling [Fig.~\ref{figSM:FitResultsAllClusters}(a)]. Segmenting the nuclei using a Hessian filter on the membrane channel, we then calculated the shortest distance from the NE to coordinates along each filament, neglecting all filaments from nuclei with major segmentation or tracing errors and applying selection criteria on the total length of the filaments ($>$\SI{1.1}{\micro\metre}) and their proximity to the NE ($<$\SI{0.3}{\micro\metre} over the nearest \SI{1.1}{\micro\meter} to the MTOC). This analysis allowed the estimation of the average area-fraction of NE-proximal MT filaments on the nuclear envelope, $\phi=\SI{0.21(0.03)}{}$ (used in \citeMTFigThree{e}).
Applying a further selection criteria on the total length of the profiles ($>$\SI{3.4}{\micro\metre}, resulting in 110 individual height profiles), we clustered the profiles into nine groups using agglomerative clustering, considering a region of interest (ROI) defined as the segment between \SI{0.2}{\micro\meter} and \SI{3.8}{\micro\meter} from the MTOC. To this end, we computed the dissimilarity matrix for the dataset, in which the dissimilarity score for a pair of profiles is given by 
\begin{equation}
  \text{Dissimilarity Score} =  \frac{ \sum_k^{\text{Min}(N_{pi},N_{pj})} |(h_{i,k} - h_{j,k}) |}{\text{Min}(N_{pi},N_{pj})}
\end{equation}
where $k$ indexes the pixels and $N_{pi}$ denotes the total length of the $i$th profile $h_i$.
For the subsequent parameter fitting, we excluded clusters containing fewer than five filaments or which had a dissimilarity score larger than 4000. 

We identified the coordinates of approximately all NPCs on the NE surface by thresholding the NPC channel [Fig.~\ref{figSM:FitResultsAllClusters}(a)], from which we estimated their average spacing over all the nuclei as $\bar{d} = \SI{440(50)}{\nano\meter}$, and over the nuclei that contribute filaments to each cluster (weighted by the number of filaments contributed) as given in~\ref{tabSM:FittedData}. To compute the 1D line-densities along the MT filaments, we counted NPCs within a \SI{160}{\nano\meter} interval around each filament's shortest-distance projection line.

\paragraph{\label{secSM:NPC-MTinteractionEnergy} Parameter fitting}
We approximated the interaction energy between an NPC and its binding site on a nearby MT by a chemical interaction energy $\epsilon_{\rm{c}} = 25 k_{\rm{B}}T$ \cite{Li2016} and an elastic Hookean contribution, where we represented the combined elastic contributions arising from local deformations of the nuclear envelope, stretching of the linker complex, and deflection of the microtubule with the effective elastic interaction term $\lambda (\hat{h}(s)-h_0)^2/2$, in which we included the resting length of the effective spring $h_0 \approx \SI{84(15)}{\nano\meter}$, estimated by measuring the minimum NPC-MT separation distance within the high-density regions of our ROIs from electron microscopy images (Table~\ref{tabSM:lestimatesfromEM},\cite{Shah2024}). We estimated the effective spring constant $\lambda$ and the effective particle size $d_0$ for the MT-NPC system by minimizing the objective function
\begin{equation}\label{eq:CostFunctForFit}
    J(\rho,\hat{\rho};\lambda,d_0) = \sum_i (\hat{\rho}_i - \rho(\hat{s}_i;\lambda,d_0))^2,
\end{equation}
in which $\hat{\rho}$ denotes the measured density at the arc-length position $\hat{s}$, and $\rho$ was computed by evaluating \citeMTEqEquilibriumFermiDiracSolution{} at $\hat{s}_i$. We fitted our parameters independently for each cluster, performing a rational-exponential reparameterization of the measured height profiles to obtain a differentiable representation of the data $\hat{h}$ [Fig.~\ref{figSM:FitResultsAllClusters}(b),~\citeMTFigThree{c}], and calculating the average NPC separation $\bar{d}$ and average number of filaments per nuclear area $\zeta/A$ independently for each cluster. The resultant estimates from each cluster are summarized, along with the corresponding $\bar{d}$ and $\zeta/A$ values, in Table~\ref{tabSM:FittedData}. While the effective particle sizes were well-constrained across the different MT profiles, the cost functions featured shallow regions extended in the direction of the spring constant parameter in some cases, showing that the fit quality has limited sensitivity to variations in this parameter within the explored range. Nonetheless, we obtained consistent estimates for the spring constants from the five distinct MT datasets.
Note that using the effective elastic interaction to account for the binding interactions between NPCs and MTs neglects any changes in $h$ due to binding of the nuclear pore complexes beyond what is accounted for through the effective spring constant $\lambda$, i.e. we assume that the MT network maintains an overall fixed morphology. 

\begin{table}[]
    \centering
    \begin{tabular}{c|c}
          Effective spring rest length $h_0$ [\SI{}{\nano\meter}] & Effective particle size $d_0$ [\SI{}{\nano\meter}]\\
          \hline
          71 & 266\\
          90 & 236\\
          89 & 222\\
          110 & 183\\
          75 & 156\\
          80 & 161\\
          87 & 163\\
          81 & 141\\
          71 & 224\\
          85 & 249\\
          85 & 235\\
          88 & 186\\
          65 & 228\\
          70 & 178\\
          94 & 187\\
          83 & 135\\
          85 & 151\\
          88 & 169\\
          91 & 177\\
          83 & 229\\
          110 & 200\\
          91 & 162\\
          108 & 190\\
          57 & 170\\
          66 & 184\\
    \end{tabular}
    \caption{Parameter estimates from electron tomograms of \emph{Sph. arctica} nuclei \cite{Shah2024}. 
    }
    \label{tabSM:lestimatesfromEM}
\end{table}

To assess the validity of our fit, we used electron tomograms from~\cite{Shah2024} to independently measure the effective NPC particle size $d_0$. Using Fiji~\cite{Schindelin2012}, we measured the minimum distance between NPCs in the high-density region of our ROI, obtaining $d_0\approx \SI{190 (40)}{\nano\meter}$ (mean $\pm$ standard deviation, Table~\ref{tabSM:lestimatesfromEM}), close to our fitted estimates [\citeMTFigThree{d}].

The original and post-processed image data is available at ~\cite[DOI: 10.6019/S-BIAD2081]{ImageDataset}.

\begin{table}[]
    \centering
    \begin{tabular}{c|c|c|c|c|c|c}
          Cluster & \begin{tabular}{@{}c@{}}Average \\ particle\\ separation \\ $\bar{d}~[\SI{}{\nano\meter}]$\end{tabular} & \begin{tabular}{@{}c@{}}Average number \\ of filaments per\\ nuclear area\\ $\zeta/A$~[\SI{}{\per\micro\meter\squared}]\end{tabular}  & \begin{tabular}{@{}c@{}}Fitted effective \\particle size \\$d_0$~[\SI{}{\nano\meter}]\end{tabular}  
          &
          \begin{tabular}{@{}c@{}}95\% CI \\$d_0$~[\SI{}{\nano\meter}]\end{tabular} 
         & \begin{tabular}{@{}c@{}}Fitted effective \\spring constant \\ $\lambda$~[\SI{}{\pico\newton\per\nano\meter}] \end{tabular}
          & \begin{tabular}{@{}c@{}}95\% CI  \\$\lambda$~[\SI{}{\pico\newton\per\nano\meter}]\end{tabular} 
          \\
         \hline
         1 & 443 & 0.27 & 310&\{350,
   260\}&0.036,   &\{0.029, $>$0.109\} \\
         2 & 451 & 0.27 & 310&\{340, 280\}&0.024 &\{0.020, $>$0.109\} \\
         3 & 437 & 0.30 & 230&\{260, 200\}&0.090  &\{0.084, $>$0.109\} \\
         4 & 452 & 0.30 & 200&\{240, 180\}&0.030&\{0.028, 0.032\} \\
         5 & 466 & 0.28 & 240&\{250, 220\}&0.026&\{0.025, 0.059\} 
    \end{tabular}
    \caption{Parameter estimates obtained by measurement or by minimizing the cost function Eq.~\ref{eq:CostFunctForFit} for five different proximity profiles of microtubules.}
    \label{tabSM:FittedData}
\end{table}


\end{document}